\documentclass[10pt,twocolumn,letterpaper]{article}

\usepackage[pagenumbers]{iccv} 

\usepackage{graphicx}
\graphicspath{{./figures/}}
\usepackage{amsmath}
\usepackage{amssymb}
\usepackage{booktabs}
\usepackage{multirow}
\usepackage{booktabs}
\usepackage{amsfonts}
\usepackage{float}
\usepackage{makecell}
\usepackage{flushend, cuted}
\usepackage{array}
\usepackage[ruled,linesnumbered]{algorithm2e}
\usepackage[accsupp]{axessibility}  

\definecolor{iccvblue}{rgb}{0.21,0.49,0.74}
\usepackage[pagebackref,breaklinks,colorlinks,allcolors=iccvblue]{hyperref}

\def\paperID{2900} 
\def\confName{ICCV}
\def\confYear{2025}

\title{Generalized and Efficient 2D Gaussian Splatting for \\ Arbitrary-scale Super-Resolution}

\author{Du Chen$^{1,2}$\thanks{Equal contribution.}  \quad
	Liyi Chen$^{1}$\footnotemark[1]  
	\quad Zhengqiang Zhang$^{1,2}$  \quad Lei Zhang$^{1,2}$\thanks{Corresponding author. This work is supported by the PolyU-OPPO Joint Innovative Research Center.} \\
	$^1$The Hong Kong Polytechnic University \quad $^2$OPPO Research Institute\\
	{\tt\small \{csdud.chen, liyi0308.chen, zhengqiang.zhang\}@connect.polyu.hk},
	{\tt\small cslzhang@comp.polyu.edu.hk}
}

\begin{document}
	\maketitle

	\begin{abstract}
		
		Implicit Neural Representations (INR) have been successfully employed for Arbitrary-scale Super-Resolution (ASR). However, INR-based models need to query the multi-layer perceptron module numerous times and render a pixel in each query, resulting in insufficient representation capability and low computational efficiency. Recently, Gaussian Splatting (GS) has shown its advantages over INR in both visual quality and rendering speed in 3D tasks, which motivates us to explore whether GS can be employed for the ASR task. However, directly applying GS to ASR is exceptionally challenging because the original GS is an optimization-based method through overfitting each single scene, while in ASR we aim to learn a single model that can generalize to different images and scaling factors. We overcome these challenges by developing two novel techniques. Firstly, to generalize GS for ASR, we elaborately design an architecture to predict the corresponding image-conditioned Gaussians of the input low-resolution image in a feed-forward manner. Each Gaussian can fit the shape and direction of an area of complex textures, showing powerful representation capability. Secondly, we implement an efficient differentiable 2D GPU/CUDA-based scale-aware rasterization to render super-resolved images by sampling discrete RGB values from the predicted continuous Gaussians. Via end-to-end training, our optimized network, namely GSASR, can perform ASR for any image and unseen scaling factors. Extensive experiments validate the effectiveness of our proposed method. The code and models are available at \url{https://github.com/ChrisDud0257/GSASR}.
		\vspace{5pt}
	\end{abstract}

	\begin{figure}[t!]
		\centering
		\includegraphics[width=1\linewidth]{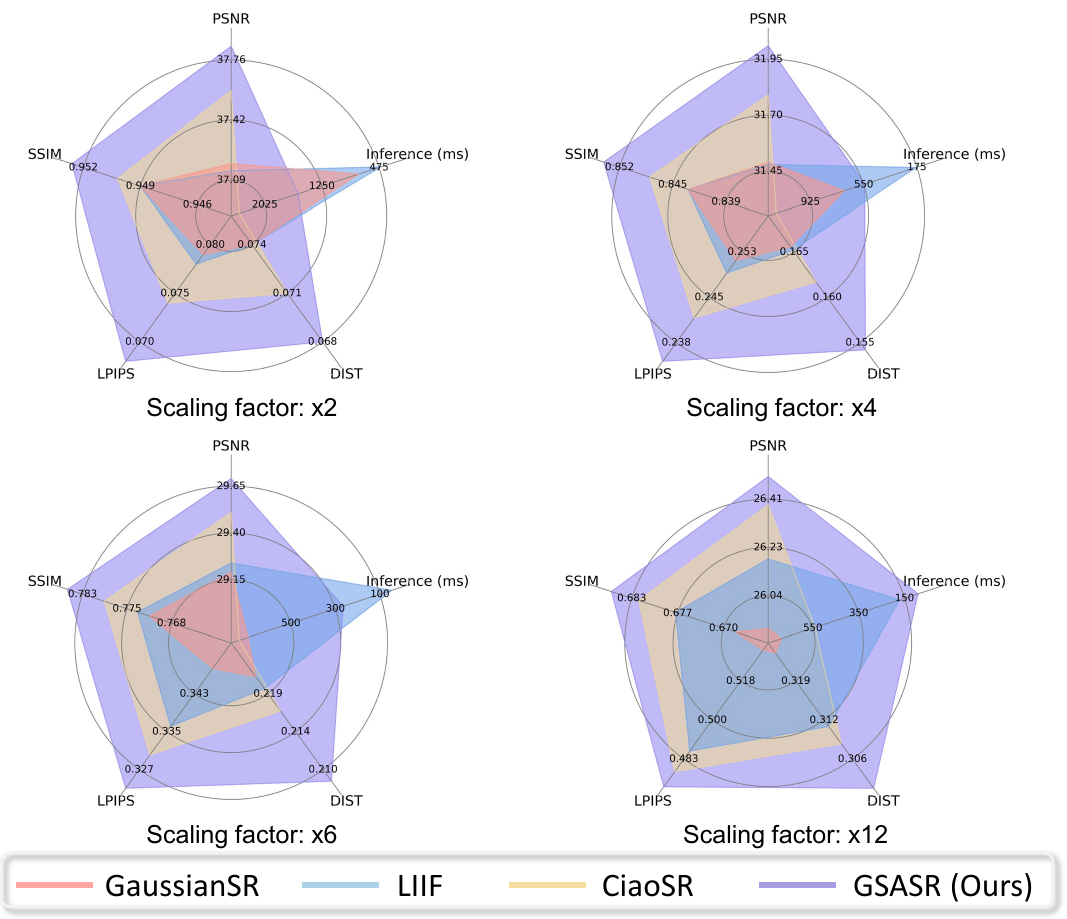}
		\caption{\footnotesize{Comparison between GSASR and representative ASR methods on DIV2K. GSASR outperforms previous methods in visual quality with competitive inference time across both in-distribution ($\times$2, $\times$4) and out-of-distribution ($\times$6, $\times$12) scaling factors.
		}}
		\label{fig:teaser}
		\vspace{-0.5cm}
	\end{figure}
	
	\vspace{-0.5cm}
	\section{Introduction}
	
	Image Super-Resolution (SR) has been a challenging inverse problem in computer vision for many years, and a variety of deep learning-based (SR) methods have been developed in the past decade \cite{wang2021real, zhang2021designing, liang2022details, chen2023human, wang2024exploiting, wang2024sinsr, yue2023resshift, wu2024one, wu2024seesr, yu2024scaling, chen2024ssl, chen2025toward, yi2025fine}. However, most of the existing SR models \cite{zhang2018image, zhang2018residual, wang2018esrgan, liang2021swinir, zhang2022efficient, chen2023activating, dong2016accelerating, li2019feedback, zhang2024transcending} are designed to operate only under integer or fixed scaling factors, \ie, $\times 2, \times 3, \times 4$, due to the fact that they perform resolution upsampling by processing discrete pixels with de-convolutional layers \cite{dong2016accelerating} or pixel-shuffle layers \cite{shi2016real}. As a result, it is hard for them to perform SR with decimal or even arbitrary scaling factors. Although the existing SR models could achieve arbitrary-scale zooming via adopting bicubic or bilinear interpolation operations, they would inevitably lead to unsatisfied or over-smoothing super-resolved results. 
	
	Inspired by the development of Implicit Neural Representations (INR) \cite{mildenhall2020nerf}, Chen \etal presented LIIF \cite{chen2021learning} to address the problem of Arbitrary-scale SR (ASR) by making use of the continuity of Multi-Layer Perceptron (MLP) to query the RGB values of a high-resolution (HR) image and feeding its index to the MLP network. The following works try to improve the ASR performance with local feature estimators \cite{lee2022local}, normalization flow \cite{yao2023local} or scale-aware attention modules \cite{cao2023ciaosr}. 
	Most of the existing INR-based ASR models \cite{chen2021learning, lee2022local, yao2023local, cao2023ciaosr} rely heavily on MLP layers to predict RGB pixels. However, this implicit function lacks enough representation capacity, as it renders the RGB color of a single pixel in each query. INR-based models need to query the MLP layers numerous times, resulting in very low computational efficiency, especially for those models \cite{cao2023ciaosr} who scale up the MLP layers and employ complex attention modules.
	
	Recently, Gaussian Splatting (GS) \cite{kerbl20233d} has shown its advantages over INR-based methods, such as Neural Radiance Field (NeRF) \cite{mildenhall2020nerf}, in terms of both accuracy and complexity, thanks to its high computational efficiency and powerful explicit representation capabilities. GS has been successfully applied to tasks such as 3D reconstruction \cite{guedon2024sugar, charatan2024pixelsplat}, representation \cite{navaneet2024compgs, girish2024eagles}, editing \cite{wang2024gaussianeditor, chen2024gaussianeditor}, generation \cite{liu2024humangaussian, chen2024text}, \etc. 
	While some studies have explored the use of GS for low-level vision tasks, \eg, image deblurring \cite{chen2024deblur, zhao2024bad}, image compression \cite{zhang2024gaussianimage}, video super-resolution \cite{yang2024mob}, these models overfit on the repetitive patterns from multi-frames in a video clip or multi-view cameras, or overfit on a single image to complete the image compression task. None of them could be applied to the image ASR task, since overfitting a model on a single image cannot meet the generalization requirements of ASR, where a single model should be used to super-resolve any input low-resolution (LR) image with any scales to its high-resolution (HR) counterpart.
	
	Instead of overfitting the model to a specified LR image, we switch from the optimization-based manner of original 3D GS \cite{kerbl20233d} to a learning-based method and present \textbf{GSASR} (\textbf{G}aussian \textbf{S}platting based \textbf{A}rbitrary-scale \textbf{S}uper-\textbf{R}esolution). Specifically, we first obtain the deep image features by encoding the input LR image with a general SR backbone \cite{lim2017enhanced, zhang2018image}. Secondly, inspired by DETR \cite{carion2020end}, we introduce learnable Gaussian embeddings to output 2D Gaussians conditioned on the extracted image features. To enhance interaction among Gaussian embeddings, we perform window self-attention \cite{liu2021swin} so that they can adapt to each other. Then, the Gaussian embeddings are projected to predict the corresponding properties of 2D Gaussians, including the opacity, position, standard deviation, correlation coefficient, and peak RGB values. These 2D Gaussians depict the image in the continuous Gaussian space. To render 2D Gaussians to obtain the super-resolved image with arbitrary scale, we implement an efficient differentiable GPU/CUDA-based rasterization, which receives the predicted 2D Gaussians with an upsampling scale vector as conditions. The scale vector controls the sampling density, and all Gaussians are processed efficiently in parallel.
	
	The advantages of our proposed GSASR are threefold. (1) Different from previous INR-based models, which query each HR pixel exclusively and lack interaction among pixels, GSASR employs 2D Gaussians to perform region-level representation, providing a much more powerful representation capability. (2) In GSASR, the Gaussians' centers are learnable and input adaptive so that the Gaussians can concentrate on areas with complex textures. In addition, the standard deviation and correlation coefficient of Gaussians indicate their shape and direction, allowing Gaussians to fit the different shapes of different objects and scenes. (3) Taking advantage of our developed efficient scale-aware 2D rasterization techniques, GSASR is much faster than state-of-the-art INR-based models. As shown in Fig.~\ref{fig:teaser} and Tab.~\ref{tab:computational cost}, under the scaling factor $\times 12$, GSASR could super-resolve an image to $720 \times 720$ resolution within only $91$ ms, while the state-of-the-art INR-based model such as CiaoSR \cite{cao2023ciaosr} will cost nearly $540$ ms.

	\begin{figure*}[t!]
		\centering
		\includegraphics[width=1\linewidth]{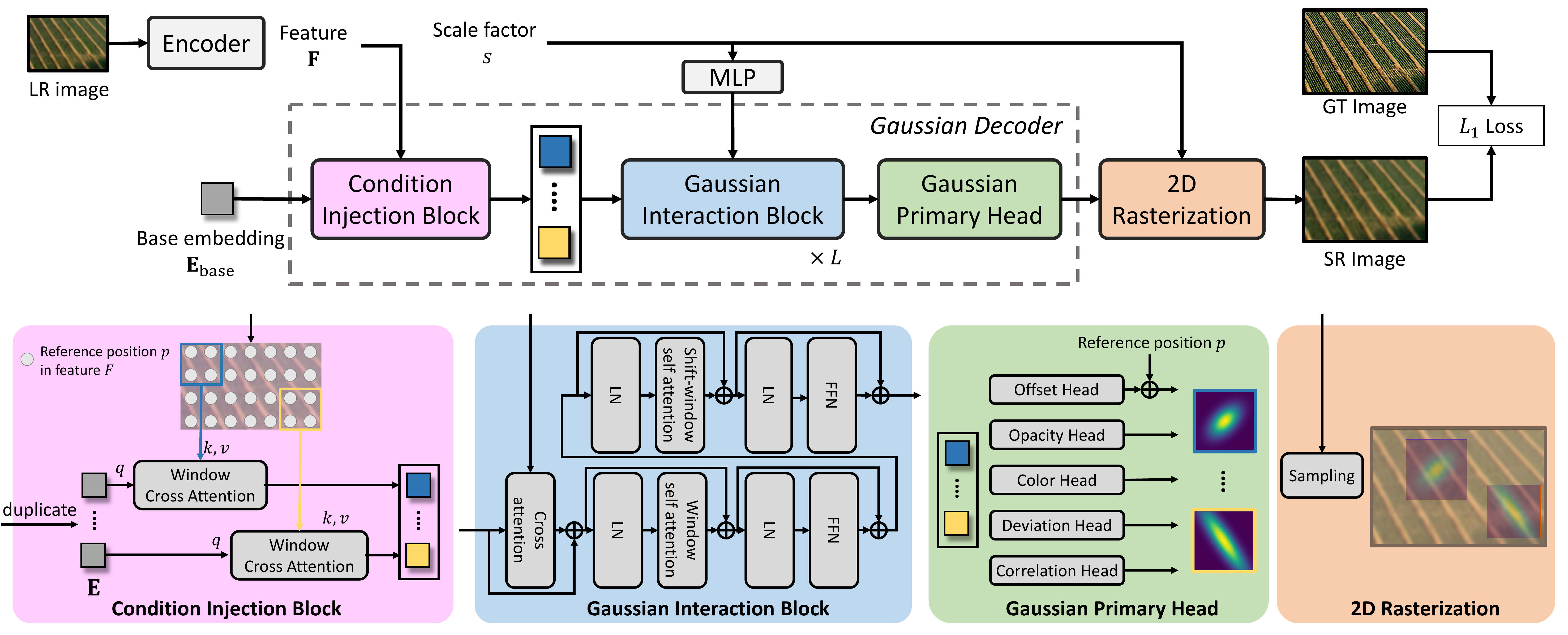}
		\caption{{Overview of GSASR.} In the training phase, an LR image is fed into the encoder to extract image features, conditioned on which the learnable Gaussian embeddings are passed through the Condition Injection Block and Gaussian Interaction Block to output 2D Gaussians. These 2D Gaussians are then rendered into an SR image of a specified resolution through differential rasterization.}
		\label{fig:framework}
		\vspace{-1.5em}
	\end{figure*}
	
	\section{Related Work}
	
	SR aims to increase the resolution of a given LR image. 
	While different methods differ from each other on specific designs, they all rely on upsampling operators to magnify the resolution of the original LR image, such as hand-crafted interpolation operators, convolutional layers \cite{dong2016accelerating, shi2016real}, and implicit neural representations \cite{chen2021learning}.
	
	\noindent \textbf{Fixed-scale SR Methods}. 
	In the early stage, deep neural network-based SR models \cite{dong2014learning, kim2016accurate, kim2016deeply, lai2017deep} usually adopt interpolation operators to magnify the resolution. They first upsample the resolution of an LR image to the size of its HR counterpart, and then feed it into networks for detail enhancement. Such a scheme is straightforward but suffers from high computational cost. FSRCNN \cite{dong2016accelerating} is proposed to apply de-convolutional layers to upsample the resolution of the deep features at the end of the network, so as to accelerate the inference speed. ESPCN \cite{shi2016real} performs upsampling by rearranging the LR feature maps into a higher-resolution space through pixel-shuffle layers. However, since the pixels are discrete in feature domain, the de-convolution or pixel-shuffle layers cannot perform continuous sampling to reconstruct SR images with arbitrary scales. 
	
	\noindent \textbf{Arbitrary-scale SR Methods}. The interpolation operators can be used to continuously predict neighborhood pixels at any magnification factors, yet they will sacrifice the image quality and obtain blurry results. Meta-SR \cite{hu2019meta} designs a meta-upscaling module that leverages DNN to predict HR details at arbitrary-scales from LR neighborhood embeddings. Wang \etal \cite{wang2021learning} performed arbitrary upsampling by introducing dynamic filters generated based on the scaling factor. Inspired by the success of INR-based methods \cite{mildenhall2020nerf} in 3D tasks, LIIF \cite{chen2021learning} introduces INR to complete the ASR task by learning an implicit mapping function between a pixel's RGB value and its index. ITSRN \cite{yang2021implicit} makes use of a transformer block and implicit position embedding to predict the super-resolved images. LTE \cite{lee2022local} employs an implicit function to describe image features in Fourier domain. SRNO \cite{wei2023super} adopts a kernel integral operator to fit the implicit function. 
	LINF \cite{yao2023local} estimates a local implicit function with normalizing flow to model the naturalness of textures. CLIT \cite{chen2023cascaded} employs a cascaded cross-scale attention block with frequency encoding layers to sense neighborhood features. CiaoSR \cite{cao2023ciaosr} introduces scale-aware attention-in-attention modules to extend the receptive field to promote the ASR performance. 
	LMF \cite{he2024latent} introduces a latent modulated function to compress the feature and accelerate the rendering process at the price of representation accuracy.

	While the recent work GaussianSR \cite{hu2025gaussiansr} shares a similar motivation to introduce GS for the ASR task, our work has two  key differences from it.  First, GaussianSR places the centers of Gaussians at the pixel positions of the LR image, heavily limiting the representation capability of Gaussians. 
	In contrast, GSASR predicts all Gaussian attributes in a free and implicit manner to unleash its expressive potential. Second, GaussianSR is slow due to its inefficient use of GS. Instead, GSASR employs an efficient differentiable 2D GPU/CUDA-based scale-aware rasterization to achieve ASR effectively and efficiently.
	

	\section{Method}
	In this section, we first describe the concept and formulation of 2D Gaussian Splatting (GS), then present the model architecture and training details of our GSASR method.
	
	\subsection{2D GS Representation and Rasterization}
	\textbf{2D GS Representation}. Our key insight is to represent an image using continuous Gaussians, which can fit various frequency components of an image due to the powerful expressive properties of GS. 
	Specifically, considering an LR image $I_{LR}$ of size $H\times W$, we represent it using $N$ 2D Gaussians, denoted by $\mathcal{G} = \{G_1, G_2, ..., G_N\}$. Each Gaussian maintains properties of opacity $\alpha~\in [0,1]$, center position $\mu=\{\mu_{x}~\in \mathbb{R}, \mu_{y}~\in \mathbb{R}\}$, standard deviation $\sigma=\{\sigma_{x}~\in (0,\infty ), \sigma_{y}~\in (0,\infty )\}$, correlation coefficient $\rho~\in [-1, 1]$, and normalized peak RGB value $c~\in [0,1]^3$. We define the RGB color at position $(x, y)$ contributed by Gaussian $G$ as $G(x,  y)$ (here we omit the index of Gaussians for simplicity of expression):
	\begin{equation}
		\label{eq1}
		G(x,y) = \alpha \cdot c \cdot f(x, y),
	\end{equation}

	The calculation of $f(x,  y)$ is defined following typical 2D Gaussian distribution: 
	\vspace{-0.8em}
	\begin{equation}
		\vspace{-0.8em}
		\begin{split}
			\label{eq2}
			f(x,y)= &{(2\pi\sigma_x\sigma_y\sqrt{1-\rho^{2}})^{-1}}exp[{-\frac{1}{2(1-\rho^2)}} \times
			\\
			&{(\frac{\Delta x^2}{\sigma_x^2}-\frac{2\rho \Delta x \Delta y}{\sigma_x\sigma_y}+\frac{\Delta y^2}{\sigma_y^2})}],
		\end{split}
	\end{equation}
	
	where $\Delta x = x - \mu_x$ and $\Delta y = y - \mu_y$ decide the position of Gaussians on an image, and the standard deviation $\sigma_x$, $\sigma_y$ and the correlation coefficient $\rho$ indicate the shape and direction of Gaussian.
	The pixel color of the LR image at position $(x\in [0,W-1], y\in [0, H-1])$ is represented as the summation of all $N$ Gaussians $\mathcal{G}$:
	\vspace{-0.6em}
	\begin{equation}
		\label{eq3}
		I_{LR}(x, y)=\sum_{i=1}^N G_i(x, y),
		\vspace{-0.6em}
	\end{equation}
	
	\begin{algorithm}[t!]
		\SetKwInput{KwIn}{Input}
		\SetKwInput{KwOut}{Output}
		\SetKwInput{KwRet}{Return}
		\SetKwProg{For}{For}{ do}{end}
		\SetKwProg{If}{If}{ then}{end}
		\caption{2D GS Scale-aware Rasterization}\label{alg:alg1}
		
		\KwIn{$N$ 2D Gaussians $\{G_1, G_2, \ldots, G_N\}$; LR image of size ($H$, $W$); scale factor $s$; rasterization ratio $r$.}
		\KwOut{Rendered image $I_{SR}$}
		
		Initialize $I_{SR}$ as an ($sH$, $sW$, 3) array of zeros.
		
		\For{each $G_i$ in $\{G_1, G_2, \ldots, G_N\}$}{
			Initialize $\alpha$, $\mu_x$, $\mu_y$, $\sigma_x$, $\sigma_y$, $\rho$, $c$ from $G_i$. 
			\\
			\For{each pixel $(x,y)$ in $I_{SR}$}{
				\If{$|x-\mu_x| < rsH$ and $|y-\mu_y| < rsW$}{
					Obtain $f(x/s,y/s)$ using Eq. \ref{eq2}\;
					Obtain $G_i(x/s,y/s)$ using Eq. \ref{eq1}\;
					$I_{SR}(x,y;s) \mathrel{+}= G_i(x/s,y/s)$.
				}
			}
		}
	\end{algorithm}

	Each Gaussian function $G_{i}$ is infinite and continuous on the 2D space, enabling discretely sampling at arbitrary positions. With a given scaling factor $s$, to obtain an SR image $I_{SR}$ of size $sH \times sW$, we sample the color at position $(x\in [0,sW-1], y\in [0, sH-1])$ with the following scale-aware rasterization:
	\vspace{-0.8em}
	\begin{equation}
		\label{eq4}
		I_{SR}(x, y; s)=\sum_{i=1}^N G_i(\frac{x}{s}, \frac{y}{s}),
		\vspace{-0.8em}
	\end{equation}

	\noindent \textbf{Fast Rasterization.} 
	In Eq.~\ref{eq4}, simply rendering an SR image by querying each pixel from all 2D Gaussians leads to a complexity of $\mathcal{O}(s^2HWN)$, which is too high for high-resolution images.
	Actually, a Gaussian generally focuses on a local area and its contribution to pixel values decays rapidly with the increase of distance. 
	Therefore, we introduce a rasterization ratio $r \leq 1$ to control the rendering range of each Gaussian. Specifically, we handle all Gaussians in parallel and only render the pixels that are close enough to the Gaussian centers, greatly reducing the computational complexity to $\mathcal{O}(r^2s^2HWN)$ and making our algorithm efficient. To achieve a better balance between performance and efficiency, we set the ratio as $r=0.1$, which could significantly speed up the rasterization process and avoid missing major Gaussian responses. Note that our rasterization process is differentiable, which can be seamlessly integrated into neural network training for end-to-end optimization. Algorithm \ref{alg:alg1} describes the detailed steps of our scale-aware rasterization process. This algorithm is implemented via CUDA C++, which is GPU-friendly and achieves fast speed and low memory requirements.

	\begin{table*}[ht]
		\centering
		\caption{Quantitative comparison between representative ASR models and our GSASR. All models use the same EDSR-backbone \cite{lim2017enhanced} as the feature extraction encoder and are tested on DIV2K and LSDIR \cite{li2023lsdir} datasets \cite{timofte2017ntire} with scaling factors $\times2$, $\times3$, $\times4$, $\times6$, $\times8$, $\times12$, $\times16$, $\times18$, $\times24$, $\times30$. The best results are highlighted in \textcolor{red}{red}. PSNR/SSIM metrics are computed on the Y channel of Ycbcr space.}
		\vspace{-0.2cm}
		\scalebox{1.0}{
			\resizebox{\linewidth}{!}{
				\begin{tabular}{c|c|ccccccccc|ccccccccc} 
					\toprule[2pt]
					\multirow{3}{*}{Scale} & \multirow{3}{*}{Metrics} & \multicolumn{18}{c}{Backbone: EDSR-baseline} \\ 
					\cline{3-20}
					&  & \multicolumn{9}{c|}{Testing Dataset: DIV2K} & \multicolumn{9}{c}{Testing Dataset: LSDIR} \\
					&  & \begin{tabular}[c]{@{}c@{}}Meta\\-SR\end{tabular} & LIIF & LTE & SRNO & LINF & LMF & \begin{tabular}[c]{@{}c@{}}Ciao\\-SR\end{tabular} & \begin{tabular}[c]{@{}c@{}}Gaussian\\-SR\end{tabular} & GSASR & \begin{tabular}[c]{@{}c@{}}Meta\\-SR\end{tabular} & LIIF & LTE & SRNO & LINF & LMF & \begin{tabular}[c]{@{}c@{}}Ciao\\-SR\end{tabular} & \begin{tabular}[c]{@{}c@{}}Gaussian\\-SR\end{tabular} & GSASR \\ 
					\midrule[1pt] \midrule[1pt]
					\multirow{4}{*}{$\times$2} & PSNR & 36.02~ & 36.05 & 36.10 & 36.27 & 36.21 & 36.21 & 36.42 & 36.10 & \textcolor{red}{36.65} & 31.36~ & 31.43 & 31.50 & 31.65 & 31.49 & 31.61 & 31.78 & 31.47 & \textcolor{red}{32.14} \\
					& SSIM & 0.9455~ & 0.9458 & 0.9461 & 0.9474 & 0.9461 & 0.9469 & 0.9476 & 0.9459 & \textcolor{red}{0.9495} & 0.9162~ & 0.9170 & 0.9177 & 0.9197 & 0.9175 & 0.9193 & 0.9208 & 0.9174 & \textcolor{red}{0.9251} \\
					& LPIPS & 0.0889~ & 0.0879 & 0.0869 & 0.0833 & 0.0887 & 0.0832 & 0.0835 & 0.0888 & \textcolor{red}{0.0767} & 0.0984~ & 0.0963 & 0.0947 & 0.0916 & 0.0973 & 0.0916 & 0.0890 & 0.097 & \textcolor{red}{0.0823} \\
					& DISTS & 0.0571~ & 0.0567 & 0.0567 & 0.0547 & 0.0564 & 0.0555 & 0.0543 & 0.0569 & \textcolor{red}{0.0514} & 0.0692~ & 0.0688 & 0.0682 & 0.0662 & 0.0686 & 0.0667 & 0.0649 & 0.0685 & \textcolor{red}{0.0612} \\ 
					\midrule[1pt]
					\multirow{4}{*}{$\times$4} & PSNR & 33.36~ & 30.43 & 30.47 & 30.57 & 30.51 & 30.56 & 30.67 & 30.46 & \textcolor{red}{30.89} & 26.13~ & 26.21 & 26.26 & 26.36 & 26.25 & 26.33 & 26.42 & 26.23 & \textcolor{red}{26.65} \\
					& SSIM & 0.8367~ & 0.8388 & 0.8395 & 0.8415 & 0.8396 & 0.8416 & 0.8431 & 0.8389 & \textcolor{red}{0.8486} & 0.7577~ & 0.7614 & 0.7627 & 0.7666 & 0.7621 & 0.7656 & 0.7681 & 0.7615 & \textcolor{red}{0.7774} \\
					& LPIPS & 0.2723~ & 0.2662 & 0.2647 & 0.2616 & 0.2680 & 0.2607 & 0.2585 & 0.2684 & \textcolor{red}{0.2518} & 0.3074~ & 0.2978 & 0.2957 & 0.2899 & 0.2998 & 0.2921 & 0.2865 & 0.3007 & \textcolor{red}{0.2777} \\
					& DISTS & 0.1394~ & 0.1403 & 0.1397 & 0.1384 & 0.1401 & 0.1379 & 0.1370 & 0.1406 & \textcolor{red}{0.1301} & 0.1666~ & 0.1678 & 0.1664 & 0.1647 & 0.1675 & 0.1647 & 0.1631 & 0.1679 & \textcolor{red}{0.1554} \\ 
					\midrule[1pt]
					\multirow{4}{*}{$\times$8} & PSNR & 26.72~ & 26.87 & 26.93 & 27.00 & 26.91 & 26.97 & 27.04 & 26.76 & \textcolor{red}{27.22} & 23.23~ & 23.32 & 23.37 & 23.43 & 23.34 & 23.40 & 23.47 & 23.27 & \textcolor{red}{23.58} \\
					& SSIM & 0.7135~ & 0.7207 & 0.7218 & 0.7243 & 0.7207 & 0.7235 & 0.7256 & 0.7155 & \textcolor{red}{0.7321} & 0.6032~ & 0.6123 & 0.6137 & 0.6171 & 0.6118 & 0.6156 & 0.6194 & 0.6059 & \textcolor{red}{0.6269} \\
					& LPIPS & 0.4365~ & 0.4212 & 0.4321 & 0.4261 & 0.4285 & 0.4264 & 0.4169 & 0.4445 & \textcolor{red}{0.4077} & 0.5020~ & 0.4812 & 0.4939 & 0.4848 & 0.4868 & 0.4883 & 0.4703 & 0.5068 & \textcolor{red}{0.4611} \\
					& DISTS & 0.2305~ & 0.2337 & 0.2348 & 0.2330 & 0.2341 & 0.2328 & 0.2314 & 0.2386 & \textcolor{red}{0.2214} & 0.2643~ & 0.2673 & 0.2671 & 0.2640 & 0.2677 & 0.2654 & 0.2619 & 0.2713 & \textcolor{red}{0.2518} \\ 
					\midrule[1pt]
					\multirow{4}{*}{$\times$16} & PSNR & 24.00~ & 24.13 & 24.20 & 24.25 & 24.14 & 24.23 & 24.27 & 23.80 & \textcolor{red}{24.38} & 21.22~ & 21.29 & 21.34 & 21.38 & 21.31 & 21.32 & 21.40 & 21.12 & \textcolor{red}{21.42} \\
					& SSIM & 0.6307~ & 0.6402 & 0.6407 & 0.6423 & 0.6394 & 0.6418 & 0.6443 & 0.6306 & \textcolor{red}{0.6473} & 0.5120~ & 0.5221 & 0.5227 & 0.5245 & 0.5209 & 0.5213 & 0.5271 & 0.5122 & \textcolor{red}{0.5296} \\
					& LPIPS & 0.5924~ & 0.5754 & 0.5912 & 0.5847 & 0.5876 & 0.5861 & 0.5666 & 0.6325 & \textcolor{red}{0.5563} & 0.6687~ & 0.6529 & 0.6691 & 0.6617 & 0.6622 & 0.6738 & 0.6386 & 0.7101 & \textcolor{red}{0.6280} \\
					& DISTS & 0.3297~ & 0.3357 & 0.3398 & 0.3367 & 0.3412 & 0.3371 & 0.3336 & 0.3582 & \textcolor{red}{0.3242} & 0.3638~ & 0.3683 & 0.3701 & 0.3671 & 0.3727 & 0.3687 & 0.3636 & 0.3866 & \textcolor{red}{0.3538} \\ 
					\midrule[1pt]
					\multirow{4}{*}{$\times$24} & PSNR & 22.63~ & 22.74 & 22.81 & 22.84 & 22.74 & 22.83 & 22.87 & 22.41 & \textcolor{red}{22.90} & 20.25~ & 20.31 & 20.35 & 20.39 & 20.32 & 20.34 & \textcolor{red}{20.40} & 20.11 & 20.38 \\
					& SSIM & 0.6027~ & 0.6108 & 0.6112 & 0.6119 & 0.6101 & 0.6117 & 0.6141 & 0.6037 & \textcolor{red}{0.6150} & 0.4854~ & 0.4938 & 0.4938 & 0.4845 & 0.4930 & 0.4924 & 0.4967 & 0.4862 & \textcolor{red}{0.4973} \\
					& LPIPS & 0.6512~ & 0.6444 & 0.6600 & 0.6539 & 0.6545 & 0.6561 & 0.6352 & 0.7042 & \textcolor{red}{0.6299} & 0.7234~ & 0.7214 & 0.7374 & 0.7292 & 0.7261 & 0.7421 & 0.7067 & 0.7812 & \textcolor{red}{0.7028} \\
					& DISTS & 0.4022~ & 0.4009 & 0.4063 & 0.4026 & 0.4106 & 0.4036 & 0.3980 & 0.4380 & \textcolor{red}{0.3877} & 0.4292~ & 0.4274 & 0.4308 & 0.4269 & 0.4348 & 0.4297 & 0.4222 & 0.4579 & \textcolor{red}{0.4117} \\ 
					\midrule[1pt]
					\multirow{4}{*}{$\times$30} & PSNR & 21.97~ & 22.07 & 22.12 & 22.16 & 22.07 & 22.03 & 22.18 & 21.75 & \textcolor{red}{22.19} & 19.75~ & 19.80 & 19.84 & 19.86 & 19.81 & 19.83 & \textcolor{red}{19.87} & 19.60~ & 19.82 \\
					& SSIM & 0.5982~ & 0.5998 & 0.6000 & 0.6003 & 0.5994 & 0.5966 & 0.6021 & 0.5943 & \textcolor{red}{0.6025} & 0.4754~ & 0.4824 & 0.4823 & 0.4824 & 0.4817 & 0.4818 & 0.4839 & 0.4766 & \textcolor{red}{0.4842} \\
					& LPIPS & 0.7018~ & 0.6773 & 0.6914 & 0.6862 & 0.6859 & 0.7137 & 0.6682 & 0.7317 & \textcolor{red}{0.6648} & 0.7386~ & 0.7537 & 0.7676 & 0.7610 & 0.7575 & 0.7645 & \textcolor{red}{0.7284} & 0.8084 & 0.7368 \\
					& DISTS & 0.4474~ & 0.4375 & 0.4445 & 0.4400 & 0.4500 & 0.4476 & 0.4339 & 0.4821 & \textcolor{red}{0.4232} & 0.4702~ & 0.4603 & 0.4645 & 0.4597 & 0.4702 & 0.4608 & 0.4535 & 0.4986 & \textcolor{red}{0.4417} \\
					\bottomrule[2pt]
				\end{tabular}
			}
		} 
		\label{tab:results on DIV2K and LSDIR with EDSR}
		\vspace{-1.7em}
	\end{table*}

	\subsection{GSASR: Architecture and Modules}
	
	Conventional 3D GS methods \cite{kerbl20233d} directly optimize 3D Gaussians to fit a specific scene. As a result, they lack the generalization capability for the ASR tasks, where the model should be generalized to unseen images. To overcome this limitation, we introduce a novel network architecture, which conditions 2D Gaussians on LR input images, generating content-aware Gaussian representations with enhanced generalization capabilities. Our proposed method, namely GSASR, is illustrated in Figure.~\ref{fig:framework}. It begins with an off-the-shelf encoder (such as EDSR-Baseline \cite{lim2017enhanced}, RDN \cite{zhang2018residual}) that extracts image features $\mathbf{F} \in \mathbb{R}^{H\times W \times C}$ from the LR input $\mathbf{I}  \in \mathbb{R}^{H\times W \times 3}$. The deep features $\mathbf{F}$ are input to the subsequent Gaussian generation process. The key of GSASR lies in the proposed \textit{Gaussian decoder}, which receives image features $\mathbf{F}$ and learnable Gaussian embeddings as inputs, and outputs 2D Gaussians $\mathcal{G}$ to depict the input image in continuous Gaussian space. Our Gaussian decoder consists of \textit{a condition injection block}, \textit{a Gaussian interaction block}, and \textit{a Gaussian primary head}. Finally, the Gaussian decoder output is rasterized into an SR image.
	
	\begin{figure}[t]
		\centering
		\includegraphics[width=1.0\linewidth]{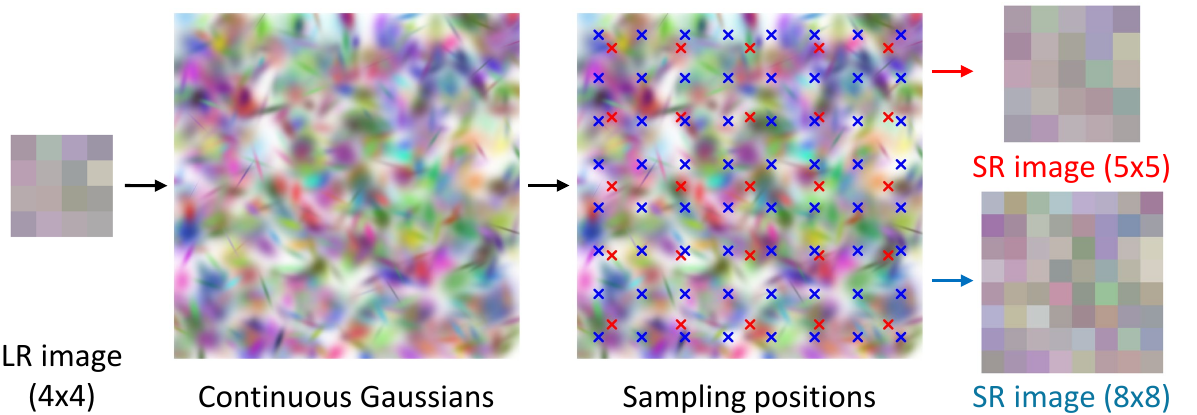}
		\vspace{-1.5em}
		\caption{{Demonstration of the sampling process in 2D rasterization.} A larger upsampling scale factor employs smaller sampling intervals from continuous Gaussians. }
		\label{fig:sampling}
		\vspace{-1.8em}
	\end{figure}
	
	\noindent \textbf{Gaussian Embedding.} In Gaussian decoder, we aim to represent the LR image through $N$ Gaussian embeddings $\mathbf{E} \in \mathbb{R}^{N\times d}$ conditioned on $\mathbf{F}$, where $d$ is the embedding dimension. As for the number of Gaussian embeddings $N$, a straightforward idea is to set it to a fixed number so that the $N$ independent learnable Gaussian embeddings interact with image features to output Gaussians $\mathcal{G}$. However, this approach will either be ineffective or lead to low computational efficiency. As the LR images exhibit variable sizes in the test phase, rendering a fixed number of Gaussians will either waste computational resources when handling LR image with very small size, or lead to deficient representation when processing larger LR images. 
	Another possible strategy is to set $N \propto s$ under different scaling factor $s$. However, it requires the Gaussian decoder to handle varying sizes of Gaussian embeddings, while the size of LR features is fixed in the training stage. In addition, as the scaling factor increases, more Gaussians are introduced, leading to longer inference time and higher GPU memory cost.
	
	
	To address these issues, we set $N  = m \times (H\times W)$ so that it is proportional to the LR image size,  where $m$ is a hyper-parameter controlling the density of Gaussians. In this way, we could employ more Gaussians to represent larger inputs, while still keep efficiency under ultra-high scaling factors. To be more specific, we split LR feature $\mathbf{F}\in \mathbb{R}^{C \times H \times W}$ into $\frac{H}{k}\times \frac{W}{k}$ windows with window size $k\times k$. The content in each window is fitted by a learnable Gaussian embedding $\mathbf{E}_{base}\in \mathbb{R}^{mk^2\times d}$, which is randomly initialized and optimized during training. To dynamically support arbitrary LR feature size $H\times W$, we duplicate $\mathbf{E}_{base}$ for $\frac{H}{k}\times \frac{W}{k}$ times to obtain $\mathbf{E} \in \mathbb{R}^{N\times d}$ and cover all windows, then we assign the $N$ Gaussian embeddings to different reference positions $p \in \mathbb{R}^{N\times 2}$, which are obtained by sampling $N$ points at equal intervals from an $H\times W$ image. According to reference positions, each $\mathbf{E}_{base}$ will interact with LR features in the window, as shown in Fig.~\ref{fig:framework}.

	\noindent \textbf{Condition Injection Block.}
	In this block, each Gaussian embedding $\mathbf{E}_{i}, i \in [1, N]$ aims to learn associated features.
	As shown in Fig.~\ref{fig:framework}, we perform window-based cross-attention~\cite{liu2021swin} to incorporate feature $\mathbf{F}$ into Gaussian embeddings $\mathbf{E}$. Such a technique is widely used in Stable Diffusion~\cite{sd}. The formulation is as follows:
	\vspace{-0.5em}
	\begin{equation}
		\text{Attention}(Q, K, V) = \text{SoftMax}\left(\frac{QK^\top}{\sqrt{d}} + B\right)V,
		\vspace{-0.5em}
	\end{equation}
	where $Q$ is from the Gaussian embeddings $\mathbf{E}$, $K$ and $V$ are from image features $\mathbf{F}$. The $i$-th embedding $\mathbf{E}_i$ at position $p_i$ interacts with the $j$-th pixel feature within the window. $B$ is inherited from Swin Transformer \cite{liu2021swin}, serving as a learnable position encoding to indicate the relative position between Gaussian embeddings and LR features. The window size of window-based cross attention is set to the same $k$ value as that of the partitioned window LR features.
	
	\noindent \textbf{Gaussian Interaction Block.} In the condition injection block, each Gaussian embedding $\mathbf{E}_i$ is processed independently, resulting in a lack of information exchange among different Gaussian embeddings. To address this issue, we stack $L$ Gaussian interaction blocks to enhance the interaction among Gaussian embeddings. The Gaussian interaction block is designed based on the Swin Transformer architecture \cite{liu2021swin}, which iteratively stacks the window attention layers and shifted window attention layers. In attention layers, we perform self-attention for embeddings whose reference positions are located in the same window, similar to the condition injection block. Besides, we introduce a cross-attention layer to make embeddings scale-aware so that the Gaussian embeddings can learn to adjust the properties according to the given scale factor. The window size of window attention layer is set to the same $k$ value as that of the partitioned window LR features.
	
	\begin{figure*}[t]
		\small
		\centering
		\begin{minipage}{0.19\textwidth}
			\includegraphics[width=1\linewidth]{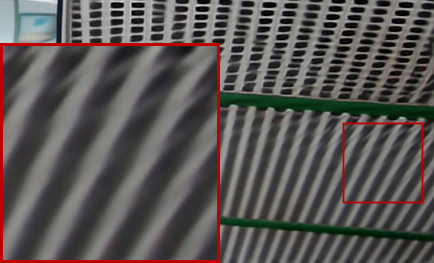}\\
			\centering{Meta-SR}\\
			\includegraphics[width=1\linewidth]{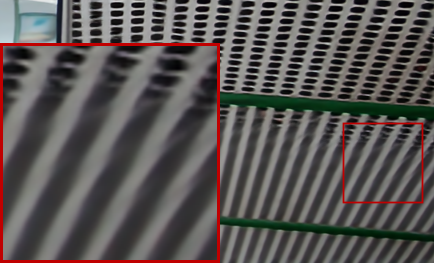 }
			\centering{LMF}\\
		\end{minipage}
		\begin{minipage}{0.19\textwidth}
			\includegraphics[width=1\linewidth]{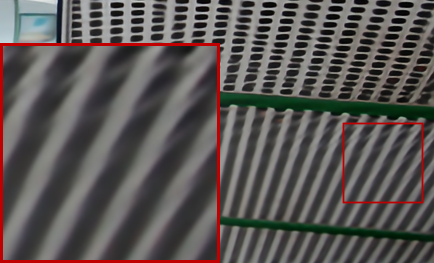}\\
			\centering{LIIF}\\
			\includegraphics[width=1\linewidth]{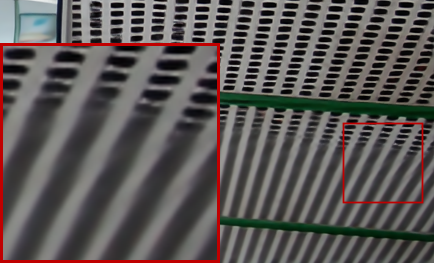 }
			\centering{CiaoSR}\\
		\end{minipage}
		\begin{minipage}{0.19\textwidth}
			\includegraphics[width=1\linewidth]{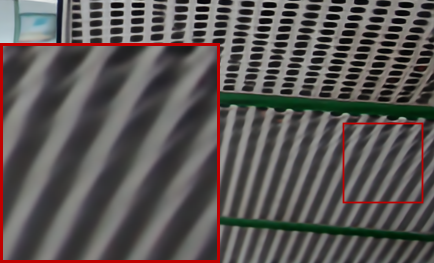}\\
			\centering{LTE}\\
			\includegraphics[width=1\linewidth]{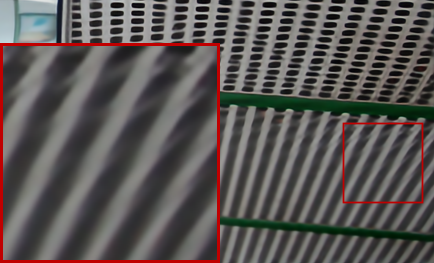 }
			\centering{GaussianSR}\\
		\end{minipage}
		\begin{minipage}{0.19\textwidth}
			\includegraphics[width=1\linewidth]{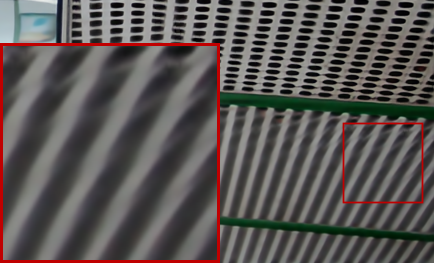}\\
			\centering{SRNO}\\
			\includegraphics[width=1\linewidth]{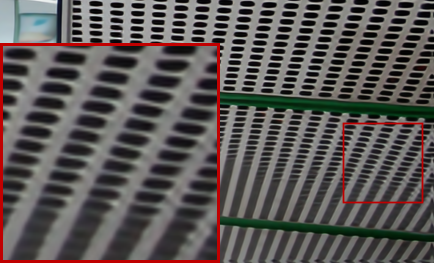 }
			\centering{GSASR (Ours)}\\
		\end{minipage}
		\begin{minipage}{0.19\textwidth}
			\includegraphics[width=1\linewidth]{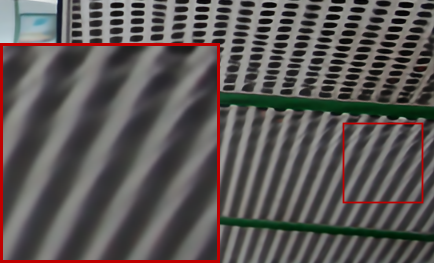}\\
			\centering{LINF}\\
			\includegraphics[width=1\linewidth]{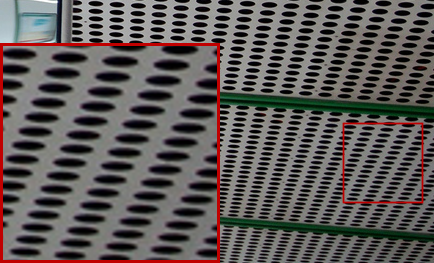}
			\centering{GT} \\
		\end{minipage}
		\vspace{-0.7em}
		\caption{Visualization of GSASR and the competing methods under $\times 4$ scaling factor with EDSR \cite{lim2017enhanced} feature extraction backbone. The competing methods result in blurry details, while GSASR generates much clear textures.}
		\label{fig:visualization1}
		\vspace{-0.7em}
	\end{figure*}
	
	\begin{figure*}[ht]
		\vspace{-0.2em}
		\small
		\centering
		\begin{minipage}{0.19\textwidth}
			\includegraphics[width=1\linewidth]{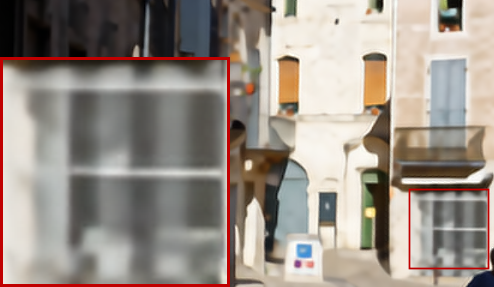}\\
			\centering{Meta-SR}\\
			\includegraphics[width=1\linewidth]{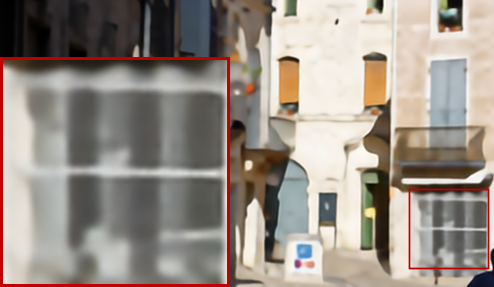 }
			\centering{LMF}\\
		\end{minipage}
		\begin{minipage}{0.19\textwidth}
			\includegraphics[width=1\linewidth]{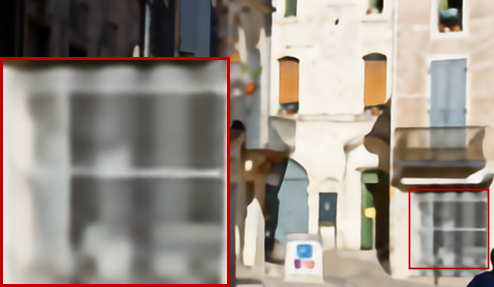}\\
			\centering{LIIF}\\
			\includegraphics[width=1\linewidth]{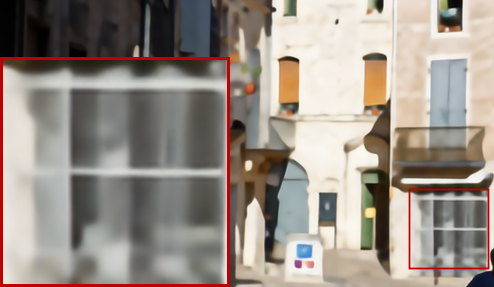 }
			\centering{CiaoSR}\\
		\end{minipage}
		\begin{minipage}{0.19\textwidth}
			\includegraphics[width=1\linewidth]{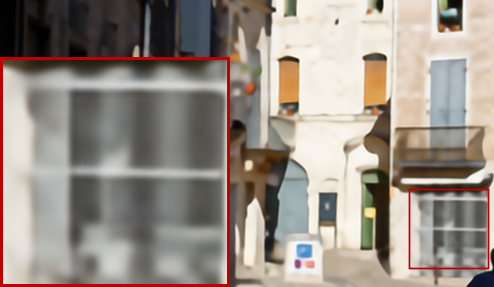}\\
			\centering{LTE}\\
			\includegraphics[width=1\linewidth]{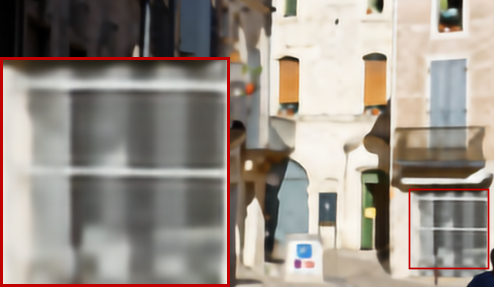 }
			\centering{GaussianSR}\\
		\end{minipage}
		\begin{minipage}{0.19\textwidth}
			\includegraphics[width=1\linewidth]{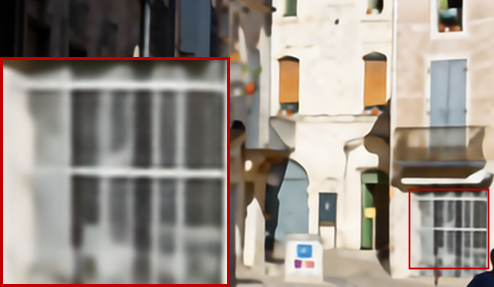}\\
			\centering{SRNO}\\
			\includegraphics[width=1\linewidth]{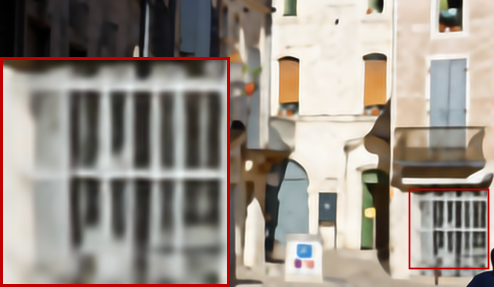 }
			\centering{GSASR (Ours)}\\
		\end{minipage}
		\begin{minipage}{0.19\textwidth}
			\includegraphics[width=1\linewidth]{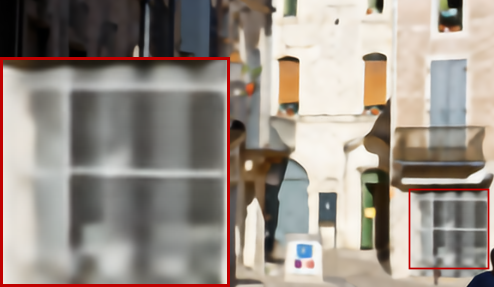}\\
			\centering{LINF}\\
			\includegraphics[width=1\linewidth]{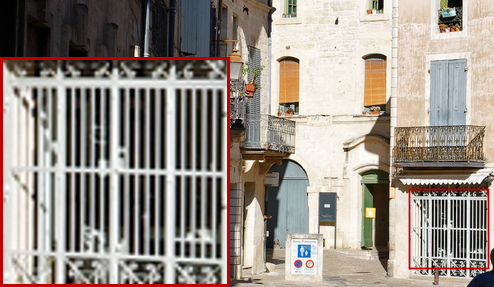}
			\centering{GT} \\
		\end{minipage}
		\vspace{-0.2cm}
		\caption{Visualization of GSASR and the competing methods under $\times 6$ scaling factor with RDN \cite{zhang2018residual} feature extraction backbone. The competing methods produce blurry textures, while GSASR generates much clearer contents.}
		\label{fig:visualization2}
		\vspace{-1.8em}
	\end{figure*}

	\noindent \textbf{Gaussian Primary Head.} We employ five disentangled  heads to convert Gaussian embeddings (output from Gaussian Interaction Block) $\mathbf{E}$ into the corresponding five properties of 2D Gaussians $\mathcal{G}$. To be more specific, the $i$-th Gaussian embedding $\mathbf{E}_i$ is fed into five lightweight MLP blocks, each of which contains three linear layers and two ReLU activation functions, to output its opacity $\alpha_i$, color $c_i$, standard deviation $\sigma_i$, reference offset $o_i$, and correlation $\rho_i$. Note that the position is obtained by adding the reference position and predicted offset $\mu_i = p_i + o_i$. 
	The five property heads are decoupled so that different heads can be adapted to predict specific properties.
	We apply the tanh function to ensure $\rho \in [-1, 1]$, and use the sigmoid function for $\{\alpha, c, \sigma \}$ to ensure that their values are physically meaningful. No activation function is applied to $o$ since the positions of Gaussians in 2D images are expected to be free. Finally, the 2D Gaussians 
	$\mathcal{G}$ are calculated from Eq.~\ref{eq1}, Eq.~\ref{eq2} with the predicted five properties $\{\alpha, \mu, c, \sigma, \rho\}$.
	
	\begin{table*}[ht]
		\centering
		\caption{Comparison of computational costs. We report the results using EDSR-backbone \cite{lim2017enhanced} as image encoder. Apart from the PSNR/SSIM/LPIPS/DISTS metrics, we also report the average inference time (ms) and the GPU memory usage (MB). PSNR/SSIM are calculated on the Y channel of Ycbcr space. The inference time/GPU memory cost is computed for the whole SR pipeline, including the encoder, decoder and rendering parts. The best results are highlighted in \textcolor{red}{red}.}
		\vspace{-0.5em}
		\scalebox{1.0}{
			\resizebox{0.9\linewidth}{!}{
				\setlength{\extrarowheight}{-2.8pt} 
				\begin{tabular}{c|c|ccccccccc} 
					\toprule[2pt]
					\multirow{3}{*}{Scale} & \multirow{3}{*}{\begin{tabular}[c]{@{}c@{}}Computational Cost\\and Performance\end{tabular}} & \multicolumn{9}{c}{Backbone: EDSR-baseline} \\
					&  & \multicolumn{9}{c}{Testing GT Size: 720 * 720} \\ 
					&  & Meta-SR & LIIF & LTE & SRNO & LINF & LMF & CiaoSR & GaussianSR & GSASR \\ 
					\midrule \midrule
					\multirow{6}{*}{$\times$2} & PSNR & 37.09 & 37.14 & 37.16 & 37.34 & 37.42 & 37.29 & 37.59 & 37.18 & \textcolor{red}{37.84} \\
					& SSIM & 0.9475 & 0.9487 & 0.9491 & 0.9501 & 0.9490 & 0.9498 & 0.9502 & 0.9489 & \textcolor{red}{0.9522} \\
					& LPIPS & 0.0787 & 0.0777 & 0.0768 & 0.0743 & 0.0785 & 0.0736 & 0.0744 & 0.0785 & \textcolor{red}{0.0676} \\
					& DISTS & 0.0741 & 0.0737 & 0.0736 & 0.0713 & 0.0731 & 0.0724 & 0.0710 & 0.0738 & \textcolor{red}{0.0676} \\
					& Inference Time & 186 & 454 & 126 & 107 & \textcolor{red}{86} & 209 & 23603 & 754 & 1573 \\
					& GPU Memory & 670 & 548 & \textcolor{red}{490} & 6301 & 3573 & 4005 & 49152 & 5200 & 13367 \\ 
					\midrule
					\multirow{6}{*}{$\times$3} & PSNR & 33.39 & 33.44 & 33.45 & 33.59 & 33.63 & 33.56 & 33.76 & 33.47 & \textcolor{red}{34.04} \\
					& SSIM & 0.8921 & 0.8931 & 0.8935 & 0.8953 & 0.8936 & 0.8953 & 0.8955 & 0.8932 & \textcolor{red}{0.9004} \\
					& LPIPS & 0.1813 & 0.1778 & 0.1756 & 0.1729 & 0.1783 & 0.1701 & 0.1718 & 0.1780 & \textcolor{red}{0.1625} \\
					& DISTS & 0.1258 & 0.1257 & 0.1255 & 0.1230 & 0.1255 & 0.1236 & 0.1201 & 0.1262 & \textcolor{red}{0.1182} \\
					& Inference Time & 161 & 438 & 118 & 114 & \textcolor{red}{89} & 147 & 1998 & 717 & 806 \\
					& GPU Memory & 493 & 570 & \textcolor{red}{333} & 6282 & 3412 & 1798 & 10002 & 5138 & 6000 \\ 
					\midrule
					\multirow{6}{*}{$\times$4} & PSNR & 31.38 & 31.48 & 31.48 & 31.65 & 31.62 & 31.60 & 31.79 & 31.49 & \textcolor{red}{32.01} \\
					& SSIM & 0.8417 & 0.8437 & 0.8444 & 0.8472 & 0.8444 & 0.8466 & 0.8482 & 0.8437 & \textcolor{red}{0.8536} \\
					& LPIPS & 0.2545 & 0.2490 & 0.2472 & 0.2429 & 0.2506 & 0.2438 & 0.2416 & 0.2511 & \textcolor{red}{0.2344} \\
					& DISTS & 0.1638 & 0.1652 & 0.1646 & 0.1626 & 0.1653 & 0.1930 & 0.1617 & 0.1656 & \textcolor{red}{0.1542} \\
					& Inference Time & \textcolor{red}{47} & 182 & 114 & 95 & 66 & 68 & 1165 & 686 & 543 \\
					& GPU Memory & 432 & 308 & \textcolor{red}{279} & 6275 & 3357 & 1139 & 3331 & 5048 & 3420 \\ 
					\midrule
					\multirow{6}{*}{$\times$6} & PSNR & 29.08 & 29.24 & 29.25 & 29.39 & 29.33 & 29.34 & 29.51 & 29.19 & \textcolor{red}{29.69} \\
					& SSIM & 0.7686 & 0.7735 & 0.7744 & 0.7777 & 0.7738 & 0.7766 & 0.7790 & 0.7715 & \textcolor{red}{0.7851} \\
					& LPIPS & 0.3455 & 0.3362 & 0.3420 & 0.3361 & 0.3411 & 0.3374 & 0.3298 & 0.3492 & \textcolor{red}{0.3223} \\
					& DISTS & 0.2179 & 0.2199 & 0.2195 & 0.2176 & 0.2201 & 0.2179 & 0.2171 & 0.2214 & \textcolor{red}{0.2083} \\
					& Inference Time & \textcolor{red}{46} & 176 & 118 & 94 & 65 & 55 & 716 & 692 & 265 \\
					& GPU Memory & 387 & 264 & \textcolor{red}{239} & 6270 & 3316 & 912 & 1548 & 5224 & 1578 \\ 
					\midrule
					\multirow{6}{*}{$\times$8} & PSNR & 27.74 & 27.89 & 27.94 & 28.04 & 27.97 & 28.00 & 28.14 & 27.77 & \textcolor{red}{28.25} \\
					& SSIM & 0.7216 & 0.7284 & 0.7297 & 0.7326 & 0.7285 & 0.7314 & 0.7345 & 0.7235 & \textcolor{red}{0.7397} \\
					& LPIPS & 0.4132 & 0.3956 & 0.4062 & 0.3988 & 0.4025 & 0.4017 & 0.3892 & 0.4187 & \textcolor{red}{0.3810} \\
					& DISTS & 0.2543 & 0.2569 & 0.2573 & 0.2555 & 0.2576 & 0.2570 & 0.2542 & 0.2623 & \textcolor{red}{0.2448} \\
					& Inference Time & \textcolor{red}{42} & 170 & 112 & 96 & 64 & 64 & 616 & 666 & 195 \\
					& GPU Memory & 371 & 248 & \textcolor{red}{224} & 6269 & 3302 & 832 & 1503 & 5012 & 1051 \\ 
					\midrule
					\multirow{6}{*}{$\times$12} & PSNR & 26.03 & 26.18 & 26.23 & 26.32 & 26.24 & 26.28 & 26.39 & 25.91 & \textcolor{red}{26.50} \\
					& SSIM & 0.6687 & 0.6773 & 0.6783 & 0.6805 & 0.6768 & 0.6796 & 0.6825 & 0.6688 & \textcolor{red}{0.6864} \\
					& LPIPS & 0.4990 & 0.4861 & 0.5011 & 0.4941 & 0.4958 & 0.4959 & 0.4769 & 0.5314 & \textcolor{red}{0.4701} \\
					& DISTS & 0.3063 & 0.3114 & 0.3141 & 0.3113 & 0.3140 & 0.3127 & 0.3084 & 0.3241 & \textcolor{red}{0.3007} \\
					& Inference Time & \textcolor{red}{41} & 172 & 113 & 91 & 64 & 52 & 540 & 688 & 91 \\
					& GPU Memory & 360 & 237 & \textcolor{red}{214} & 6268 & 3292 & 775 & 1470 & 5214 & 472 \\
					\bottomrule[2pt]
				\end{tabular}
			}
		}
		\label{tab:computational cost}
		\vspace{-1em}
	\end{table*}
	
	\vspace{-0.35em}
	\noindent \textbf{Rendering.} In Eq.~\ref{eq4}, during rasterization, the scaling factor $s$ serves as a condition to determine the sampling interval in 2D Gaussians $\mathcal{G}$. As illustrated in Fig.~\ref{fig:sampling}, since the rasterization is parameter-free, for different scaling factors $s$, it will adaptively adjust the sampling density to render images with different sizes. Larger scaling factors lead to smaller sampling intervals, vice versa. Therefore, GSASR is equipped with powerful generalization capability to arbitrary scaling factors, which are even unseen in the training process. 
	In the training stage, the rendered SR image after rasterization is compared with the HR ground truth to calculate $L_1$ loss. During back propagation, the gradient will pass through the differentiable rasterization to optimize the model backbone, guiding the framework to convert the input LR image into the corresponding image-conditioned 2D Gaussians $\mathcal{G}$ in a learning-based manner.
	
	\section{Experiments}
	\vspace{-0.15em}
	\subsection{Experimental Setup}
	\vspace{-0.15em}
	
	\noindent{\textbf{Implementation Details}}. We utilize the widely-used DIV2K \cite{timofte2017ntire} dataset as training set. We follow \cite{chen2021learning, cao2023ciaosr} to fix LR patches to $48 \times 48$. The scaling factor $s$ is randomly selected within the range $[1.0, 4.0]$. We first crop $48s \times 48s$ GT patch from the original full-size GT image, and then apply bicubic down-sampling \cite{cao2023ciaosr} to obtain the corresponding LR patches. Following LIIF \cite{chen2021learning}, we employ EDSR \cite{lim2017enhanced} and RDN \cite{zhang2018residual} as the image encoder backbones to extract features from the LR images.
	We train GSASR on $4$ NVIDIA A100 GPUs for $500,000$ iterations with batch size 64. The initial learning rate is $2e^{-4}$, and halves at $250,000$, $400,000$, $450,000$, $475,000$ iterations. The Adam \cite{kingma2014adam} optimizer is utilized. To speed up the convergence of the training progress, we set the number of warm-up iterations to $2,000$. The window size $k$ in window LR features, condition injection block and Gaussian interaction block is set to $12$, the proportion parameter $m$ in Gaussian embedding is set to $16$, the dimension $d$ of Gaussian embedding is set to $180$, and we stack $L=6$ Gaussian interaction blocks.

	
	\noindent{\textbf{Comparison Methods}}. We compare our proposed GSASR with Meta-SR \cite{hu2019meta}, LIIF \cite{chen2021learning}, LTE \cite{lee2022local}, SRNO \cite{wei2023super}, LINF \cite{yao2023local}, CiaoSR \cite{cao2023ciaosr}, LMF \cite{he2024latent} and GaussianSR \cite{hu2025gaussiansr}.
	
	\noindent{\textbf{Evaluation Protocols}}. We evaluate the competing models on Set5 \cite{bevilacqua2012low}, Set14 \cite{zeyde2010single}, DIV2K100 \cite{timofte2017ntire}, Urban100 \cite{huang2015single}, BSDS100 \cite{martin2001database}, Manga109 \cite{matsui2017sketch}, General100 \cite{dong2016accelerating}, and LSDIR \cite{li2023lsdir} datasets under scaling factors $\times2$, $\times3$, $\times4$, $\times6$, $\times8$, $\times12$. To validate the generalization on higher scaling factors, we further report the results with $\times16$, $\times18$, $\times24$, $\times30$ scaling factors on DIV2K100 \cite{timofte2017ntire} and LSDIR \cite{li2023lsdir} datasets. The PSNR, SSIM \cite{wang2004image}, LPIPS \cite{zhang2018unreasonable} and DISTS \cite{ding2020image} metrics are used for performance evaluation. Note that some existing methods \cite{chen2021learning, cao2023ciaosr} calculate PSNR on RGB channels for DIV2K \cite{timofte2017ntire}, but on Y channel of YCbCr space for other datasets. To unify the testing protocols, we calculate PSNR/SSIM on the Y channel across all datasets and all competing methods for comparison. For a fair comparison, we download all competing models from their official websites, utilize the same data to generate SR results and the same evaluation codes to compute the metrics.
	
	For the comparison of computational cost, we crop $100$ GT patches with $720 \times 720$ resolution from DIV2K \cite{timofte2017ntire}, and use bicubic interpolation to generate the corresponding LR images with scaling factors $\times2$, $\times3$, $\times4$, $\times6$, $\times8$, $\times12$. We report the average inference time (ms) and GPU memory usage (MB) on a single NVIDIA A100 GPU. The computational cost is calculated over the full SR process, including encoder, GS decoder and rendering.
	
	\begin{figure*}[t]
		\small
		\centering
		\begin{minipage}{0.3\textwidth}
			\includegraphics[width=1\linewidth]{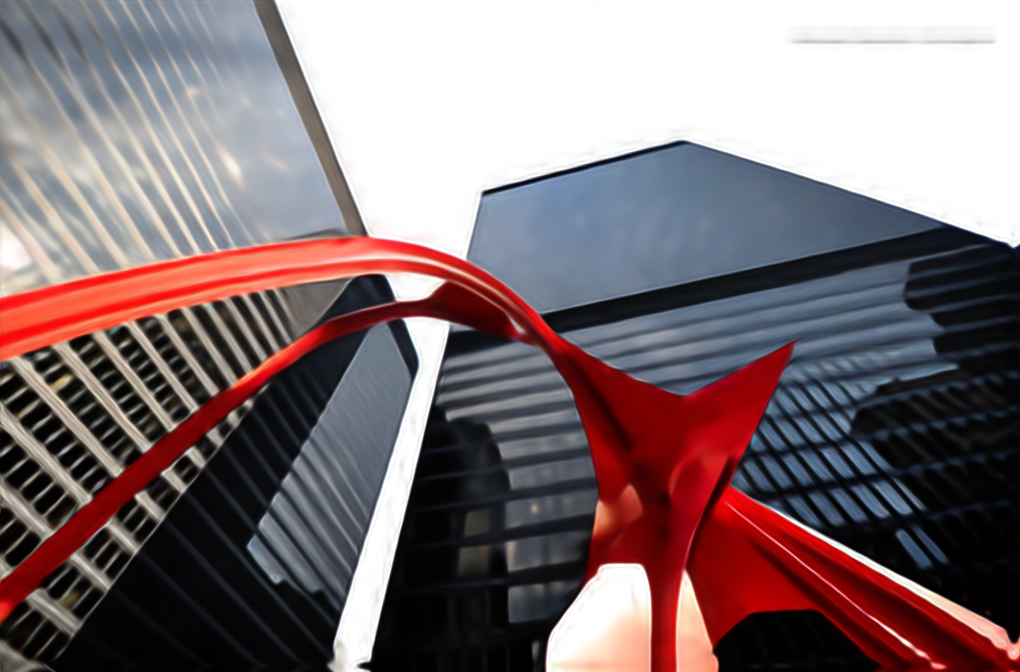}\\
			\centering{SR Image}\\
		\end{minipage}
		\begin{minipage}{0.3\textwidth}
			\includegraphics[width=1\linewidth]{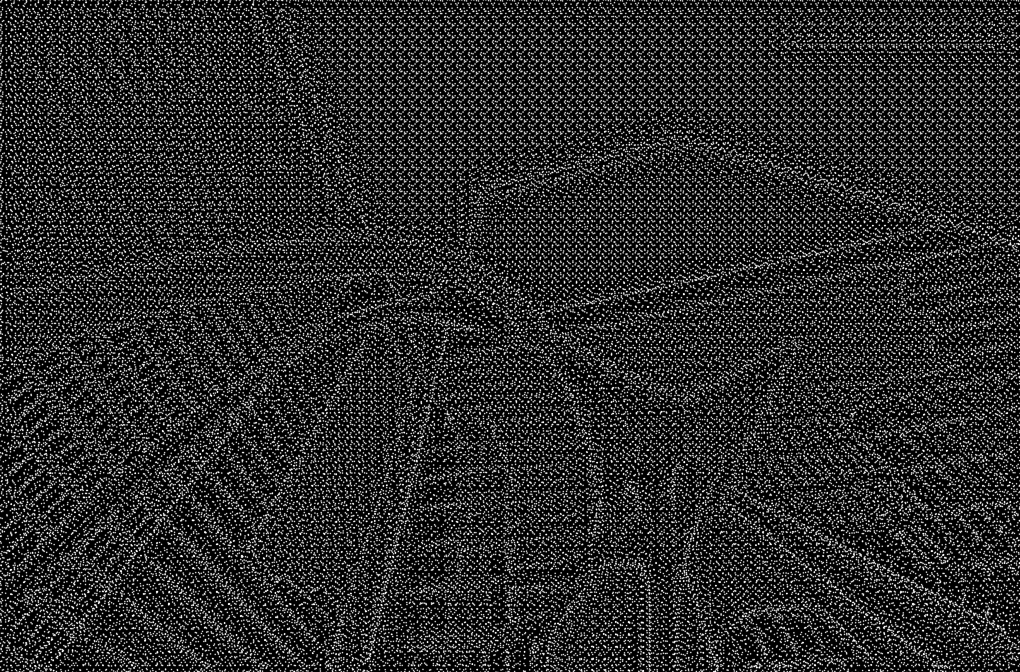}\\
			\centering{Position Distribution of Gaussians}\\
		\end{minipage}
		\begin{minipage}{0.3\textwidth}
			\includegraphics[width=1\linewidth]{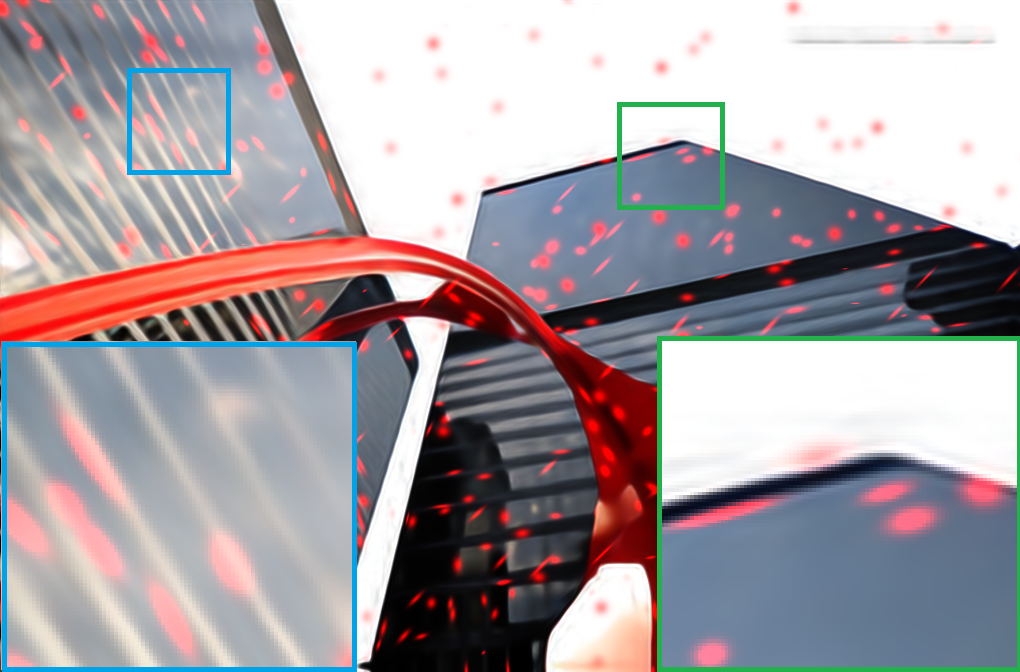}\\
			\centering{Colorized Gaussians}\\
		\end{minipage}
		\vspace{-0.8em}
		\caption{Demonstration of the rich expressiveness of Gaussians for ASR. The position distribution is obtained by setting $\{\sigma, \rho, c\}$ to fixed values. In the middle image, one could observe that Gaussians are evenly distributed in regions with simple textures or flatten area (such as the sky), while their positions are adjusted in regions with complex textures to fit details. (\textbf{Please zoom in for better observation}.) In the right image, we randomly select parts of Gaussians and highlight their colors to red. One can see that 2D Gaussians can learn to fit the different object shapes ($e.g.,$ the edge of window). }
		\label{fig:visualization of position and shape of Gaussians}
		\vspace{-2em}
	\end{figure*}

	\subsection{Experiment Results}
	
	\textbf{Quantitative Results}. Table~\ref{tab:results on DIV2K and LSDIR with EDSR} shows the numerical results of our proposed GSASR and other comparison methods with EDSR-baseline \cite{lim2017enhanced} encoder. One could see that GSASR outperforms existing methods in terms of both fidelity metrics (PSNR and SSIM \cite{wang2004image}) and perceptual quality metrics (LPIPS \cite{zhang2018unreasonable} and DISTS \cite{ding2020image}) under scaling factors from $\times 2$ to $\times 30$. It obtains significant improvements over the methods based on INR, exhibiting greater potential for ASR. Compared with the recent GaussianSR method \cite{hu2025gaussiansr}, which also employs GS for feature representation, GSASR shows clear advantages. GSASR predicts the position of Gaussians in an end-to-end manner, while GaussianSR \cite{hu2025gaussiansr} simply replaces the position parameters with the fixed position of the RGB pixels of the LR image and lacks sufficient representation capability. Besides, GaussianSR lacks GPU/CUDA-based rasterization to accelerate the inference speed and it fixes the number of Gaussians to $100$, no matter how large the scaling factor is. Therefore, the performance of GaussianSR drops a lot under high magnification factors. In contrast, we present an efficient 2D GPU/CUDA-based rasterization to embed more Gaussians to promote ASR performance. As a result, although GSASR is trained with a scaling factor from $1.0$ to $4.0$, it could not only obtain excellent performance for in-distribution scaling factors (\ie, scaling factor less than $4$), but also show strong capability in dealing with out-of-distribution situations (\ie, scaling factor larger than $4$). 
	
	Due to the limited space, we present more experimental results across more datasets (Set5 \cite{bevilacqua2012low}, Set14 \cite{zeyde2010single}, Urban100 \cite{huang2015single}, BSDS100 \cite{martin2001database}, Manga109 \cite{matsui2017sketch}, General100 \cite{dong2016accelerating}), more encoder backbones (EDSR-baseline \cite{lim2017enhanced}, RDN \cite{zhang2018residual}), and more protocols (PSNR/SSIM on RGB channels) in the \textbf{supplementary file}.
	
	\noindent\textbf{Qualitative Results}.
	Fig.~\ref{fig:visualization1} and Fig.~\ref{fig:visualization2} show the qualitative comparison of competing methods. We can see that the compared methods mostly produce blurry details (\eg, the holes in Fig.~\ref{fig:visualization1}), while GSASR generates much clearer edges and details, owing to the strong representation ability of 2D Gaussians. More visualization results can be found in the \textbf{supplementary file}. To explain why GSASR works well for the ASR task, we visualize the positions together with the shape of Gaussians in Fig.~\ref{fig:visualization of position and shape of Gaussians}. One could see that the positions of Gaussians tend to cluster in areas with complex textures (such as the windows), while they are uniformly distributed in flat areas (such as the sky). This validates that our proposed strategy is highly suitable for restoring complex texture details. Meanwhile, from the blue and green boxes in Fig.~\ref{fig:visualization of position and shape of Gaussians}, one could find that the Gaussians will adaptively adjust their orientations and scales based on the direction and shape of the textures.
	
	\noindent\textbf{Computational Costs}.
	Table~\ref{tab:computational cost} shows that GSASR surpasses all competing methods in most of the fidelity (PSNR, SSIM) and perceptual (LPIPS, DISTS) measures. In terms of speed, GSASR is nearly twice as fast as the state-of-the-art method CiaoSR \cite{cao2023ciaosr} under $\times 3, \times 4$ scaling factors. The speed advantage becomes more significant when handling with larger scaling factors. 
	Thanks to the efficient 2D GPU/CUDA-based rasterization, GSASR could render an image with high speed without sacrificing the reconstruction fidelity. We further compare the rendering cost between our CUDA-based rasterization and the Pytorch-based one in GaussianSR \cite{hu2025gaussiansr} in the \textbf{supplementary file}. The computational cost of the RDN backbone can also be found in the \textbf{supplementary file}. 
	
	\noindent\textbf{Ablation Study}.
	\label{ablation study}
	We conduct ablation studies on (1) the number of Gaussians $N$, (2) the functionality of the reference position $p$, (3) the rasterization ratio $r$, (4) the dimension $d$ of Gaussian embedding, (5) the window size $k$ in partitioned window LR features, condition injection block and Gaussian interaction block, (6) the functionality of the learnable standard deviation parameters $o$, and (7) the functionality of learnable scaling parameters $\sigma$. Due to the limited space, details and results of all thoes ablation studies are presented in the \textbf{supplementary file}.
	
	Finally, we discuss and explore the performance of GSASR with larger backbone in the \textbf{supplementary file}.
	
	\vspace{-0.15cm}
	\section{Conclusion and Limitation}
	\vspace{-0.15cm}
	We presented GSASR, a brandly-new 2D GS-based ASR model. To adapt GS to the ASR task, we first elaborately designed an architecture to convert an LR input to image-conditioned Gaussians, then implemented an efficient differentiable scale-aware 2D GPU/CUDA-based rasterization to render images with both high quality and fast speed. Through sampling values from Gaussians, we could render an output image with arbitrary magnification factors. Extensive experiments demonstrated that our GSASR model has much more powerful representation capability, together with more friendly computational costs, than implicit neural function based methods. 
	
	\noindent\textbf{Limitations}. GSASR still has some limitations. First, its performance depends on the number of Gaussians, especially under ultra-high scaling factors. While employing more Gaussians could bring better performance, it will sacrifice inference speed. Second, to convert an image into Gaussians, a large number of parameters are used to fit the complex mapping function. How to reduce the number of parameters needs further investigation.

	{
		\small
		\bibliographystyle{ieeenat_fullname}
		\bibliography{main}
	}

\end{document}


\maketitle
	
	\noindent In this supplementary file, we provide the following materials:
	\begin{itemize}
		\item More quantitative results on other testing datasets (please refer to Section 4.2 in the main paper);
		\item More quantitative results on RGB channels (please refer to Section 4.2 in the main paper);
		\item More qualitative results (please refer to Section 4.2 in the main paper);
		\item Rendering cost comparison between GSASR and GaussianSR (please refer to Section 4.2 in the main paper);
		\item More results on the computational costs (please refer to Section 4.2 in the main paper);
		\item Ablation study (please refer to Section 4.2 in the main paper).
		\item Performance of GSASR with larger backbone (please refer to Section 4.2 in the main paper).

	\end{itemize}

	\section{Quantitative Results on More Testing Benchmarks}
	
	\begin{table*}[h]
		\centering
		\caption{Quantitative comparison between representative ASR models and our GSASR. All models use the same RDN \cite{zhang2018residual} backbone as the feature extraction encoder, and are tested on DIV2K and LSDIR \cite{li2023lsdir} datasets \cite{timofte2017ntire} with scaling factors $\times2$, $\times3$, $\times4$, $\times6$, $\times8$, $\times12$, $\times16$, $\times18$, $\times24$, $\times30$. The best results are highlighted in \textcolor{red}{red}. The PSNR and SSIM metrics are computed on the Y channel of Ycbcr space.}
		\vspace{-0.2cm}
		\scalebox{1.0}{
			\resizebox{\linewidth}{!}{
				\begin{tabular}{c|c|ccccccccc|ccccccccc} 
					\toprule[2pt]
					\multirow{3}{*}{Scale} & \multirow{3}{*}{Metrics} & \multicolumn{18}{c}{Backbone: RDN} \\ 
					\cline{3-20}
					&  & \multicolumn{9}{c|}{Testing Dataset: DIV2K} & \multicolumn{9}{c}{Testing Dataset: LSDIR} \\
					&  & \begin{tabular}[c]{@{}c@{}}Meta\\-SR\end{tabular} & LIIF & LTE & SRNO & LINF & LMF & \begin{tabular}[c]{@{}c@{}}Ciao\\-SR\end{tabular} & \begin{tabular}[c]{@{}c@{}}Gaussian\\-SR\end{tabular} & GSASR & \begin{tabular}[c]{@{}c@{}}Meta\\-SR\end{tabular} & LIIF & LTE & SRNO & LINF & LMF & \begin{tabular}[c]{@{}c@{}}Ciao\\-SR\end{tabular} & \begin{tabular}[c]{@{}c@{}}Gaussian\\-SR\end{tabular} & GSASR \\ 
					\midrule[1pt] \midrule[1pt]
					\multirow{4}{*}{$\times$2} & PSNR & 36.54 & 36.37 & 36.41 & 36.53 & 36.53 & 36.48 & 36.67 & 36.54~ & \textcolor{red}{36.73} & 31.97 & 31.87 & 31.96 & 32.10 & 31.94 & 32.02 & 32.16 & 31.95~ & \textcolor{red}{32.26} \\
					& SSIM & 0.9483 & 0.9481 & 0.9484 & 0.9493 & 0.9483 & 0.9487 & 0.9494 & 0.9484~ & \textcolor{red}{0.9500} & 0.9224 & 0.9221 & 0.9228 & 0.9247 & 0.9223 & 0.9234 & 0.9247 & 0.9228~ & \textcolor{red}{0.9261} \\
					& LPIPS & 0.0816 & 0.0828 & 0.0816 & 0.0783 & 0.0821 & 0.0806 & 0.0797 & 0.0829~ & \textcolor{red}{0.0757} & 0.0872 & 0.0876 & 0.0861 & 0.0832 & 0.0874 & 0.0866 & 0.0839 & 0.0875~ & \textcolor{red}{0.0812} \\
					& DISTS & 0.0542 & 0.0543 & 0.0535 & 0.0528 & 0.0541 & 0.0540 & 0.0524 & 0.0543~ & \textcolor{red}{0.0509} & 0.0644 & 0.0645 & 0.0635 & 0.0623 & 0.0645 & 0.0639 & 0.0615 & 0.0640~ & \textcolor{red}{0.0603} \\ 
					\midrule[1pt]
					\multirow{4}{*}{$\times$3} & PSNR & 32.79 & 32.68 & 32.73 & 32.83 & 32.77 & 32.78 & 32.89 & 32.74~ & \textcolor{red}{32.97} & 28.34 & 28.26 & 28.34 & 28.45 & 28.30 & 28.37 & 28.46 & 28.33~ & \textcolor{red}{28.56} \\
					& SSIM & 0.8935 & 0.8934 & 0.8938 & 0.8951 & 0.8936 & 0.8947 & 0.8952 & 0.8939~ & \textcolor{red}{0.8970} & 0.8399 & 0.8398 & 0.8409 & 0.8435 & 0.8398 & 0.8421 & 0.8432 & 0.8407~ & \textcolor{red}{0.8462} \\
					& LPIPS & 0.1850 & 0.1842 & 0.1837 & 0.1817 & 0.1846 & 0.1807 & 0.1821 & 0.1847~ & \textcolor{red}{0.1768} & 0.1997 & 0.1983 & 0.1979 & 0.1947 & 0.1995 & 0.1954 & 0.1953 & 0.1993~ & \textcolor{red}{0.1908} \\
					& DISTS & 0.1000 & 0.0998 & 0.0987 & 0.0980 & 0.0999 & 0.0984 & 0.0965 & 0.0998~ & \textcolor{red}{0.0955} & 0.1192 & 0.1188 & 0.1176 & 0.1159 & 0.1194 & 0.1169 & 0.1139 & 0.1189~ & \textcolor{red}{0.1139} \\ 
					\midrule[1pt]
					\multirow{4}{*}{$\times$4} & PSNR & 30.78 & 30.71 & 30.75 & 30.85 & 30.77 & 30.81 & 30.91 & 30.76~ & \textcolor{red}{30.96} & 26.53 & 26.48 & 26.54 & 26.64 & 26.51 & 26.58 & 26.66 & 26.53~ & \textcolor{red}{26.73} \\
					& SSIM & 0.8455 & 0.8449 & 0.8459 & 0.8478 & 0.8453 & 0.8467 & 0.8481 & 0.8457~ & \textcolor{red}{0.8500} & 0.7724 & 0.7714 & 0.7734 & 0.7767 & 0.7719 & 0.7744 & 0.7770 & 0.7727~ & \textcolor{red}{0.7801} \\
					& LPIPS & 0.2568 & 0.2566 & 0.2558 & 0.2525 & 0.2574 & 0.2546 & 0.2525 & 0.2570~ & \textcolor{red}{0.2505} & 0.2831 & 0.2838 & 0.2822 & 0.2771 & 0.2840 & 0.2817 & 0.2768 & 0.2837~ & \textcolor{red}{0.2752} \\
					& DISTS & 0.1341 & 0.1354 & 0.1341 & 0.1330 & 0.1355 & 0.1341 & 0.1327 & 0.1347~ & \textcolor{red}{0.1288} & 0.1581 & 0.1603 & 0.1589 & 0.1567 & 0.1609 & 0.1584 & 0.1563 & 0.1595~ & \textcolor{red}{0.1533} \\ 
					\midrule[1pt]
					\multirow{4}{*}{$\times$6} & PSNR & 28.40 & 28.44 & 28.49 & 28.57 & 28.48 & 28.52 & 28.61 & 28.46~ & \textcolor{red}{28.65} & 24.55 & 24.59 & 24.64 & 24.70 & 24.61 & 24.66 & 24.73 & 24.60~ & \textcolor{red}{24.77} \\
					& SSIM & 0.7715 & 0.7739 & 0.7749 & 0.7772 & 0.7739 & 0.7757 & 0.7777 & 0.7727~ & \textcolor{red}{0.7800} & 0.6756 & 0.6791 & 0.6806 & 0.6841 & 0.6787 & 0.6815 & 0.6853 & 0.6772~ & \textcolor{red}{0.6881} \\
					& LPIPS & 0.3505 & 0.3468 & 0.3551 & 0.3510 & 0.3523 & 0.3534 & 0.3452 & 0.3584~ & \textcolor{red}{0.3443} & 0.3960 & 0.3915 & 0.4020 & 0.3964 & 0.3961 & 0.4009 & 0.3864 & 0.4043~ & \textcolor{red}{0.3867} \\
					& DISTS & 0.1837 & 0.1882 & 0.1878 & 0.1861 & 0.1890 & 0.1873 & 0.1867 & 0.1880~ & \textcolor{red}{0.1817} & 0.2110 & 0.2166 & 0.2161 & 0.2131 & 0.2181 & 0.2154 & 0.2131 & 0.2167~ & \textcolor{red}{0.2093} \\ 
					\midrule[1pt]
					\multirow{4}{*}{$\times$8} & PSNR & 26.97 & 27.07 & 27.13 & 27.20 & 27.11 & 27.16 & 27.24 & 26.99~ & \textcolor{red}{27.28} & 23.44 & 23.49 & 23.54 & 23.60 & 23.52 & 23.57 & 23.63 & 23.45~ & \textcolor{red}{23.63} \\
					& SSIM & 0.7222 & 0.7274 & 0.7286 & 0.7308 & 0.7272 & 0.7293 & 0.7317 & 0.7228~ & \textcolor{red}{0.7339} & 0.6151 & 0.6221 & 0.6230 & 0.6264 & 0.6211 & 0.6240 & 0.6281 & 0.6156~ & \textcolor{red}{0.6299} \\
					& LPIPS & 0.4133 & 0.4072 & 0.4214 & 0.4166 & 0.4163 & 0.4188 & 0.4075 & 0.4290~ & \textcolor{red}{0.4060} & 0.4679 & 0.4627 & 0.4798 & 0.4729 & 0.4711 & 0.4772 & \textcolor{red}{0.4592} & 0.4875~ & 0.4593 \\
					& DISTS & 0.2181 & 0.2251 & 0.2268 & 0.2248 & 0.2273 & 0.2260 & 0.2244 & 0.2285~ & \textcolor{red}{0.2192} & 0.2489 & 0.2562 & 0.2569 & 0.2537 & 0.2588 & 0.2563 & 0.2528 & 0.2601~ & \textcolor{red}{0.2490} \\ 
					\midrule[1pt]
					\multirow{4}{*}{$\times$12} & PSNR & 25.23 & 25.39 & 25.45 & 25.52 & 25.41 & 25.48 & 25.55 & 25.12~ & \textcolor{red}{25.56} & 22.13 & 22.21 & 22.26 & 22.31 & 22.22 & 22.28 & \textcolor{red}{22.33} & 22.06~ & 22.30 \\
					& SSIM & 0.6653 & 0.6739 & 0.6748 & 0.6768 & 0.6730 & 0.6754 & 0.6783 & 0.6648~ & \textcolor{red}{0.6794} & 0.5494 & 0.5597 & 0.5605 & 0.5629 & 0.5576 & 0.5608 & 0.5651 & 0.5484~ & \textcolor{red}{0.5656} \\
					& LPIPS & 0.5001 & 0.4981 & 0.5180 & 0.5132 & 0.5120 & 0.5153 & 0.4982 & 0.5455~ & \textcolor{red}{0.4947} & 0.5671 & 0.5685 & 0.5899 & 0.5823 & 0.5803 & 0.5884 & 0.5635 & 0.6174~ & \textcolor{red}{0.5621} \\
					& DISTS & 0.2703 & 0.2809 & 0.2857 & 0.2835 & 0.2855 & 0.2844 & 0.2811 & 0.2921~ & \textcolor{red}{0.2759} & 0.3055 & 0.3138 & 0.3165 & 0.3133 & 0.3187 & 0.3160 & 0.3116 & 0.3252~ & \textcolor{red}{0.3077} \\ 
					\midrule[1pt]
					\multirow{4}{*}{$\times$16} & PSNR & 24.12 & 24.30 & 24.36 & 24.42 & 24.30 & 24.38 & 24.44 & 23.93~ & \textcolor{red}{24.44} & 19.34 & 21.40 & 21.45 & 21.49 & 21.41 & 21.45 & \textcolor{red}{21.50} & 21.20~ & 21.47 \\
					& SSIM & 0.6346 & 0.6446 & 0.6452 & 0.6467 & 0.6433 & 0.6456 & 0.6484 & 0.6340~ & \textcolor{red}{0.6487} & 0.4813 & 0.5273 & 0.5277 & 0.5294 & 0.5253 & 0.5269 & 0.5318 & 0.5156~ & \textcolor{red}{0.5318} \\
					& LPIPS & 0.5606 & 0.5580 & 0.5788 & 0.5739 & 0.5751 & 0.5760 & 0.5572 & 0.6166~ & \textcolor{red}{0.5537} & 0.6762 & 0.6343 & 0.6556 & 0.6487 & 0.6488 & 0.6598 & 0.6281 & 0.6932~ & \textcolor{red}{0.6261} \\
					& DISTS & 0.3117 & 0.3243 & 0.3312 & 0.3288 & 0.3324 & 0.3295 & 0.3268 & 0.3455~ & \textcolor{red}{0.3211} & 0.3845 & 0.3566 & 0.3605 & 0.3575 & 0.3638 & 0.3596 & 0.3562 & 0.3745~ & \textcolor{red}{0.3512} \\ 
					\midrule[1pt]
					\multirow{4}{*}{$\times$18} & PSNR & 23.69 & 23.87 & 23.93 & 23.99 & 23.87 & 23.96 & \textcolor{red}{24.01} & 23.48~ & 23.98 & 20.99 & 21.09 & 21.14 & 21.17 & 21.10 & 21.13 & \textcolor{red}{21.18} & 20.88~ & 21.14 \\
					& SSIM & 0.6244 & 0.6346 & 0.6349 & 0.6362 & 0.6333 & 0.6353 & \textcolor{red}{0.6380} & 0.6239~ & 0.6379 & 0.5056 & 0.5165 & 0.5168 & 0.5181 & 0.5147 & 0.5160 & \textcolor{red}{0.5205} & 0.5056~ & 0.5203 \\
					& LPIPS & 0.5776 & 0.5805 & 0.6013 & 0.5968 & 0.5957 & 0.5987 & 0.5796 & 0.6418~ & \textcolor{red}{0.5766} & 0.6489 & 0.6575 & 0.6490 & 0.6720 & 0.6696 & 0.6822 & 0.6511 & 0.7188~ & \textcolor{red}{0.6497} \\
					& DISTS & 0.3303 & 0.3419 & 0.3499 & 0.3476 & 0.3513 & 0.3480 & 0.3455 & 0.3676~ & \textcolor{red}{0.3391} & 0.3633 & 0.3733 & 0.3784 & 0.3752 & 0.3814 & 0.3764 & 0.3740 & 0.3944~ & \textcolor{red}{0.3682} \\ 
					\midrule[1pt]
					\multirow{4}{*}{$\times$24} & PSNR & 22.71 & 22.87 & 22.92 & 22.97 & 22.87 & 22.93 & \textcolor{red}{22.99} & 22.49~ & 22.96 & 19.26 & 20.39 & 20.44 & 20.47 & 20.41 & 20.43 & \textcolor{red}{20.48} & 20.16~ & 20.42 \\
					& SSIM & 0.6041 & 0.6139 & 0.6138 & 0.6145 & 0.6128 & 0.6136 & \textcolor{red}{0.6166} & 0.6048~ & 0.6162 & 0.4728 & 0.4967 & 0.4965 & 0.4973 & 0.4953 & 0.4955 & \textcolor{red}{0.4995} & 0.4873~ & 0.4985 \\
					& LPIPS & 0.6273 & 0.6301 & 0.6505 & 0.6461 & 0.6423 & 0.6487 & 0.6292 & 0.6939~ & \textcolor{red}{0.6275} & 0.7112 & 0.7073 & 0.7274 & 0.7207 & 0.7143 & 0.7316 & \textcolor{red}{0.7005} & 0.7701~ & 0.7006 \\
					& DISTS & 0.3830 & 0.3865 & 0.3978 & 0.3955 & 0.4008 & 0.3961 & 0.3932 & 0.4224~ & \textcolor{red}{0.3851} & 0.4205 & 0.4145 & 0.4220 & 0.4185 & 0.4260 & 0.4212 & 0.4172 & 0.4434~ & \textcolor{red}{0.4095} \\ 
					\midrule[1pt]
					\multirow{4}{*}{$\times$30} & PSNR & 17.25 & 22.18 & 22.23 & 22.27 & 32.18 & 22.18 & \textcolor{red}{22.28} & 21.81~ & 22.23 & 19.78 & 19.86 & 19.91 & 19.93 & 19.89 & 19.90 & \textcolor{red}{19.94} & 19.64~ & 19.86 \\
					& SSIM & 0.5546 & 0.6016 & 0.6017 & 0.6024 & 0.6010 & 0.5995 & \textcolor{red}{0.6041} & 0.5947~ & 0.6032 & 0.4758 & 0.4842 & 0.4840 & 0.4844 & 0.4835 & 0.4837 & \textcolor{red}{0.4861} & 0.4770~ & 0.4851 \\
					& LPIPS & 0.6817 & 0.6652 & 0.6833 & 0.6799 & 0.6757 & 0.7009 & \textcolor{red}{0.6627} & 0.7247~ & 0.6633 & 0.7164 & 0.7418 & 0.7598 & 0.7540 & 0.7468 & 0.7570 & \textcolor{red}{0.7269} & 0.8009~ & 0.7358 \\
					& DISTS & 0.4405 & 0.4215 & 0.4359 & 0.4332 & 0.4394 & 0.4361 & 0.4301 & 0.4650~ & \textcolor{red}{0.4213} & 0.4509 & 0.4454 & 0.4562 & 0.4520 & 0.4605 & 0.4536 & 0.4495 & 0.4823~ & \textcolor{red}{0.4407} \\
					\bottomrule[2pt]
				\end{tabular}
			}
		}
		\label{tab:results on DIV2K and LSDIR with RDN}
	\end{table*}
	
	\begin{table*}[h]
		\centering
		\caption{Quantitative results of  representative ASR models and our proposed GSASR. All models use the same EDSR-backbone \cite{lim2017enhanced} as the feature extraction encoder, and are tested on Urban100 \cite{huang2015single} and Manga109 \cite{matsui2017sketch} with scaling factors $\times2$, $\times3$, $\times4$, $\times6$, $\times8$, $\times12$. The best results are highlighted in \textcolor{red}{red}. The PSNR and SSIM \cite{wang2004image} metrics are computed on the Y channel of Ycbcr space.}
		\vspace{-0.2cm}
		\scalebox{1.0}{
			\resizebox{\linewidth}{!}{
				\setlength{\extrarowheight}{2.0pt} 
				\begin{tabular}{c|c|ccccccccc|ccccccccc} 
					\toprule[2pt]
					\multirow{3}{*}{Scale} & \multirow{3}{*}{Metrics} & \multicolumn{18}{c}{Backbone: EDSR-baseline} \\ 
					\cline{3-20}
					&  & \multicolumn{9}{c|}{Testing Dataset: Urban100} & \multicolumn{9}{c}{Testing Dataset: Manga109} \\
					&  & \begin{tabular}[c]{@{}c@{}}Meta\\-SR\end{tabular} & LIIF & LTE & SRNO & LINF & LMF & \begin{tabular}[c]{@{}c@{}}Ciao\\-SR\end{tabular} & \begin{tabular}[c]{@{}c@{}}Gaussian\\-SR\end{tabular} & GSASR & \begin{tabular}[c]{@{}c@{}}Meta\\-SR\end{tabular} & LIIF & LTE & SRNO & LINF & LMF & \begin{tabular}[c]{@{}c@{}}Ciao\\-SR\end{tabular} & \begin{tabular}[c]{@{}c@{}}Gaussian\\-SR\end{tabular} & GSASR \\ 
					\midrule[1pt] \midrule[1pt]
					\multirow{4}{*}{$\times$2} & PSNR & 32.05~ & 32.12~ & 32.26 & 32.56 & 32.12 & 32.48 & 32.79~ & 32.22~ & \textcolor{red}{33.27~} & 38.41~ & 38.55~ & 38.52 & 38.92 & 38.72 & 38.75 & 39.16~ & 38.55~ & \textcolor{red}{39.06~} \\
					& SSIM & 0.9280~ & 0.9288~ & 0.9299 & 0.9324 & 0.9283 & 0.9323 & 0.9344~ & 0.9296~ & \textcolor{red}{0.9389~} & 0.9769~ & 0.9770~ & 0.9770 & 0.9778 & 0.9773 & 0.9774 & 0.9781~ & 0.9771~ & \textcolor{red}{0.9782~} \\
					& LPIPS & 0.0659~ & 0.0643~ & 0.0629 & 0.0597 & 0.0660 & 0.0603 & 0.0580~ & 0.0590~ & \textcolor{red}{0.0525~} & 0.0239~ & 0.0238~ & 0.0229 & 0.0229 & 0.0239 & 0.0227 & 0.0224~ & 0.0237~ & \textcolor{red}{0.0209~} \\
					& DISTS & 0.0712~ & 0.0717~ & 0.0711 & 0.0692 & 0.0717 & 0.0696 & 0.0676~ & 0.0712~ & \textcolor{red}{0.0634~} & 0.0226~ & 0.0229~ & 0.0234 & 0.0227 & 0.0231 & 0.0228 & 0.0222~ & 0.0233~ & \textcolor{red}{0.0215~} \\ 
					\midrule[1pt]
					\multirow{4}{*}{$\times$3} & PSNR & 28.10~ & 28.20~ & 28.31 & 28.54 & 28.21 & 28.59 & 28.67~ & 28.27~ & \textcolor{red}{29.17~} & 33.49~ & 33.46~ & 33.54 & 33.88 & 33.57 & 33.73 & 33.92~ & 33.55~ & \textcolor{red}{34.10~} \\
					& SSIM & 0.8520~ & 0.8544~ & 0.8561 & 0.8599 & 0.8538 & 0.8615 & 0.8627~ & 0.8553~ & \textcolor{red}{0.8733~} & 0.9442~ & 0.9449~ & 0.9454 & 0.9470 & 0.9453 & 0.9467 & 0.9474~ & 0.9452~ & \textcolor{red}{0.9500~} \\
					& LPIPS & 0.1591~ & 0.1542~ & 0.1513 & 0.1465 & 0.1570 & 0.1437 & 0.1408~ & 0.1541~ & \textcolor{red}{0.1320~} & 0.0669~ & 0.0652~ & 0.0643 & 0.0624 & 0.0652 & 0.0618 & \textcolor{red}{0.0601~} & 0.0656~ & 0.0604~ \\
					& DISTS & 0.1277~ & 0.1290~ & 0.1278 & 0.1243 & 0.1291 & 0.1250 & 0.1181~ & 0.1288~ & \textcolor{red}{0.1156~} & 0.0505~ & 0.0528~ & 0.0536 & 0.0515 & 0.0522 & 0.0515 & \textcolor{red}{0.0477~} & 0.0544~ & 0.0487~ \\ 
					\midrule[1pt]
					\multirow{4}{*}{$\times$4} & PSNR & 25.94~ & 26.14~ & 26.24 & 26.48 & 26.16 & 26.50 & 26.69~ & 26.19~ & \textcolor{red}{27.01~} & 30.37~ & 30.54~ & 30.58 & 30.83 & 30.52 & 30.76 & 30.99~ & 30.54~ & \textcolor{red}{31.16~} \\
					& SSIM & 0.7824~ & 0.7885~ & 0.7910 & 0.7976 & 0.7878 & 0.7985 & 0.8031~ & 0.7893~ & \textcolor{red}{0.8142~} & 0.9063~ & 0.9097~ & 0.9108 & 0.9137 & 0.9104 & 0.9130 & 0.9160~ & 0.9094~ & \textcolor{red}{0.9194~} \\
					& LPIPS & 0.2367~ & 0.2271~ & 0.2223 & 0.2136 & 0.2308 & 0.2151 & 0.2078~ & 0.2283~ & \textcolor{red}{0.1987~} & 0.1112~ & 0.1072~ & 0.1064 & 0.1017 & 0.1079 & 0.1032 & 0.0991~ & 0.1085~ & \textcolor{red}{0.0990~} \\
					& DISTS & 0.1683~ & 0.1738~ & 0.1718 & 0.1679 & 0.1748 & 0.1696 & 0.1659~ & 0.1730~ & \textcolor{red}{0.1552~} & 0.0735~ & 0.0814~ & 0.0831 & 0.0804 & 0.0818 & 0.0804 & 0.0802~ & 0.0831~ & \textcolor{red}{0.0749~} \\ 
					\midrule[1pt]
					\multirow{4}{*}{$\times$6} & PSNR & 23.57~ & 23.78~ & 23.84 & 24.07 & 23.79 & 24.08 & 24.23~ & 23.77~ & \textcolor{red}{24.51~} & 26.26~ & 26.73~ & 26.85 & 27.05 & 26.71 & 26.95 & 27.16~ & 26.63~ & \textcolor{red}{27.36~} \\
					& SSIM & 0.6726~ & 0.6850~ & 0.6872 & 0.6956 & 0.6836 & 0.6968 & 0.7029~ & 0.6814~ & \textcolor{red}{0.7165~} & 0.8214~ & 0.8378~ & 0.8401 & 0.8455 & 0.8376 & 0.9428 & 0.8486~ & 0.8333~ & \textcolor{red}{0.8539~} \\
					& LPIPS & 0.3540~ & 0.3355~ & 0.3431 & 0.3292 & 0.3425 & 0.3327 & 0.3117~ & 0.3499~ & \textcolor{red}{0.2975~} & 0.2024~ & 0.1853~ & 0.1883 & 0.1784 & 0.1859 & 0.1840 & 0.1693~ & 0.1942~ & \textcolor{red}{0.1680~} \\
					& DISTS & 0.2312~ & 0.2354~ & 0.2320 & 0.2258 & 0.2374 & 0.2300 & 0.2250~ & 0.2364~ & \textcolor{red}{0.2111~} & 0.1182~ & 0.1255~ & 0.1240 & 0.1208 & 0.1265 & 0.1215 & 0.1235~ & 0.1271~ & \textcolor{red}{0.1169~} \\ 
					\midrule[1pt]
					\multirow{4}{*}{$\times$8} & PSNR & 22.27~ & 22.45~ & 22.53 & 22.69 & 22.45 & 22.73 & 22.83~ & 22.36~ & \textcolor{red}{23.09~} & 24.06~ & 24.53~ & 24.63 & 24.78 & 24.51 & 24.71 & 24.93~ & 24.26~ & \textcolor{red}{25.09~} \\
					& SSIM & 0.6004~ & ¥0.6169 & 0.6190 & 0.6269 & 0.6141 & 0.6286 & 0.6344~ & 0.6070~ & \textcolor{red}{0.6480~} & 0.7529~ & 0.7786~ & 0.7810 & 0.7871 & 0.7779 & 0.7837 & 0.7933~ & 0.7657~ & \textcolor{red}{0.7981~} \\
					& LPIPS & 0.4536~ & ¥0.4201 & 0.4341 & 0.4168 & 0.4301 & 0.4205 & 0.3932~ & 0.4530~ & \textcolor{red}{0.3754~} & 0.2989~ & 0.2603~ & 0.2677 & 0.2533 & 0.2595 & 0.2620 & 0.2343~ & 0.2868~ & \textcolor{red}{0.2304~} \\
					& DISTS & 0.2765~ & ¥0.2795 & 0.2763 & 0.2698 & 0.2831 & 0.2748 & 0.2666~ & 0.2883~ & \textcolor{red}{0.2515~} & 0.1576~ & 0.1630~ & 0.1603 & 0.1559 & 0.1640 & 0.1584 & 0.1565~ & 0.1714~ & \textcolor{red}{0.1480~} \\ 
					\midrule[1pt]
					\multirow{4}{*}{$\times$12} & PSNR & 20.77~ & 20.89~ & 20.96 & 21.10 & 20.88 & 21.12 & 21.19~ & 20.68~ & \textcolor{red}{21.39~} & 21.71~ & 22.05~ & 22.13 & 22.22 & 22.02 & 22.17 & 22.34~ & 21.63~ & \textcolor{red}{22.43~} \\
					& SSIM & 0.5191~ & 0.5370~ & 0.5383 & 0.5454 & 0.5337 & 0.5460 & 0.5530~ & 0.5217~ & \textcolor{red}{0.5628~} & 0.6655~ & 0.6967~ & 0.6982 & 0.7029 & 0.6953 & 0.7001 & 0.7124~ & 0.6745~ & \textcolor{red}{0.7147~} \\
					& LPIPS & 0.5818~ & 0.5540~ & 0.5729 & 0.5544 & 0.5676 & 0.5405 & 0.5204~ & 0.6198~ & \textcolor{red}{0.4994~} & 0.4324~ & 0.3931~ & 0.4046 & 0.3897 & 0.3900 & 0.3973 & 0.3540~ & 0.4508~ & \textcolor{red}{0.3433~} \\
					& DISTS & 0.3397~ & 0.3421~ & 0.3399 & 0.3333 & 0.3479 & 0.3388 & 0.3278~ & 0.3646~ & \textcolor{red}{0.3135~} & 0.2201~ & 0.2252~ & 0.2215 & 0.2168 & 0.2271 & 0.2205 & 0.2145~ & 0.2466~ & \textcolor{red}{0.2014~} \\
					\bottomrule[2pt]
			\end{tabular}}
		}
		
		\label{tab:results on Urban100 and Manga109 with EDSR-baseline}
	\end{table*}

	\begin{table*}[h]
		\centering
		\caption{Quantitative results of  representative ASR models and our proposed GSASR. All models use the same RDN backbone \cite{zhang2018image} as the feature extraction encoder, and are tested on Urban100 \cite{huang2015single} and Manga109 \cite{matsui2017sketch} with scaling factors $\times2$, $\times3$, $\times4$, $\times6$, $\times8$, $\times12$. The best results are highlighted in \textcolor{red}{red}. The PSNR and SSIM \cite{wang2004image} metrics are computed on the Y channel of Ycbcr space.}
		\scalebox{1.0}{
			\resizebox{\linewidth}{!}{
				\setlength{\extrarowheight}{2.0pt} 
				\begin{tabular}{c|c|ccccccccc|ccccccccc} 
					\bottomrule[2pt]
					\multirow{3}{*}{Scale} & \multirow{3}{*}{Metrics} & \multicolumn{18}{c}{Backbone: RDN} \\ 
					\cline{3-20}
					&  & \multicolumn{9}{c|}{Testing Dataset: Urban100} & \multicolumn{9}{c}{Testing Dataset: Manga109} \\
					&  & \begin{tabular}[c]{@{}c@{}}Meta\\-SR\end{tabular} & LIIF & LTE & SRNO & LINF & LMF & \begin{tabular}[c]{@{}c@{}}Ciao\\-SR\end{tabular} & \begin{tabular}[c]{@{}c@{}}Gaussian\\-SR\end{tabular} & GSASR & \begin{tabular}[c]{@{}c@{}}Meta\\-SR\end{tabular} & LIIF & LTE & SRNO & LINF & LMF & \begin{tabular}[c]{@{}c@{}}Ciao\\-SR\end{tabular} & \begin{tabular}[c]{@{}c@{}}Gaussian\\-SR\end{tabular} & GSASR \\ 
					\midrule[1pt] \midrule[1pt]
					\multirow{4}{*}{$\times$2} & PSNR & 33.04 & 32.84~ & 33.00 & 33.27 & 32.87 & 33.08 & 33.30~ & 32.96~ & \textcolor{red}{33.53~} & 39.34 & 39.04~ & 39.08 & 39.35 & 39.23 & 39.24 & \textcolor{red}{39.55~} & 39.16~ & 39.17~ \\
					& SSIM & 0.9363 & 0.9353~ & 0.9365 & 0.9390 & 0.9350 & 0.9371 & 0.9388~ & 0.9363~ & \textcolor{red}{0.9406~} & 0.9783 & 0.9782~ & 0.9781 & 0.9785 & 0.9784 & 0.9784 & \textcolor{red}{0.9789~} & 0.9781~ & 0.9784~ \\
					& LPIPS & 0.0552 & 0.0569~ & 0.0552 & 0.0518 & 0.0569 & 0.0557 & 0.0534~ & 0.0563~ & \textcolor{red}{0.0507~} & 0.0227 & 0.0231~ & 0.0223 & 0.0218 & 0.0228 & 0.0229 & 0.0218~ & 0.0230~ & \textcolor{red}{0.0209~} \\
					& DISTS & 0.0666 & 0.0666~ & 0.0660 & 0.0640 & 0.0671 & 0.0659 & 0.0638~ & 0.0665~ & \textcolor{red}{0.0617~} & 0.0224 & 0.0218~ & 0.0221 & 0.0218 & 0.0222 & 0.0222 & 0.0216~ & 0.0222~ & \textcolor{red}{0.0213~} \\ 
					\midrule[1pt]
					\multirow{4}{*}{$\times$3} & PSNR & 28.94 & 28.81~ & 28.96 & 29.12 & 28.82 & 29.11 & 29.17~ & 28.93~ & \textcolor{red}{29.35~} & 34.40 & 34.11~ & 34.26 & 34.58 & 34.25 & 34.36 & \textcolor{red}{34.61~} & 34.29~ & 34.26~ \\
					& SSIM & 0.8677 & 0.8664~ & 0.8686 & 0.8714 & 0.8658 & 0.8709 & 0.8716~ & 0.8680~ & \textcolor{red}{0.8760~} & 0.9492 & 0.9487~ & 0.9492 & 0.9507 & 0.9489 & 0.9499 & \textcolor{red}{0.9509~} & 0.9493~ & 0.9507~ \\
					& LPIPS & 0.1381 & 0.1382~ & 0.1373 & 0.1332 & 0.1399 & 0.1335 & 0.1334~ & 0.1384~ & \textcolor{red}{0.1294~} & 0.0622 & 0.0620~ & 0.0612 & 0.0598 & 0.0617 & 0.0598 & \textcolor{red}{0.0592~} & 0.0616~ & 0.0596~ \\
					& DISTS & 0.1194 & 0.1194~ & 0.1186 & 0.1158 & 0.1208 & 0.1167 & 0.1134~ & 0.1196~ & \textcolor{red}{0.1126~} & 0.0509 & 0.0504~ & 0.0508 & 0.0496 & 0.0502 & 0.0498 & \textcolor{red}{0.0483~} & 0.0509~ & 0.0488~ \\ 
					\midrule[1pt]
					\multirow{4}{*}{$\times$4} & PSNR & 26.71 & 26.67~ & 26.80 & 26.97 & 26.69 & 26.94 & 27.10~ & 26.77~ & \textcolor{red}{27.15~} & 31.33 & 31.15~ & 31.27 & 31.56 & 31.19 & 31.41 & \textcolor{red}{31.61~} & 31.27~ & 31.31~ \\
					& SSIM & 0.8055 & 0.8041~ & 0.8074 & 0.8119 & 0.8039 & 0.8104 & 0.8142~ & 0.8064~ & \textcolor{red}{0.8177~} & 0.9177 & 0.9169~ & 0.9180 & 0.9208 & 0.9173 & 0.9195 & \textcolor{red}{0.9216~} & 0.9180~ & 0.9208~ \\
					& LPIPS & 0.2062 & 0.2077~ & 0.2047 & 0.1987 & 0.2090 & 0.2020 & 0.1966~ & 0.2069~ & \textcolor{red}{0.1953~} & 0.1001 & 0.1003~ & 0.0995 & 0.0970 & 0.1006 & 0.0980 & \textcolor{red}{0.0957~} & 0.0999~ & 0.0973~ \\
					& DISTS & 0.1562 & 0.1612~ & 0.1600 & 0.1563 & 0.1636 & 0.1589 & 0.1559~ & 0.1610~ & \textcolor{red}{0.1515~} & 0.0747 & 0.0778~ & 0.0786 & 0.0773 & 0.0788 & 0.0773 & 0.0775~ & 0.0782~ & \textcolor{red}{0.0747~} \\ 
					\midrule[1pt]
					\multirow{4}{*}{$\times$6} & PSNR & 24.07 & 24.19~ & 24.27 & 24.42 & 24.18 & 24.39 & 24.58~ & 24.16~ & \textcolor{red}{24.63~} & 27.09 & 27.30~ & 27.48 & 27.70 & 27.31 & 27.52 & \textcolor{red}{27.70~} & 27.24~ & 27.59~ \\
					& SSIM & 0.6966 & 0.7028~ & 0.7058 & 0.7113 & 0.7010 & 0.7102 & 0.7173~ & 0.6996~ & \textcolor{red}{0.7214~} & 0.8426 & 0.8507~ & 0.8531 & 0.8572 & 0.8502 & 0.8537 & \textcolor{red}{0.8585~} & 0.8470~ & 0.8582~ \\
					& LPIPS & 0.3158 & 0.3099~ & 0.3220 & 0.3148 & 0.3155 & 0.3169 & 0.2950~ & 0.3241~ & \textcolor{red}{0.2943~} & 0.1720 & 0.1687~ & 0.1746 & 0.1888 & 0.1718 & 0.1732 & \textcolor{red}{0.1616~} & 0.1758~ & 0.1638~ \\
					& DISTS & 0.2084 & 0.2176~ & 0.2160 & 0.2110 & 0.2217 & 0.2149 & 0.2111~ & 0.2191~ & \textcolor{red}{0.2064~} & \textcolor{red}{0.1063} & 0.1184~ & 0.1171 & 0.1160 & 0.1209 & 0.1155 & 0.1201~ & 0.1175~ & 0.1141\textcolor{red}{~} \\ 
					\midrule[1pt]
					\multirow{4}{*}{$\times$8} & PSNR & 22.64 & 22.78~ & 22.86 & 23.01 & 22.77 & 22.97 & 23.13~ & 22.64~ & \textcolor{red}{23.19~} & 24.64 & 25.02~ & 25.11 & 25.30 & 24.99 & 25.16 & \textcolor{red}{25.40~} & 24.62~ & 25.27~ \\
					& SSIM & 0.6213 & 0.6338~ & 0.6362 & 0.6423 & 0.6305 & 0.6408 & 0.6484~ & 0.6234~ & \textcolor{red}{0.6522~} & 0.7746 & 0.7947~ & 0.7961 & 0.8014 & 0.7925 & 0.7968 & \textcolor{red}{0.8057~} & 0.7794~ & 0.8034~ \\
					& LPIPS & 0.4022 & 0.3903~ & 0.4107 & 0.4011 & 0.4010 & 0.4037 & 0.3733~ & 0.4249~ & \textcolor{red}{0.3723~} & 0.2501 & 0.2342~ & 0.2480 & 0.2390 & 0.2391 & 0.2460 & \textcolor{red}{0.2218~} & 0.2631~ & 0.2244~ \\
					& DISTS & 0.2512 & 0.2591~ & 0.2586 & 0.2520 & 0.2657 & 0.2580 & 0.2507~ & 0.2704~ & \textcolor{red}{0.2460~} & \textcolor{red}{0.1371} & 0.1503~ & 0.1503 & 0.1480 & 0.1550 & 0.1492 & 0.1503~ & 0.1575~ & 0.1433~ \\ 
					\midrule[1pt]
					\multirow{4}{*}{$\times$12} & PSNR & 21.00 & 21.15~ & 21.22 & 21.35 & 21.12 & 21.33 & 21.44~ & 20.84~ & \textcolor{red}{21.47~} & 22.03 & 22.38~ & 22.45 & 22.58 & 22.32 & 22.48 & \textcolor{red}{22.67~} & 21.76~ & 22.58~ \\
					& SSIM & 0.5317 & 0.5499~ & 0.5514 & 0.5571 & 0.5462 & 0.5561 & 0.5637~ & 0.5319~ & \textcolor{red}{0.5670~} & 0.6804 & 0.7115~ & 0.7116 & 0.7165 & 0.7073 & 0.7121 & \textcolor{red}{0.7247~} & 0.6834~ & 0.7209~ \\
					& LPIPS & 0.5332 & 0.5198~ & 0.5477 & 0.5374 & 0.5365 & 0.5403 & 0.5005~ & 0.5929~ & \textcolor{red}{0.4927~} & 0.3830 & 0.3597~ & 0.3809 & 0.3686 & 0.3670 & 0.3772 & \textcolor{red}{0.3352~} & 0.4281~ & 0.3370~ \\
					& DISTS & 0.3144 & 0.3209~ & 0.3216 & 0.3149 & 0.3304 & 0.3216 & 0.3114~ & 0.3485~ & \textcolor{red}{0.3072~} & 0.1972 & 0.2076~ & 0.2082 & 0.2038 & 0.2156 & 0.2077 & 0.2043~ & 0.2329~ & \textcolor{red}{0.1958~} \\
					\bottomrule[2pt]
				\end{tabular}
			}
		}
		\label{tab:results on Urban100 and Manga109 with RDN}
	\end{table*}

	\begin{table*}[h]
		\centering
		\caption{Quantitative results of representative ASR models and our proposed GSASR. All models use the same EDSR-backbone \cite{lim2017enhanced} as the feature extraction encoder, and are tested on BSDS100 \cite{martin2001database} and General100 \cite{dong2016accelerating} with scaling factors $\times2$, $\times3$, $\times4$, $\times6$, $\times8$, $\times12$. The best results are highlighted in \textcolor{red}{red}. The PSNR and SSIM \cite{wang2004image} metrics are computed on the Y channel of Ycbcr space.}
		\scalebox{1.0}{
			\resizebox{\linewidth}{!}{
				\setlength{\extrarowheight}{2.0pt} 
				\begin{tabular}{c|c|ccccccccc|ccccccccc} 
					\toprule[2pt]
					\multirow{3}{*}{Scale} & \multirow{3}{*}{Metrics} & \multicolumn{18}{c}{Backbone: EDSR-baseline} \\ 
					\cline{3-20}
					&  & \multicolumn{9}{c|}{Testing Dataset: BSDS100} & \multicolumn{9}{c}{Testing Dataset: General100} \\
					&  & \begin{tabular}[c]{@{}c@{}}Meta\\-SR\end{tabular} & LIIF & LTE & SRNO & LINF & LMF & \begin{tabular}[c]{@{}c@{}}Ciao\\-SR\end{tabular} & \begin{tabular}[c]{@{}c@{}}Gaussian\\-SR\end{tabular} & GSASR & \begin{tabular}[c]{@{}c@{}}Meta\\-SR\end{tabular} & LIIF & LTE & SRNO & LINF & LMF & \begin{tabular}[c]{@{}c@{}}Ciao\\-SR\end{tabular} & \begin{tabular}[c]{@{}c@{}}Gaussian\\-SR\end{tabular} & GSASR \\ 
					\midrule[1pt] \midrule[1pt]
					\multirow{4}{*}{$\times$2} & PSNR & 32.13~ & 32.14~ & 32.17~ & 32.23~ & 32.16~ & 32.21~ & 32.28~ & 32.17~ & \textcolor{red}{32.39~} & 38.01~ & 38.04~ & 38.10~ & 38.25~ & 38.27~ & 38.18~ & 38.47~ & 38.10~ & \textcolor{red}{38.61~} \\
					& SSIM & 0.8994~ & 0.8994~ & 0.8997~ & 0.9007~ & 0.8989~ & 0.9004~ & 0.9007~ & 0.8998~ & \textcolor{red}{0.9023~} & 0.9611~ & 0.9612~ & 0.9615~ & 0.9622~ & 0.9613~ & 0.9619~ & 0.9624~ & 0.9615~ & \textcolor{red}{0.9632~} \\
					& LPIPS & 0.1484~ & 0.1479~ & 0.1462~ & 0.1452~ & 0.1520~ & 0.1430~ & 0.1462~ & 0.1480~ & \textcolor{red}{0.1370~} & 0.0441~ & 0.0435~ & 0.0429~ & 0.0424~ & 0.0447~ & 0.0416~ & 0.0423~ & 0.0440~ & \textcolor{red}{0.0385~} \\
					& DISTS & 0.1043~ & 0.1045~ & 0.1045~ & 0.1031~ & 0.1044~ & 0.1043~ & 0.1019~ & 0.1042~ & \textcolor{red}{0.1004~} & 0.0587~ & 0.0589~ & 0.0590~ & 0.0574~ & 0.0585~ & 0.0583~ & 0.0573~ & 0.0590~ & \textcolor{red}{0.0544~} \\ 
					\midrule[1pt]
					\multirow{4}{*}{$\times$3} & PSNR & 29.07~ & 29.08~ & 29.12~ & 29.17~ & 29.12~ & 29.16~ & 29.19~ & 29.12~ & \textcolor{red}{29.32~} & 33.82~ & 33.87~ & 33.93~ & 34.07~ & 33.99~ & 34.01~ & 34.14~ & 33.92~ & \textcolor{red}{34.38~} \\
					& SSIM & 0.8055~ & 0.8062~ & 0.8065~ & 0.8076~ & 0.8056~ & 0.8081~ & 0.8076~ & 0.8065~ & \textcolor{red}{0.8121~} & 0.9111~ & 0.9118~ & 0.9122~ & 0.9135~ & 0.9120~ & 0.9133~ & 0.9136~ & 0.9121~ & \textcolor{red}{0.9169~} \\
					& LPIPS & 0.2884~ & 0.2819~ & 0.2800~ & 0.2793~ & 0.2867~ & 0.2738~ & 0.2782~ & 0.2819~ & \textcolor{red}{0.2710~} & 0.1164~ & 0.1128~ & 0.1121~ & 0.1108~ & 0.1142~ & 0.1076~ & 0.1090~ & 0.1137~ & \textcolor{red}{0.1040~} \\
					& DISTS & 0.1637~ & 0.1639~ & 0.1639~ & 0.1632~ & 0.1639~ & 0.1630~ & 0.1597~ & 0.1644~ & \textcolor{red}{0.1592~} & 0.1060~ & 0.1067~ & 0.1069~ & 0.1044~ & 0.1059~ & 0.1043~ & 0.1026~ & 0.1075~ & \textcolor{red}{0.1000} \\ 
					\midrule[1pt]
					\multirow{4}{*}{$\times$4} & PSNR & 27.54~ & 27.59~ & 27.61~ & 27.66~ & 27.61~ & 27.65~ & 27.70~ & 27.60~ & \textcolor{red}{27.81~} & 31.34~ & 31.47~ & 31.53~ & 31.66~ & 31.55~ & 31.58~ & 31.56~ & 31.49~ & \textcolor{red}{31.90~} \\
					& SSIM & 0.7355~ & 0.7373~ & 0.7379~ & 0.7403~ & 0.7373~ & 0.7397~ & 0.7411~ & 0.7375~ & \textcolor{red}{0.7460~} & 0.8616~ & 0.8637~ & 0.8643~ & 0.8669~ & 0.8640~ & 0.8657~ & 0.8667~ & 0.8638~ & \textcolor{red}{0.8720~} \\
					& LPIPS & 0.3795~ & 0.3717~ & 0.3703~ & 0.3635~ & 0.3743~ & 0.3664~ & 0.3620~ & 0.3736~ & \textcolor{red}{0.3597~} & 0.1829~ & 0.1776~ & 0.1769~ & 0.1733~ & 0.1791~ & 0.1730~ & 0.1723~ & 0.1794~ & \textcolor{red}{0.1654~} \\
					& DISTS & 0.2025~ & 0.2037~ & 0.2042~ & 0.2040~ & 0.2040~ & 0.2032~ & 0.2033~ & 0.2045~ & \textcolor{red}{0.1982~} & 0.1457~ & 0.1479~ & 0.1479~ & 0.1456~ & 0.1479~ & 0.1459~ & 0.1470~ & 0.1487~ & \textcolor{red}{0.1383~} \\ 
					\midrule[1pt]
					\multirow{4}{*}{$\times$6} & PSNR & 25.74~ & 25.83~ & 25.86~ & 25.90~ & 25.85~ & 25.89~ & 25.95~ & 25.82~ & \textcolor{red}{26.04~} & 28.39~ & 28.64~ & 28.70~ & 28.78~ & 28.69~ & 28.70~ & 28.89~ & 28.59~ & \textcolor{red}{29.02~} \\
					& SSIM & 0.6440~ & 0.6489~ & 0.6495~ & 0.6522~ & 0.6489~ & 0.6512~ & 0.6536~ & 0.6465~ & \textcolor{red}{0.6592~} & 0.7793~ & 0.7858~ & 0.7868~ & 0.7904~ & 0.7858~ & 0.7869~ & 0.7922~ & 0.7834~ & \textcolor{red}{0.7977~} \\
					& LPIPS & 0.4948~ & 0.4838~ & 0.4919~ & 0.4827~ & 0.4888~ & 0.4869~ & 0.4763~ & 0.4982~ & \textcolor{red}{0.4702~} & 0.2822~ & 0.2705~ & 0.2764~ & 0.2695~ & 0.2747~ & 0.2734~ & 0.2622~ & 0.2833~ & \textcolor{red}{0.2539~} \\
					& DISTS & 0.2586~ & 0.2603~ & 0.2605~ & 0.2587~ & 0.2596~ & 0.2591~ & 0.2588~ & 0.2601~ & \textcolor{red}{0.2532~} & 0.2026~ & 0.2056~ & 0.2060~ & 0.2032~ & 0.2054~ & 0.2039~ & 0.2033~ & 0.2075~ & \textcolor{red}{0.1948~} \\ 
					\midrule[1pt]
					\multirow{4}{*}{$\times$8} & PSNR & 24.68~ & 24.78~ & 24.81~ & 24.87~ & 24.80~ & 24.60~ & 24.89~ & 24.72~ & \textcolor{red}{24.99~} & 26.64~ & 26.91~ & 26.97~ & 27.04~ & 26.95~ & 26.96~ & 27.16~ & 26.73~ & \textcolor{red}{27.22~} \\
					& SSIM & 0.5894~ & 0.5963~ & 0.5966~ & 0.5998~ & 0.5960~ & 0.5864~ & 0.6013~ & 0.5910~ & \textcolor{red}{0.6065~} & 0.7202~ & 0.7300~ & 0.7311~ & 0.7345~ & 0.7296~ & 0.7315~ & 0.7370~ & 0.7233~ & \textcolor{red}{0.7417~} \\
					& LPIPS & 0.5713~ & 0.5526~ & 0.5667~ & 0.5564~ & 0.5602~ & 0.5771~ & 0.5457~ & 0.5770~ & \textcolor{red}{0.5402~} & 0.3648~ & 0.3400~ & 0.3524~ & 0.3428~ & 0.3449~ & 0.3480~ & 0.3304~ & 0.3647~ & \textcolor{red}{0.3220~} \\
					& DISTS & 0.2965~ & 0.2992~ & 0.3010~ & 0.2981~ & 0.2990~ & 0.2945~ & 0.2976~ & 0.3015~ & \textcolor{red}{0.2913~} & 0.2414~ & 0.2467~ & 0.2478~ & 0.2445~ & 0.2466~ & 0.2454~ & 0.2439~ & 0.2519~ & \textcolor{red}{0.2348~} \\ 
					\midrule[1pt]
					\multirow{4}{*}{$\times$12} & PSNR & 23.42~ & 23.51~ & 23.54~ & 23.57~ & 23.53~ & 23.56~ & 23.61~ & 23.37~ & \textcolor{red}{23.62~} & 24.45~ & 24.69~ & 24.77~ & 24.83~ & 24.71~ & 24.76~ & \textcolor{red}{24.94~} & 24.33~ & 24.85~ \\
					& SSIM & 0.5355~ & 0.5437~ & 0.5437~ & 0.5457~ & 0.5431~ & 0.5447~ & 0.5479~ & 0.5358~ & \textcolor{red}{0.5512~} & 0.6454~ & 0.6590~ & 0.6602~ & 0.6622~ & 0.6581~ & 0.6607~ & 0.6662~ & 0.6472~ & \textcolor{red}{0.6693~} \\
					& LPIPS & 0.6580~ & 0.6475~ & 0.6664~ & 0.6568~ & 0.6575~ & 0.6598~ & 0.6397~ & 0.6941~ & \textcolor{red}{0.6302~} & 0.4735~ & 0.4527~ & 0.4694~ & 0.4629~ & 0.4578~ & 0.4624~ & 0.4374~ & 0.5033~ & \textcolor{red}{0.4252~} \\
					& DISTS & 0.3482~ & 0.3512~ & 0.3548~ & 0.3521~ & 0.3529~ & 0.3521~ & 0.3498~ & 0.3608~ & \textcolor{red}{0.3419~} & 0.2936~ & 0.3006~ & 0.3042~ & 0.3017~ & 0.3020~ & 0.3007~ & 0.2986~ & 0.3140~ & \textcolor{red}{0.2869~} \\
					\bottomrule[2pt]
				\end{tabular}
			}
		}
		\label{tab:results on BSDS100 and General100 with EDSR}
	\end{table*}

	\begin{table*}[h]
		\centering
		\caption{Quantitative results of  representative ASR models and our proposed GSASR. All models use the same RDN backbone \cite{zhang2018residual} as the feature extraction encoder, and are tested on BSDS100 \cite{martin2001database} and General100 \cite{dong2016accelerating} with scaling factors $\times2$, $\times3$, $\times4$, $\times6$, $\times8$, $\times12$. The best results are highlighted in \textcolor{red}{red}. The PSNR and SSIM \cite{wang2004image} metrics are computed on the Y channel of Ycbcr space.}
		\scalebox{1.0}{
			\resizebox{\linewidth}{!}{
				\setlength{\extrarowheight}{2.0pt} 
				\begin{tabular}{c|c|ccccccccc|ccccccccc} 
					\toprule[2pt]
					\multirow{3}{*}{Scale} & \multirow{3}{*}{Metrics} & \multicolumn{18}{c}{Backbone: RDN} \\ 
					\cline{3-20}
					&  & \multicolumn{9}{c|}{Testing Dataset: BSDS100} & \multicolumn{9}{c}{Testing Dataset: General100} \\
					&  & \begin{tabular}[c]{@{}c@{}}Meta\\-SR\end{tabular} & LIIF & LTE & SRNO & LINF & LMF & \begin{tabular}[c]{@{}c@{}}Ciao\\-SR\end{tabular} & \begin{tabular}[c]{@{}c@{}}Gaussian\\-SR\end{tabular} & GSASR & \begin{tabular}[c]{@{}c@{}}Meta\\-SR\end{tabular} & LIIF & LTE & SRNO & LINF & LMF & \begin{tabular}[c]{@{}c@{}}Ciao\\-SR\end{tabular} & \begin{tabular}[c]{@{}c@{}}Gaussian\\-SR\end{tabular} & GSASR \\ 
					\midrule[1pt] \midrule[1pt]
					\multirow{4}{*}{$\times$2} & PSNR & 32.37~ & 32.28~ & 32.32~ & 32.37~ & 32.31~ & 32.34~ & 32.40~ & 32.32~ & \textcolor{red}{32.43~} & 38.63~ & 38.30~ & 38.39~ & 38.53~ & 38.58~ & 38.45~ & \textcolor{red}{38.74~} & 38.41~ & 38.72~ \\
					& SSIM & 0.9012~ & 0.9011~ & 0.9015~ & 0.9026~ & 0.9012~ & 0.9019~ & 0.9022~ & 0.9014~ & \textcolor{red}{0.9029~} & 0.9628~ & 0.9626~ & 0.9628~ & 0.9636~ & 0.9627~ & 0.9631~ & 0.9634~ & 0.9628~ & \textcolor{red}{0.9637~} \\
					& LPIPS & 0.1397~ & 0.1455~ & 0.1431~ & 0.1385~ & 0.1463~ & 0.1427~ & 0.1414~ & 0.1453~ & \textcolor{red}{0.1367~} & 0.0405~ & 0.0419~ & 0.0413~ & 0.0398~ & 0.0424~ & 0.0414~ & 0.0409~ & 0.0420~ & \textcolor{red}{0.0382~} \\
					& DISTS & 0.1026~ & 0.1016~ & 0.1017~ & 0.1011~ & 0.1013~ & 0.1025~ & 0.1004~ & 0.1020~ & \textcolor{red}{0.0992~} & 0.0572~ & 0.0566~ & 0.0560~ & 0.0555~ & 0.0566~ & 0.0568~ & 0.0556~ & 0.0566~ & \textcolor{red}{0.0540~} \\ 
					\midrule[1pt]
					\multirow{4}{*}{$\times$3} & PSNR & 29.29~ & 29.25~ & 29.28~ & 29.33~ & 29.26~ & 29.29~ & 29.34~ & 29.28~ & \textcolor{red}{29.36~} & 34.38~ & 34.21~ & 34.29~ & 34.43~ & 34.33~ & 34.34~ & \textcolor{red}{34.49~} & 34.31~ & 34.48~ \\
					& SSIM & 0.8095~ & 0.8099~ & 0.8103~ & 0.8115~ & 0.8096~ & 0.8113~ & 0.8113~ & 0.8102~ & \textcolor{red}{0.8131~} & 0.9155~ & 0.9154~ & 0.9157~ & 0.9171~ & 0.9154~ & 0.9165~ & 0.9169~ & 0.9160~ & \textcolor{red}{0.9178~} \\
					& LPIPS & 0.2761~ & 0.2782~ & 0.2750~ & 0.2717~ & 0.2797~ & 0.2708~ & 0.2747~ & 0.2774~ & \textcolor{red}{0.2706~} & 0.1075~ & 0.1077~ & 0.1075~ & 0.1050~ & 0.1089~ & 0.1048~ & 0.1053~ & 0.1080~ & \textcolor{red}{0.1032~} \\
					& DISTS & 0.1623~ & 0.1613~ & 0.1614~ & 0.1609~ & 0.1606~ & 0.1605~ & 0.1591~ & 0.1619~ & \textcolor{red}{0.1578~} & 0.1041~ & 0.1034~ & 0.1026~ & 0.1015~ & 0.1031~ & 0.1017~ & 0.1009~ & 0.1033~ & \textcolor{red}{0.0997~} \\ 
					\midrule[1pt]
					\multirow{4}{*}{$\times$4} & PSNR & 27.76~ & 27.73~ & 27.76~ & 27.81~ & 27.75~ & 27.78~ & 27.83~ & 27.76~ & \textcolor{red}{27.84~} & 31.89~ & 31.80~ & 31.99~ & 32.02~ & 31.88~ & 31.91~ & 31.85~ & 31.87~ & \textcolor{red}{32.00~} \\
					& SSIM & 0.7424~ & 0.7422~ & 0.7430~ & 0.7449~ & 0.7422~ & 0.7439~ & 0.7453~ & 0.7429~ & \textcolor{red}{0.7471~} & 0.8697~ & 0.8692~ & 0.8702~ & 0.8724~ & 0.8695~ & 0.8708~ & 0.8716~ & 0.8701~ & \textcolor{red}{0.8735~} \\
					& LPIPS & 0.3651~ & 0.3646~ & 0.3628~ & 0.3576~ & 0.3651~ & 0.3614~ & 0.3597~ & 0.3652~ & \textcolor{red}{0.3597~} & 0.1692~ & 0.1698~ & 0.1690~ & 0.1655~ & 0.1708~ & 0.1676~ & 0.1669~ & 0.1693~ & \textcolor{red}{0.1642~} \\
					& DISTS & 0.2005~ & 0.2016~ & 0.2008~ & 0.2007~ & 0.2013~ & 0.2007~ & 0.2002~ & 0.2011~ & \textcolor{red}{0.1970~} & 0.1420~ & 0.1435~ & 0.1422~ & 0.1407~ & 0.1429~ & 0.1478~ & 0.1423~ & 0.1427~ & \textcolor{red}{0.1367~} \\ 
					\midrule[1pt]
					\multirow{4}{*}{$\times$6} & PSNR & 25.93~ & 25.97~ & 25.99~ & 26.03~ & 26.00~ & 26.02~ & \textcolor{red}{26.08~} & 25.97~ & 26.07~ & 28.84~ & 28.92~ & 28.99~ & 29.10~ & 28.97~ & 29.00~ & \textcolor{red}{29.18~} & 28.94~ & 29.11~ \\
					& SSIM & 0.6521~ & 0.6549~ & 0.6556~ & 0.6575~ & 0.6548~ & 0.6562~ & 0.6587~ & 0.6529~ & \textcolor{red}{0.6607~} & 0.7910~ & 0.7936~ & 0.7954~ & 0.7979~ & 0.7940~ & 0.7952~ & 0.7996~ & 0.7928~ & \textcolor{red}{0.8000~} \\
					& LPIPS & 0.4764~ & 0.4726~ & 0.4821~ & 0.4756~ & 0.4793~ & 0.4809~ & \textcolor{red}{0.4707~} & 0.4863~ & 0.4714~ & 0.2591~ & 0.2562~ & 0.2648~ & 0.2577~ & 0.2618~ & 0.2636~ & 0.2536~ & 0.2673~ & \textcolor{red}{0.2521~} \\
					& DISTS & 0.2524~ & 0.2562~ & 0.2556~ & 0.2544~ & 0.2559~ & 0.2551~ & 0.2554~ & 0.2551~ & \textcolor{red}{0.2513~} & 0.1931~ & 0.1990~ & 0.1985~ & 0.1959~ & 0.1998~ & 0.1977~ & 0.1975~ & 0.1985~ & \textcolor{red}{0.1929~} \\ 
					\midrule[1pt]
					\multirow{4}{*}{$\times$8} & PSNR & 24.84~ & 24.90~ & 24.94~ & 24.97~ & 24.93~ & 24.80~ & 25.00~ & 24.84~ & \textcolor{red}{25.01~} & 26.97~ & 27.18~ & 27.22~ & 27.31~ & 27.21~ & 27.22~ & \textcolor{red}{27.39~} & 27.00~ & 27.30~ \\
					& SSIM & 0.5965~ & 0.6020~ & 0.6025~ & 0.6045~ & 0.6014~ & 0.5968~ & 0.6058~ & 0.5968~ & \textcolor{red}{0.6080~} & 0.7307~ & 0.7382~ & 0.7394~ & 0.7423~ & 0.7374~ & 0.7397~ & 0.7440~ & 0.7322~ & \textcolor{red}{0.7442~} \\
					& LPIPS & 0.5460~ & 0.5409~ & 0.5578~ & 0.5509~ & 0.5521~ & 0.5707~ & \textcolor{red}{0.5409~} & 0.5650~ & 0.5416~ & 0.3319~ & 0.3232~ & 0.3406~ & 0.3315~ & 0.3327~ & 0.3383~ & 0.3214~ & 0.3482~ & \textcolor{red}{0.3200~} \\
					& DISTS & \textcolor{red}{0.2894~} & 0.2945~ & 0.2962~ & 0.2943~ & 0.2950~ & 0.2940~ & 0.2943~ & 0.2959~ & 0.2900~ & \textcolor{red}{0.2299~} & 0.2381~ & 0.2398~ & 0.2371~ & 0.2404~ & 0.2387~ & 0.2377~ & 0.2420~ & 0.2327~ \\ 
					\midrule[1pt]
					\multirow{4}{*}{$\times$12} & PSNR & 23.50~ & 23.61~ & 23.64~ & 23.66~ & 23.62~ & 23.64~ & \textcolor{red}{23.70~} & 23.44~ & 23.66~ & 24.66~ & 24.93~ & 24.97~ & 25.03~ & 24.92~ & 24.97~ & \textcolor{red}{25.13~} & 24.50~ & 24.90~ \\
					& SSIM & 0.5395~ & 0.5476~ & 0.5478~ & 0.5490~ & 0.5465~ & 0.5480~ & 0.5512~ & 0.5389~ & \textcolor{red}{0.5528~} & 0.6529~ & 0.6666~ & 0.6671~ & 0.6693~ & 0.6648~ & 0.6675~ & \textcolor{red}{0.6725~} & 0.6533~ & 0.6717~ \\
					& LPIPS & 0.6306~ & 0.6350~ & 0.6582~ & 0.6517~ & 0.6508~ & 0.6519~ & 0.6345~ & 0.6823~ & \textcolor{red}{0.6334~} & 0.4393~ & 0.4311~ & 0.4578~ & 0.4476~ & 0.4451~ & 0.4514~ & 0.4255~ & 0.4871~ & \textcolor{red}{0.4220~} \\
					& DISTS & \textcolor{red}{0.3403~} & 0.3461~ & 0.3505~ & 0.3493~ & 0.3491~ & 0.3493~ & 0.3478~ & 0.3536~ & 0.3418~ & \textcolor{red}{0.2807~} & 0.2926~ & 0.2989~ & 0.2953~ & 0.2964~ & 0.2954~ & 0.2933~ & 0.3046~ & 0.2850~ \\
					\bottomrule[2pt]
				\end{tabular}
			}
		}
		\label{tab:results on BSDS100 and General100 with RDN}
	\end{table*}

	\begin{table*}[h]
		\centering
		\caption{Quantitative results of  representative ASR models and our proposed GSASR. All models use the same EDSR-backbone \cite{lim2017enhanced} as the feature extraction encoder, and are tested on Set5 \cite{bevilacqua2012low} and Set14 \cite{zeyde2010single} with scaling factors $\times2$, $\times3$, $\times4$, $\times6$, $\times8$, $\times12$. The best results are highlighted in \textcolor{red}{red}. The PSNR and SSIM \cite{wang2004image} metrics are computed on the Y channel of Ycbcr space.}
		\scalebox{1.0}{
			\resizebox{\linewidth}{!}{
				\setlength{\extrarowheight}{2.0pt} 
				\begin{tabular}{c|c|ccccccccc|ccccccccc} 
					\toprule[2pt]
					\multirow{3}{*}{Scale} & \multirow{3}{*}{Metrics} & \multicolumn{18}{c}{Backbone: EDSR-baseline} \\ 
					\cline{3-20}
					&  & \multicolumn{9}{c|}{Testing Dataset: Set5} & \multicolumn{9}{c}{Testing Dataset: Set14} \\
					&  & \begin{tabular}[c]{@{}c@{}}Meta\\-SR\end{tabular} & LIIF & LTE & SRNO & LINF & LMF & \begin{tabular}[c]{@{}c@{}}Ciao\\-SR\end{tabular} & \begin{tabular}[c]{@{}c@{}}Gaussian\\-SR\end{tabular} & GSASR & \begin{tabular}[c]{@{}c@{}}Meta\\-SR\end{tabular} & LIIF & LTE & SRNO & LINF & LMF & \begin{tabular}[c]{@{}c@{}}Ciao\\-SR\end{tabular} & \begin{tabular}[c]{@{}c@{}}Gaussian\\-SR\end{tabular} & GSASR \\ 
					\midrule[1pt] \midrule[1pt]
					\multirow{4}{*}{$\times$2} & PSNR & 37.86 & 37.87~ & 37.93~ & 38.03~ & 37.99~ & 37.97~ & 38.14~ & 37.91~ & \textcolor{red}{38.33~} & 33.55~ & 33.63~ & 33.68~ & 33.78~ & 33.63~ & 33.75~ & 33.90~ & 33.64~ & \textcolor{red}{34.02~} \\
					& SSIM & 0.9603 & 0.9604 & 0.9604 & 0.9609 & 0.9605~ & 0.9607~ & 0.9610~ & 0.9604~ & \textcolor{red}{0.9619} & 0.9177~ & 0.9186~ & 0.9190~ & 0.9201~ & 0.9178~ & 0.9196~ & 0.9203~ & 0.9189~ & \textcolor{red}{0.9222~} \\
					& LPIPS & 0.0887 & 0.0549 & 0.0545 & 0.055 & 0.0564~ & 0.0538~ & 0.0542~ & 0.0553~ & \textcolor{red}{0.0517~} & 0.0953~ & 0.0943~ & 0.0927~ & 0.0920~ & 0.0970~ & 0.0916~ & 0.0929~ & 0.0936~ & \textcolor{red}{0.0846~} \\
					& DISTS & 0.0819 & 0.0819 & 0.0827 & 0.0802 & 0.0814~ & 0.0803~ & 0.0797~ & 0.0833~ & \textcolor{red}{0.0766~} & 0.0845~ & 0.0839~ & 0.0844~ & 0.0830~ & 0.0841~ & 0.0829~ & 0.0821~ & 0.0845~ & \textcolor{red}{0.0793~} \\ 
					\midrule[1pt]
					\multirow{4}{*}{$\times$3} & PSNR & 34.31 & 34.35~ & 34.39~ & 34.47~ & 34.45~ & 34.52~ & 34.49~ & 34.39~ & \textcolor{red}{34.84~} & 30.27~ & 30.32~ & 30.35~ & 30.47~ & 30.33~ & 30.41~ & 30.46~ & 30.34~ & \textcolor{red}{30.65~} \\
					& SSIM & 0.9268 & 0.9273 & 0.9272 & 0.9282 & 0.9277~ & 0.9285~ & 0.9283~ & 0.9276~ & \textcolor{red}{0.9309~} & 0.8421~ & 0.8431~ & 0.8437~ & 0.8448~ & 0.8430~ & 0.8449~ & 0.8450~ & 0.8435~ & \textcolor{red}{0.8498~} \\
					& LPIPS & 0.1276 & 0.1242 & 0.1243 & 0.1234 & 0.1261~ & 0.1217~ & 0.1211~ & 0.1253~ & \textcolor{red}{0.1209~} & 0.2110~ & 0.2067~ & 0.2045~ & 0.2038~ & 0.2081~ & 0.2005~ & 0.2018~ & 0.2056~ & \textcolor{red}{0.1928~} \\
					& DISTS & 0.1292 & 0.1308 & 0.1324 & 0.1315 & 0.1299~ & 0.1317~ & \textcolor{red}{0.1284~} & 0.1329~ & 0.1291~ & 0.1322~ & 0.1311~ & 0.1320~ & 0.1303~ & 0.1302~ & 0.1297~ & 0.1271~ & 0.1327~ & \textcolor{red}{0.1267~} \\ 
					\midrule[1pt]
					\multirow{4}{*}{$\times$4} & PSNR & 32.04 & 32.20~ & 32.20~ & 32.35~ & 32.26~ & 32.30~ & 32.42~ & 32.22~ & \textcolor{red}{32.79~} & 28.51~ & 28.60~ & 28.63~ & 28.74~ & 28.62~ & 28.69~ & 28.77~ & 28.62~ & \textcolor{red}{28.88~} \\
					& SSIM & 0.8930 & 0.8955 & 0.8959 & 0.8974 & 0.8960~ & 0.8967~ & 0.8983~ & 0.8958~ & \textcolor{red}{0.9022~} & 0.7808~ & 0.7828~ & 0.7836~ & 0.7856~ & 0.7826~ & 0.7849~ & 0.7865~ & 0.7831~ & \textcolor{red}{0.7902~} \\
					& LPIPS & 0.1768 & 0.1717 & 0.1733 & 0.1738 & 0.1757~ & 0.1695~ & 0.1688~ & 0.1742~ & \textcolor{red}{0.1686~} & 0.2882~ & 0.2832~ & 0.2828~ & 0.2790~ & 0.2861~ & 0.2794~ & 0.2783~ & 0.2849~ & \textcolor{red}{0.2680~} \\
					& DISTS & 0.1559 & 0.1585 & 0.1585 & 0.158 & 0.1565~ & 0.1596~ & 0.1580~ & 0.1586~ & \textcolor{red}{0.1553~} & 0.1636~ & 0.1639~ & 0.1637~ & 0.1638~ & 0.1642~ & 0.1618~ & 0.1615~ & 0.1646~ & \textcolor{red}{0.1574~} \\ 
					\midrule[1pt]
					\multirow{4}{*}{$\times$6} & PSNR & 28.58 & 28.92~ & 28.93~ & 29.02~ & 28.90~ & 28.98~ & 29.13~ & 28.82~ & \textcolor{red}{29.39~} & 26.28~ & 26.44~ & 26.48~ & 26.53~ & 26.45~ & 26.50~ & 26.60~ & 26.41~ & \textcolor{red}{26.71~} \\
					& SSIM & 0.8204 & 0.8318 & 0.8311 & 0.8333 & 0.8314~ & 0.8331~ & 0.8357~ & 0.8282~ & \textcolor{red}{0.8418~} & 0.6897~ & 0.6964~ & 0.6972~ & 0.6989~ & 0.6960~ & 0.6968~ & 0.7003~ & 0.6936~ & \textcolor{red}{0.7058~} \\
					& LPIPS & 0.2506 & 0.2406 & 0.2487 & 0.2471 & 0.2433~ & 0.2470~ & 0.2369~ & 0.2484~ & \textcolor{red}{0.2289~} & 0.3957~ & 0.3861~ & 0.3919~ & 0.3890~ & 0.3900~ & 0.3939~ & 0.3823~ & 0.3973~ & \textcolor{red}{0.3731~} \\
					& DISTS & 0.2038 & 0.2009 & 0.204 & 0.2038 & 0.2020~ & 0.2003~ & 0.2014~ & 0.2021~ & \textcolor{red}{0.1964~} & 0.2204~ & 0.2197~ & 0.2193~ & 0.2184~ & 0.2201~ & 0.2195~ & 0.2163~ & 0.2221~ & \textcolor{red}{0.2108~} \\ 
					\midrule[1pt]
					\multirow{4}{*}{$\times$8} & PSNR & 26.69 & 26.96~ & 27.02~ & 27.05~ & 26.95~ & 27.07~ & 27.16~ & 26.77~ & \textcolor{red}{27.33~} & 24.75~ & 24.92~ & 24.96~ & 25.03~ & 24.92~ & 25.00~ & 25.09~ & 24.80~ & \textcolor{red}{25.23~} \\
					& SSIM & 0.7579 & 0.7764 & 0.777 & 0.7773 & 0.7763~ & 0.7788~ & 0.7829~ & 0.7669~ & \textcolor{red}{0.7863~} & 0.6300~ & 0.6395~ & 0.6402~ & 0.6456~ & 0.6386~ & 0.6403~ & 0.6450~ & 0.6330~ & \textcolor{red}{0.6505~} \\
					& LPIPS & 0.3280 & 0.2993 & 0.3179 & 0.3152 & 0.3030~ & 0.3092~ & 0.2920~ & 0.3198~ & \textcolor{red}{0.2857~} & 0.4674~ & 0.4459~ & 0.4593~ & 0.4515~ & 0.4526~ & 0.4595~ & 0.4395~ & 0.4683~ & \textcolor{red}{0.4359~} \\
					& DISTS & 0.2338 & 0.233 & 0.2376 & 0.238 & 0.2336~ & 0.2381~ & 0.2334~ & 0.2396~ & \textcolor{red}{0.2326~} & 0.2578~ & 0.2594~ & 0.2595~ & 0.2583~ & 0.2589~ & 0.2582~ & 0.2541~ & 0.2650~ & \textcolor{red}{0.2480~} \\ 
					\midrule[1pt]
					\multirow{4}{*}{$\times$12} & PSNR & 24.25 & 24.43~ & 24.48~ & 24.50~ & 24.47~ & 24.48~ & 24.63~ & 24.12~ & \textcolor{red}{24.67~} & 23.05~ & 23.15~ & 23.20~ & 23.20~ & 23.17~ & 23.09~ & \textcolor{red}{23.27~} & 22.92~ & 23.24~ \\
					& SSIM & 0.6656 & 0.6888 & 0.6883 & 0.6907 & 0.6884~ & 0.6896~ & 0.6985~ & 0.6705~ & \textcolor{red}{0.7055~} & 0.5616~ & 0.5738~ & 0.5741~ & 0.5758~ & 0.5718~ & 0.5706~ & 0.5788~ & 0.5621~ & \textcolor{red}{0.5813~} \\
					& LPIPS & 0.4339 & 0.4129 & 0.4363 & 0.4257 & 0.4116~ & 0.4220~ & 0.4002~ & 0.4586~ & \textcolor{red}{0.3915~} & 0.5587~ & 0.5465~ & 0.5638~ & 0.5543~ & 0.5476~ & 0.5686~ & 0.5340~ & 0.5903~ & \textcolor{red}{0.5306~} \\
					& DISTS & 0.2910 & 0.2922 & 0.2977 & 0.2986 & 0.2941~ & 0.2947~ & 0.2921~ & 0.3072~ & \textcolor{red}{0.2853~} & 0.3078~ & 0.3117~ & 0.3162~ & 0.3139~ & 0.3138~ & 0.3145~ & 0.3091~ & 0.3254~ & \textcolor{red}{0.3049~} \\
					\bottomrule[2pt]
				\end{tabular}
			}
		}
		\label{tab:results on Set5 and Set14 with EDSR}
	\end{table*}

	\begin{table*}[h]
		\centering
		\caption{Quantitative results of  representative ASR models and our proposed GSASR. All models use the same RDN backbone \cite{zhang2018residual} as the feature extraction encoder, and are tested on Set5 \cite{bevilacqua2012low} and Set14 \cite{zeyde2010single} with scaling factors $\times2$, $\times3$, $\times4$, $\times6$, $\times8$, $\times12$. The best results are highlighted in \textcolor{red}{red}. The PSNR and SSIM \cite{wang2004image} metrics are computed on the Y channel of Ycbcr space.}
		\scalebox{1.0}{
			\resizebox{\linewidth}{!}{
				\setlength{\extrarowheight}{2.0pt} 
				\begin{tabular}{c|c|ccccccccc|ccccccccc} 
					\toprule[2pt]
					\multirow{3}{*}{Scale} & \multirow{3}{*}{Metrics} & \multicolumn{18}{c}{Backbone: RDN} \\ 
					\cline{3-20}
					&  & \multicolumn{9}{c|}{Testing Dataset: Set5} & \multicolumn{9}{c}{Testing Dataset: Set14} \\
					&  & \begin{tabular}[c]{@{}c@{}}Meta\\-SR\end{tabular} & LIIF & LTE & SRNO & LINF & LMF & \begin{tabular}[c]{@{}c@{}}Ciao\\-SR\end{tabular} & \begin{tabular}[c]{@{}c@{}}Gaussian\\-SR\end{tabular} & GSASR & \begin{tabular}[c]{@{}c@{}}Meta\\-SR\end{tabular} & LIIF & LTE & SRNO & LINF & LMF & \begin{tabular}[c]{@{}c@{}}Ciao\\-SR\end{tabular} & \begin{tabular}[c]{@{}c@{}}Gaussian\\-SR\end{tabular} & GSASR \\ 
					\midrule[1pt] \midrule[1pt]
					\multirow{4}{*}{$\times$2} & PSNR & 38.24 & 38.07~ & 38.11 & 38.2 & 38.21~ & 38.13~ & 38.32~ & 38.13~ & \textcolor{red}{38.38~} & 34.03~ & 33.92~ & 34.04~ & 34.21~ & 33.93~ & 34.07~ & \textcolor{red}{34.23~} & 34.06~ & 34.16~ \\
					& SSIM & 0.9612 & 0.9611~ & 0.9612 & 0.9617 & 0.9613~ & 0.9613~ & 0.9618~ & 0.9612~ & \textcolor{red}{0.9621~} & 0.9207~ & 0.9209~ & 0.9213~ & 0.9232~ & 0.9206~ & 0.9213~ & 0.9226~ & 0.9215~ & \textcolor{red}{0.9226~} \\
					& LPIPS & 0.0536 & 0.0544~ & 0.0538 & 0.0537 & 0.0549~ & 0.0550 & 0.0538~ & 0.0547~ & \textcolor{red}{0.0510~} & 0.0899~ & 0.0892~ & 0.0889~ & 0.0879~ & 0.0907~ & 0.0895~ & 0.0882~ & 0.0902~ & \textcolor{red}{0.0845~} \\
					& DISTS & 0.0798 & 0.0796~ & 0.079 & 0.0786 & 0.0796~ & 0.0790~ & 0.0786~ & 0.0797~ & \textcolor{red}{0.0767~} & 0.0812~ & 0.0807~ & 0.0803~ & 0.0794~ & 0.0812~ & 0.0804~ & 0.0794~ & 0.0804~ & \textcolor{red}{0.0788~} \\ 
					\midrule[1pt]
					\multirow{4}{*}{$\times$3} & PSNR & 34.74 & 34.62~ & 34.66 & 34.77 & 34.70~ & 34.75~ & 34.87~ & 34.70~ & \textcolor{red}{34.90~} & 30.58~ & 30.50~ & 30.55~ & 30.67~ & 30.55~ & 30.59~ & 30.66~ & 30.57~ & \textcolor{red}{30.74~} \\
					& SSIM & 0.9296 & 0.9292~ & 0.9297 & 0.9305 & 0.9296~ & 0.9302~ & 0.9306~ & 0.9299~ & \textcolor{red}{0.9313~} & 0.8470~ & 0.8470~ & 0.8473~ & 0.8492~ & 0.8466~ & 0.8483~ & 0.8484~ & 0.8475~ & \textcolor{red}{0.8511~} \\
					& LPIPS & 0.1229 & 0.1234~ & 0.1233 & 0.122 & 0.1228~ & 0.1211~ & 0.1215~ & 0.1232~ & \textcolor{red}{0.1208~} & 0.2005~ & 0.1988~ & 0.1997~ & 0.1993~ & 0.2009~ & 0.1973~ & 0.1994~ & 0.2004~ & \textcolor{red}{0.1936~} \\
					& DISTS & 0.1312 & 0.1312~ & 0.1303 & 0.1321 & 0.1300~ & 0.1307~ & 0.1299~ & 0.1314~ & \textcolor{red}{0.1283~} & 0.1278~ & 0.1272~ & 0.1276~ & 0.1260~ & 0.1275~ & 0.1271~ & 0.1258~ & 0.1282~ & \textcolor{red}{0.1265~} \\ 
					\midrule[1pt]
					\multirow{4}{*}{$\times$4} & PSNR & 32.51 & 32.47~ & 32.56 & 32.64 & 32.49~ & 32.56~ & 32.68~ & 32.55~ & \textcolor{red}{32.83~} & 28.85~ & 28.77~ & 28.84~ & 28.93~ & 28.82~ & 28.87~ & 28.94~ & 28.82~ & \textcolor{red}{28.96~} \\
					& SSIM & 0.8989 & 0.8989~ & 0.8998 & 0.9008 & 0.8991~ & 0.8998~ & 0.9010~ & 0.8994~ & \textcolor{red}{0.9027~} & 0.7882~ & 0.7873~ & 0.7888~ & 0.7906~ & 0.7879~ & 0.7892~ & 0.7906~ & 0.7882~ & \textcolor{red}{0.7922~} \\
					& LPIPS & 0.1712 & 0.1701~ & 0.1708 & 0.1696 & 0.1721~ & 0.1717~ & 0.1689~ & 0.1719~ & \textcolor{red}{0.1688~} & 0.2765~ & 0.2766~ & 0.2762~ & 0.2744~ & 0.2770~ & 0.2763~ & 0.2746~ & 0.2786~ & \textcolor{red}{0.2709~} \\
					& DISTS & 0.1556 & 0.1557~ & 0.1558 & 0.157 & 0.1559~ & 0.1580~ & 0.1571~ & 0.1575~ & \textcolor{red}{0.1551~} & 0.1588~ & 0.1595~ & 0.1590~ & 0.1588~ & 0.1590~ & 0.1595~ & 0.1581~ & 0.1600~ & \textcolor{red}{0.1573~} \\ 
					\midrule[1pt]
					\multirow{4}{*}{$\times$6} & PSNR & 29.10 & 29.10~ & 29.25 & 29.35 & 29.20~ & 29.20~ & 29.46~ & 29.06~ & \textcolor{red}{29.52~} & 26.56~ & 26.61~ & 26.70~ & 26.75~ & 26.64~ & 26.70~ & \textcolor{red}{26.79~} & 26.60~ & 26.78~ \\
					& SSIM & 0.8326 & 0.8359~ & 0.8367 & 0.8393 & 0.8362~ & 0.8363~ & 0.8415~ & 0.8324~ & \textcolor{red}{0.8441~} & 0.6996~ & 0.7027~ & 0.7038~ & 0.7061~ & 0.7022~ & 0.7042~ & 0.7068~ & 0.7002~ & \textcolor{red}{0.7082~} \\
					& LPIPS & 0.2389 & 0.2344~ & 0.2468 & 0.2433 & 0.2383~ & 0.2461~ & 0.2325~ & 0.2457~ & \textcolor{red}{0.2265~} & 0.3782~ & 0.3753~ & 0.3849~ & 0.3821~ & 0.3810~ & 0.3875~ & 0.3757~ & 0.3886~ & \textcolor{red}{0.3724~} \\
					& DISTS & \textcolor{red}{0.1948} & 0.2011~ & 0.2049 & 0.2015 & 0.2002~ & 0.2016~ & 0.1987~ & 0.1998~ & 0.1967~ & 0.2104~ & 0.2115~ & 0.2127~ & 0.2110~ & 0.2136~ & 0.2137~ & 0.2100~ & 0.2139~ & \textcolor{red}{0.2096~} \\ 
					\midrule[1pt]
					\multirow{4}{*}{$\times$8} & PSNR & 26.97 & 27.11~ & 27.23 & 27.26 & 27.22~ & 27.25~ & 27.36~ & 27.00~ & \textcolor{red}{27.52~} & 25.02~ & 25.13~ & 25.13~ & 25.25~ & 25.14~ & 25.16~ & 25.28~ & 25.03~ & \textcolor{red}{25.28~} \\
					& SSIM & 0.7692 & 0.7810~ & 0.7828 & 0.7838 & 0.7833~ & 0.7825~ & 0.7877~ & 0.7723~ & \textcolor{red}{0.7937~} & 0.6397~ & 0.6464~ & 0.6473~ & 0.6503~ & 0.6464~ & 0.6468~ & 0.6522~ & 0.6412~ & \textcolor{red}{0.6525~} \\
					& LPIPS & 0.3036 & 0.2918~ & 0.3088 & 0.308 & 0.2956~ & 0.3077~ & 0.2871~ & 0.3133~ & \textcolor{red}{0.2788~} & 0.4376~ & 0.4347~ & 0.4518~ & 0.4482~ & 0.4406~ & 0.4526~ & 0.4353~ & 0.4555~ & \textcolor{red}{0.4329~} \\
					& DISTS & \textcolor{red}{0.2252} & 0.2304~ & 0.2348 & 0.2343 & 0.2300~ & 0.2332~ & 0.2311~ & 0.2307~ & 0.2285~ & \textcolor{red}{0.2471~} & 0.2506~ & 0.2525~ & 0.2502~ & 0.2524~ & 0.2535~ & 0.2483~ & 0.2552~ & 0.2472~ \\ 
					\midrule[1pt]
					\multirow{4}{*}{$\times$12} & PSNR & 24.49 & 24.69~ & 24.66 & 24.77 & 24.70~ & 24.67~ & \textcolor{red}{24.83~} & 24.24~ & 24.68~ & 23.18~ & 23.22~ & 23.30~ & 23.32~ & 23.28~ & 23.28~ & \textcolor{red}{23.41~} & 23.02~ & 23.27~ \\
					& SSIM & 0.6778 & 0.7011~ & 0.6984 & 0.7019 & 0.6999~ & 0.6965~ & \textcolor{red}{0.7069~} & 0.6769~ & 0.7053~ & 0.5680~ & 0.5787~ & 0.5807~ & 0.5825~ & 0.5779~ & 0.5772~ & \textcolor{red}{0.5848~} & 0.5667~ & 0.5834~ \\
					& LPIPS & 0.3952 & 0.3963~ & 0.4272 & 0.4191 & 0.4063~ & 0.4210~ & 0.3920~ & 0.4507~ & \textcolor{red}{0.3813~} & \textcolor{red}{0.5235~} & 0.5302~ & 0.5540~ & 0.5418~ & 0.5357~ & 0.5576~ & 0.5256~ & 0.5761~ & 0.5254~ \\
					& DISTS & \textcolor{red}{0.2775} & 0.2877~ & 0.2926 & 0.2898 & 0.2901~ & 0.2915~ & 0.2882~ & 0.2996~ & 0.2820~ & \textcolor{red}{0.2964~} & 0.3053~ & 0.3097~ & 0.3053~ & 0.3078~ & 0.3104~ & 0.3052~ & 0.3163~ & 0.3007~ \\
					\bottomrule[2pt]
				\end{tabular}
			}
		}
		\label{tab:results on Set5 and Set14 with RDN}
	\end{table*}

	\begin{table*}[h]
		\centering
		\caption{Quantitative comparison between representative ASR models and our GSASR on DIV2K, LSDIR, and Urban100 datasets. All models use the same EDSR-backbone as the feature extraction encoder, and are tested with scaling factors $\times2$, $\times3$, $\times4$, $\times8$, $\times8$. The best results are highlighted in \textcolor{red}{red}. The PSNR and SSIM metrics are computed on the RGB channel space.}
		\vspace{-0.2cm}
		\scalebox{1.0}{
			\resizebox{\linewidth}{!}{
				\begin{tabular}{c|c|cccc|cccc|cccc} 
					\toprule[2pt]
					\multirow{3}{*}{Scale} & \multirow{3}{*}{Metrics} & \multicolumn{12}{c}{Backbone: EDSR-baseline} \\ 
					\cline{3-14}
					&   & \multicolumn{4}{c|}{Testing Dataset: DIV2K} & \multicolumn{4}{c|}{Testing Dataset: LSDIR} & \multicolumn{4}{c}{Testing Dataset: Urban100} \\
					& & LIIF & CiaoSR & GaussianSR & GSASR & LIIF & CiaoSR & GaussianSR & GSASR & LIIF & CiaoSR & GaussianSR & GSASR \\
					\midrule[1pt] \midrule[1pt]
					\multirow{2}{*}{$\times2$} & PSNR & 34.56 & 34.92 & 34.61 & \textcolor{red}{35.13} & 29.92 & 30.28 & 29.96 & \textcolor{red}{30.63} & 30.49 & 31.14 & 30.59 & \textcolor{red}{31.61} \\
					& SSIM & 0.9369 & 0.9398 & 0.9372 & \textcolor{red}{0.9416} & 0.9086 & 0.9140 & 0.9093 & \textcolor{red}{0.9182} & 0.9187 & 0.9254 & 0.9198 & \textcolor{red}{0.9300} \\
					\midrule[1pt]
					\multirow{2}{*}{$\times3$} & PSNR & 30.91 & 31.15 & 30.96 & \textcolor{red}{31.42} & 26.45 & 26.64 & 26.49 & \textcolor{red}{26.99} & 26.65 & 27.11 & 26.72 & \textcolor{red}{27.60} \\
					& SSIM & 0.8737 & 0.8773 & 0.8741 & \textcolor{red}{0.8823} & 0.8177 & 0.8233 & 0.8184 & \textcolor{red}{0.8326} & 0.8374 & 0.8468 & 0.8385 & \textcolor{red}{0.8577} \\
					\midrule[1pt]
					\multirow{2}{*}{$\times4$} & PSNR & 28.97 & 29.22 & 29.01 & \textcolor{red}{29.43} & 24.73 & 24.94 & 24.75 & \textcolor{red}{25.18} & 24.63 & 25.17 & 24.68 & \textcolor{red}{25.48} \\
					& SSIM & 0.8190 & 0.8248 & 0.8194 & \textcolor{red}{0.8303} & 0.7440 & 0.7526 & 0.7446 & \textcolor{red}{0.7620} & 0.7666 & 0.7828 & 0.7677 & \textcolor{red}{0.7939} \\
					\midrule[1pt]
					\multirow{2}{*}{$\times8$} & PSNR & 25.40 & 25.57 & 25.29 & \textcolor{red}{25.76} & 21.81 & 21.96 & 21.75 & \textcolor{red}{22.07} & 20.99 & 21.37 & 20.90 & \textcolor{red}{21.63} \\
					& SSIM & 0.6899 & 0.6964 & 0.6843 & \textcolor{red}{0.7029} & 0.5856 & 0.5948 & 0.5787 & \textcolor{red}{0.6022} & 0.5846 & 0.6038 & 0.5735 & \textcolor{red}{0.6170} \\
					\bottomrule[2pt]
				\end{tabular}
			}
		}
		\label{tab:results on rgb channel with edsr}
	\end{table*}

	\begin{table*}[h]
		\centering
		\caption{Quantitative comparison between representative ASR models and our GSASR on DIV2K, LSDIR, and Urban100 datasets. All models use the same RDN as the feature extraction encoder, and are tested with scaling factors $\times2$, $\times3$, $\times4$, $\times8$, $\times8$. The best results are highlighted in \textcolor{red}{red}. The PSNR and SSIM metrics are computed on the RGB channel space.}
		\vspace{-0.2cm}
		\scalebox{1.0}{
			\resizebox{\linewidth}{!}{
				\begin{tabular}{c|c|cccc|cccc|cccc} 
					\toprule[2pt]
					\multirow{3}{*}{Scale} & \multirow{3}{*}{Metrics} & \multicolumn{12}{c}{Backbone:RDN} \\ 
					\cline{3-14}
					&  & \multicolumn{4}{c|}{Testing Dataset:DIV2K} & \multicolumn{4}{c|}{Testing Dataset:LSDIR} & \multicolumn{4}{c}{Testing Dataset:Urban100} \\
					&  & LIIF & CiaoSR & GaussianSR & GSASR & LIIF & CiaoSR & GaussianSR & GSASR & LIIF & CiaoSR & GaussianSR & GSASR \\ 
					\midrule[1pt] \midrule[1pt]
					\multirow{2}{*}{x2} & PSNR & 34.87 & 35.17 & 34.94 & \textcolor{red}{35.21} & 30.36 & 30.65 & 30.44 & \textcolor{red}{30.75} & 31.18 & 31.63 & 31.30 & \textcolor{red}{31.86} \\
					& SSIM & 0.9396 & 0.9417 & 0.9399 & \textcolor{red}{0.9422} & 0.9141 & 0.9180 & 0.915 & \textcolor{red}{0.9193} & 0.9256 & 0.9300 & 0.9268 & \textcolor{red}{0.9318} \\ 
					\midrule[1pt]
					\multirow{2}{*}{x3} & PSNR & 31.21 & 31.42 & 31.28 & \textcolor{red}{31.49} & 26.78 & 26.98 & 26.85 & \textcolor{red}{27.08} & 27.23 & 27.59 & 27.35 & \textcolor{red}{27.78} \\
					& SSIM & 0.8791 & 0.8817 & 0.8795 & \textcolor{red}{0.8834} & 0.8270 & 0.8317 & 0.8280 & \textcolor{red}{0.8347} & 0.8502 & 0.8562 & 0.8518 & \textcolor{red}{0.8606} \\ 
					\midrule[1pt]
					\multirow{2}{*}{x4} & PSNR & 29.25 & 29.45 & 29.30 & \textcolor{red}{29.49} & 25.00 & 25.19 & 25.05 & \textcolor{red}{25.25} & 25.15 & 25.57 & 25.24 & \textcolor{red}{25.62} \\
					& SSIM & 0.8259 & 0.8302 & 0.8267 & \textcolor{red}{0.8318} & 0.7548 & 0.7620 & 0.7562 & \textcolor{red}{0.7648} & 0.7832 & 0.7945 & 0.7856 & \textcolor{red}{0.7977} \\ 
					\midrule[1pt]
					\multirow{2}{*}{x8} & PSNR & 25.61 & 25.78 & 25.52 & \textcolor{red}{25.82} & 21.98 & 22.12 & 21.93 & \textcolor{red}{22.13} & 21.32 & 21.67 & 21.18 & \textcolor{red}{21.72} \\
					& SSIM & 0.6975 & 0.7031 & 0.6925 & \textcolor{red}{0.7050} & 0.5962 & 0.6042 & 0.5893 & \textcolor{red}{0.6054} & 0.6025 & 0.6186 & 0.5910 & \textcolor{red}{0.6215} \\
					\bottomrule[2pt]
				\end{tabular}
			}
			\label{tab:results on rgb channel with rdn}
		}
	\end{table*}

	We compare our proposed GSASR with state-of-the-art (SoTA) methods, including Meta-SR \cite{hu2019meta}, LIIF \cite{chen2021learning}, LTE \cite{lee2022local}, SRNO \cite{wei2023super}, LINF \cite{yao2023local}, CiasoSR \cite{cao2023ciaosr}, LMF \cite{he2024latent} and GaussianSR \cite{hu2025gaussiansr}. 
	In Sec. 4.1 of the main paper, we have reported the results with EDSR-baseline \cite{lim2017enhanced} on DIV2K100 \cite{timofte2017ntire} and LSDIR \cite{li2023lsdir} datasets. Here we report the results on the widely-used Set5 \cite{bevilacqua2012low}, Set14 \cite{zeyde2010single}, Urban100 \cite{huang2015single}, BSDS100 \cite{martin2001database}, Manga109 \cite{matsui2017sketch}, General100 \cite{dong2016accelerating} testing sets with both EDSR-baseline and RDN \cite{zhang2018residual}. For the performance measures, we report PSNR, SSIM~\cite{wang2004image}, LPIPS \cite{zhang2018unreasonable}, and DISTS \cite{ding2020image} metrics. The PSNR and SSIM \cite{wang2004image} metrics are computed on Y channel of Ycbcr space. Since some existing methods \cite{chen2021learning, cao2023ciaosr} calculate PSNR on RGB channel for DIV2K, while on Y channel of Ycbcr space for other datasets, to unify the testing protocol, we calculate PSNR/SSIM on Y channel of Ycbcr space across all testing datasets and all comparison methods. For a fair comparison, we download all of the competing models from their official websites, then utilize the same data to generate the SR results and calculate the metrics through the same evaluation codes. The results are presented in Tables~\ref{tab:results on DIV2K and LSDIR with RDN}, ~\ref{tab:results on Urban100 and Manga109 with EDSR-baseline}, ~\ref{tab:results on Urban100 and Manga109 with RDN}, ~\ref{tab:results on BSDS100 and General100 with EDSR}, ~\ref{tab:results on BSDS100 and General100 with RDN}, ~\ref{tab:results on Set5 and Set14 with EDSR}, ~\ref{tab:results on Set5 and Set14 with RDN}.
	
	From the quantitative results, one could find that GSASR significantly outperforms other arbitrary-scale super-resolution models in most benchmarks. GSASR drops slightly under high scaling factors (such as $\times 12$). This is because its performance depends on the number of employed Gaussians. When inferring on ultra-high scaling factors, the representation capability will be limited if the number of Gaussians is insufficient. One way to handle this issue is to increase the number of Gaussians. Nonetheless, GSASR's results on $\times 12$ ASR are still very competitive.
	
	\section{Quantitative Results on RGB Channels}
	
	We also quantitatively compare the performance of different models on RGB channels in Tables~\ref{tab:results on rgb channel with edsr} and ~\ref{tab:results on rgb channel with rdn}. One can observe that GSASR consistently achieves the best performance across all scaling factors and metrics on different datasets.
	
	\section{Qualitative Results}
	
\begin{figure*}[h]
		\small
		\centering
		\begin{minipage}{0.19\textwidth}
			\includegraphics[width=1\linewidth]{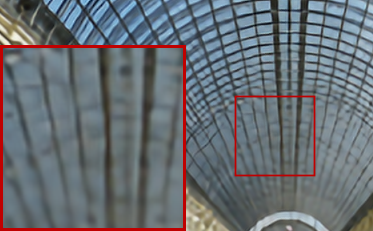}\\
			\centering{Meta-SR}\\
			\includegraphics[width=1\linewidth]{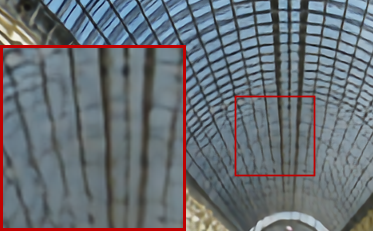 }
			\centering{LMF}\\
		\end{minipage}
		\begin{minipage}{0.19\textwidth}
			\includegraphics[width=1\linewidth]{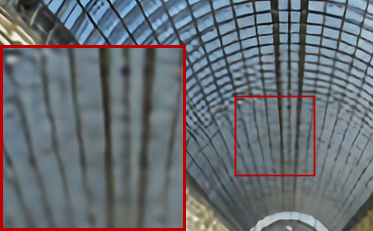}\\
			\centering{LIIF}\\
			\includegraphics[width=1\linewidth]{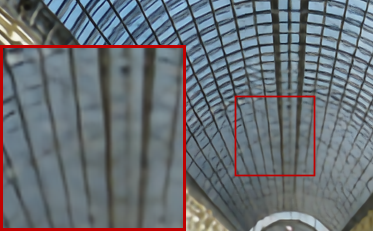 }
			\centering{CiaoSR}\\
		\end{minipage}
		\begin{minipage}{0.19\textwidth}
			\includegraphics[width=1\linewidth]{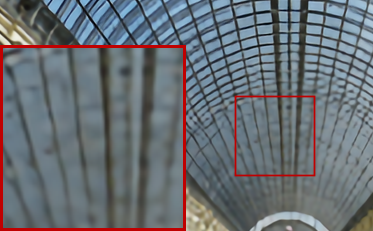}\\
			\centering{LTE}\\
			\includegraphics[width=1\linewidth]{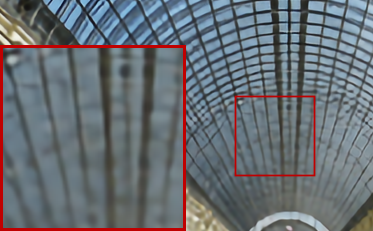 }
			\centering{GaussianSR}\\
		\end{minipage}
		\begin{minipage}{0.19\textwidth}
			\includegraphics[width=1\linewidth]{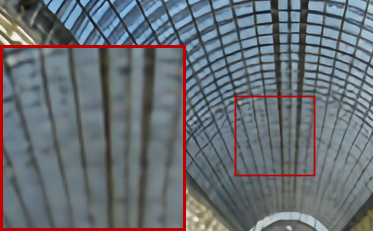}\\
			\centering{SRNO}\\
			\includegraphics[width=1\linewidth]{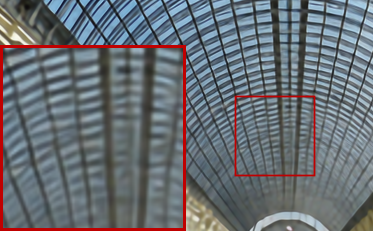 }
			\centering{GSASR (Ours)}\\
		\end{minipage}
		\begin{minipage}{0.19\textwidth}
			\includegraphics[width=1\linewidth]{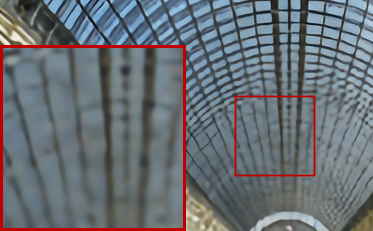}\\
			\centering{LINF}\\
			\includegraphics[width=1\linewidth]{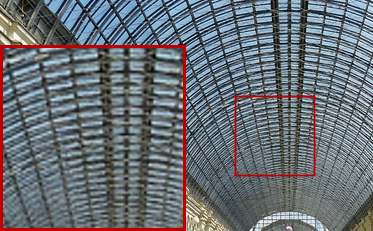}
			\centering{GT} \\
		\end{minipage}
		\captionsetup{font={small}}
		\caption{Visualization of GSASR and the competing methods under $\times 4$ scaling factor with EDSR \cite{lim2017enhanced} feature extraction backbone.}
		\label{fig:visualization3}
		
	\end{figure*}

\begin{figure*}[h]
		\small
		\centering
		\begin{minipage}{0.19\textwidth}
			\includegraphics[width=1\linewidth]{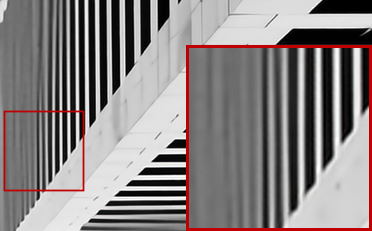}\\
			\centering{Meta-SR}\\
			\includegraphics[width=1\linewidth]{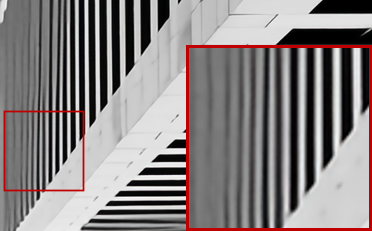 }
			\centering{LMF}\\
		\end{minipage}
		\begin{minipage}{0.19\textwidth}
			\includegraphics[width=1\linewidth]{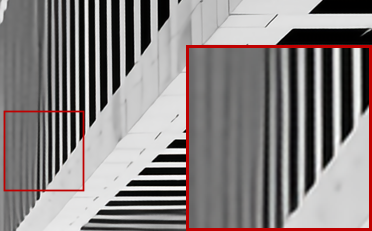}\\
			\centering{LIIF}\\
			\includegraphics[width=1\linewidth]{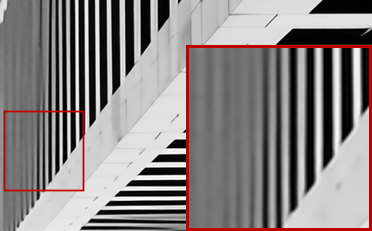 }
			\centering{CiaoSR}\\
		\end{minipage}
		\begin{minipage}{0.19\textwidth}
			\includegraphics[width=1\linewidth]{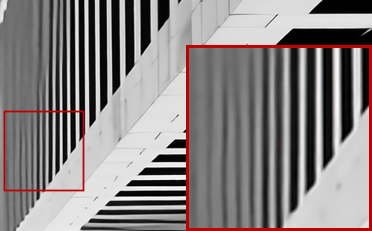}\\
			\centering{LTE}\\
			\includegraphics[width=1\linewidth]{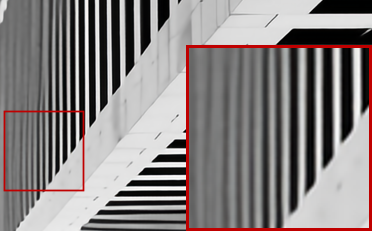 }
			\centering{GaussianSR}\\
		\end{minipage}
		\begin{minipage}{0.19\textwidth}
			\includegraphics[width=1\linewidth]{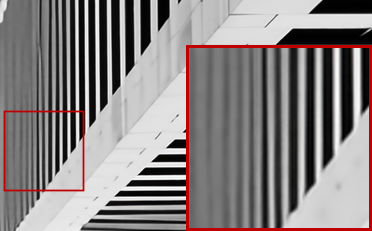}\\
			\centering{SRNO}\\
			\includegraphics[width=1\linewidth]{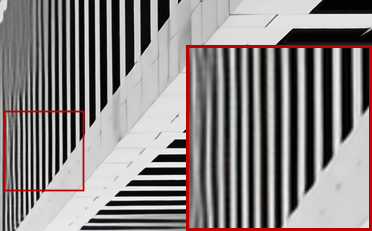 }
			\centering{GSASR (Ours)}\\
		\end{minipage}
		\begin{minipage}{0.19\textwidth}
			\includegraphics[width=1\linewidth]{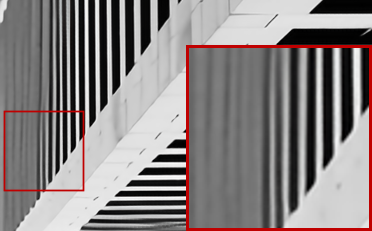}\\
			\centering{LINF}\\
			\includegraphics[width=1\linewidth]{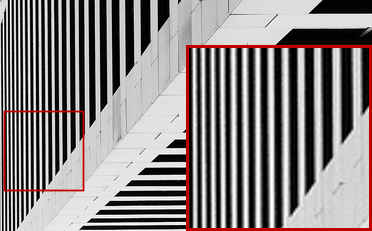}
			\centering{GT} \\
		\end{minipage}
		\captionsetup{font={small}}
		\caption{Visualization of GSASR and other methods under $\times 4$ scaling factor with RDN \cite{zhang2018residual} feature extraction backbone. }
		\label{fig:visualization4}
	\end{figure*}

\begin{figure*}[h]
		\small
		\centering
		\begin{minipage}{0.19\textwidth}
			\includegraphics[width=1\linewidth]{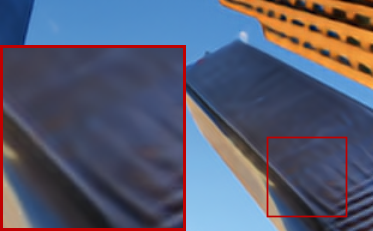}\\
			\centering{Meta-SR}\\
			\includegraphics[width=1\linewidth]{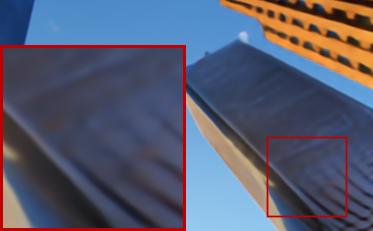 }
			\centering{LMF}\\
		\end{minipage}
		\begin{minipage}{0.19\textwidth}
			\includegraphics[width=1\linewidth]{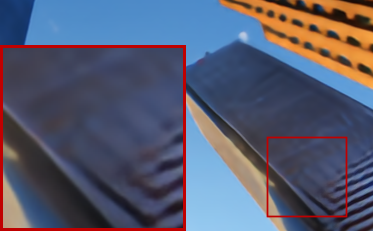}\\
			\centering{LIIF}\\
			\includegraphics[width=1\linewidth]{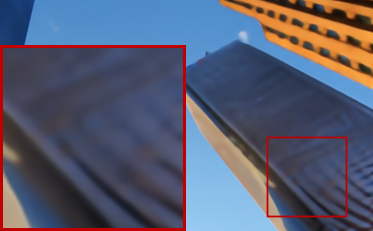 }
			\centering{CiaoSR}\\
		\end{minipage}
		\begin{minipage}{0.19\textwidth}
			\includegraphics[width=1\linewidth]{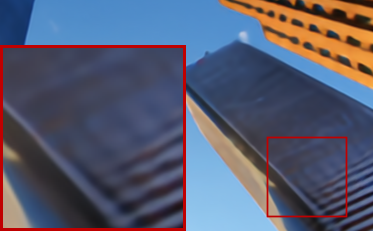}\\
			\centering{LTE}\\
			\includegraphics[width=1\linewidth]{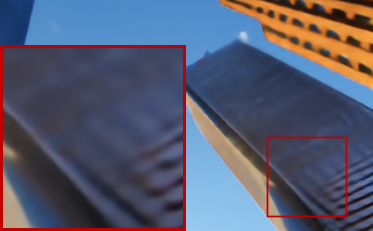 }
			\centering{GaussianSR}\\
		\end{minipage}
		\begin{minipage}{0.19\textwidth}
			\includegraphics[width=1\linewidth]{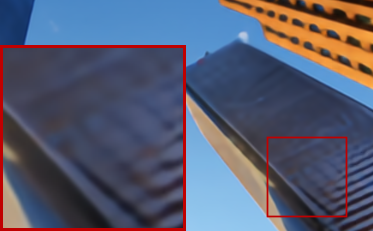}\\
			\centering{SRNO}\\
			\includegraphics[width=1\linewidth]{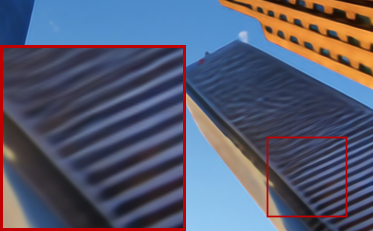 }
			\centering{GSASR (Ours)}\\
		\end{minipage}
		\begin{minipage}{0.19\textwidth}
			\includegraphics[width=1\linewidth]{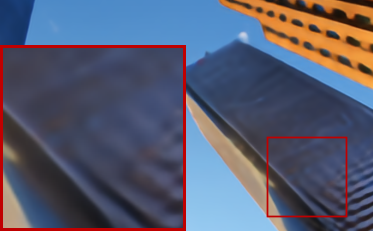}\\
			\centering{LINF}\\
			\includegraphics[width=1\linewidth]{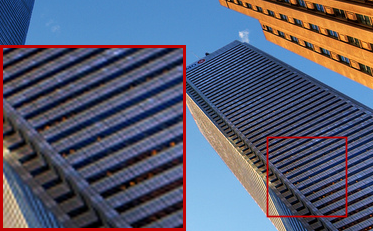}
			\centering{GT} \\
		\end{minipage}
		\captionsetup{font={small}}
		\caption{Visualization of GSASR and other methods under $\times 6$ scaling factor with EDSR \cite{lim2017enhanced} feature extraction backbone. }
		\label{fig:visualization5}
	\end{figure*}

\begin{figure*}[h]
		\small
		\centering
		\begin{minipage}{0.19\textwidth}
			\includegraphics[width=1\linewidth]{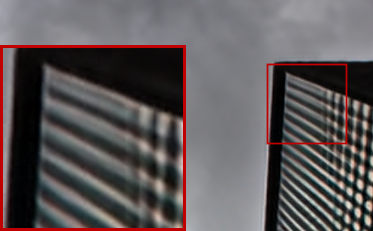}\\
			\centering{Meta-SR}\\
			\includegraphics[width=1\linewidth]{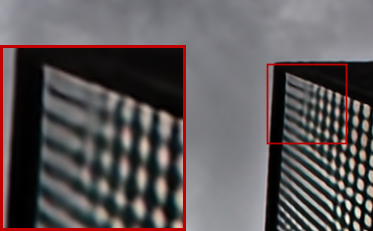 }
			\centering{LMF}\\
		\end{minipage}
		\begin{minipage}{0.19\textwidth}
			\includegraphics[width=1\linewidth]{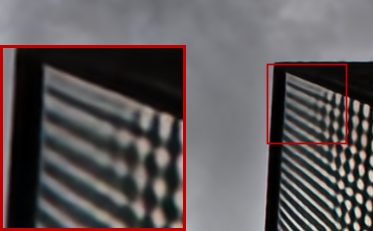}\\
			\centering{LIIF}\\
			\includegraphics[width=1\linewidth]{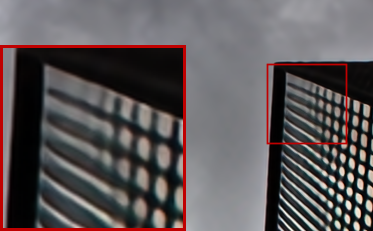 }
			\centering{CiaoSR}\\
		\end{minipage}
		\begin{minipage}{0.19\textwidth}
			\includegraphics[width=1\linewidth]{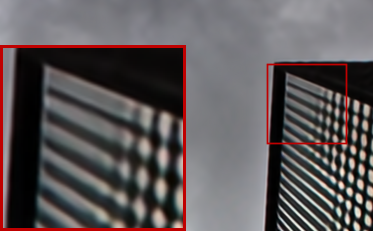}\\
			\centering{LTE}\\
			\includegraphics[width=1\linewidth]{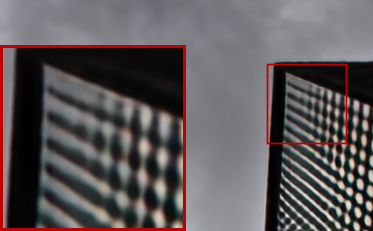 }
			\centering{GaussianSR}\\
		\end{minipage}
		\begin{minipage}{0.19\textwidth}
			\includegraphics[width=1\linewidth]{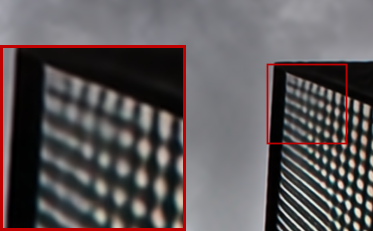}\\
			\centering{SRNO}\\
			\includegraphics[width=1\linewidth]{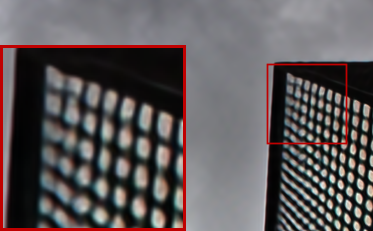 }
			\centering{GSASR (Ours)}\\
		\end{minipage}
		\begin{minipage}{0.19\textwidth}
			\includegraphics[width=1\linewidth]{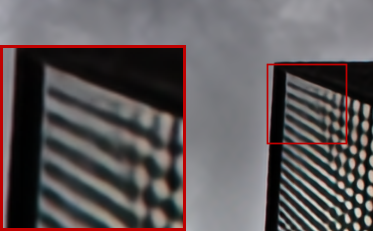}\\
			\centering{LINF}\\
			\includegraphics[width=1\linewidth]{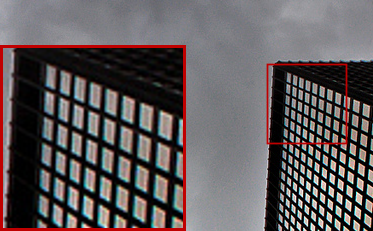}
			\centering{GT} \\
		\end{minipage}
		\captionsetup{font={small}}
		\caption{Visualization of GSASR and other methods under $\times 6$ scaling factor with RDN \cite{zhang2018residual} feature extraction backbone. }
		\label{fig:visualization6}
	\end{figure*}

\begin{figure*}[h]
		\small
		\centering
		\begin{minipage}{0.19\textwidth}
			\includegraphics[width=1\linewidth]{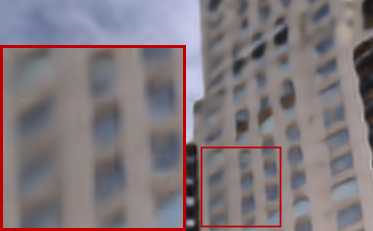}\\
			\centering{Meta-SR}\\
			\includegraphics[width=1\linewidth]{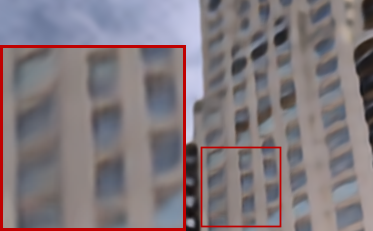 }
			\centering{LMF}\\
		\end{minipage}
		\begin{minipage}{0.19\textwidth}
			\includegraphics[width=1\linewidth]{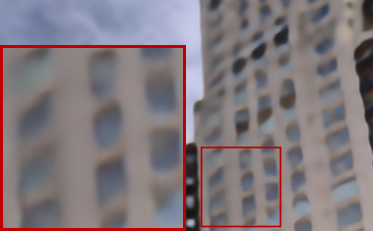}\\
			\centering{LIIF}\\
			\includegraphics[width=1\linewidth]{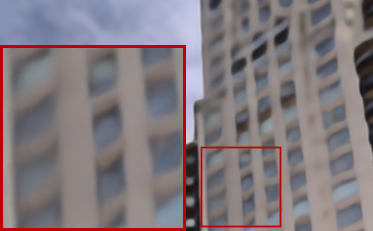 }
			\centering{CiaoSR}\\
		\end{minipage}
		\begin{minipage}{0.19\textwidth}
			\includegraphics[width=1\linewidth]{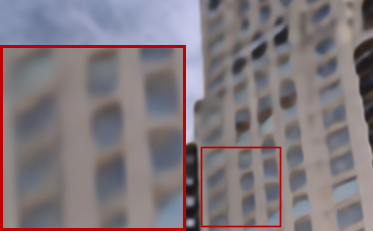}\\
			\centering{LTE}\\
			\includegraphics[width=1\linewidth]{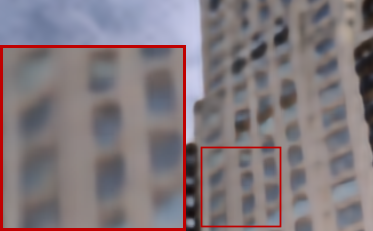 }
			\centering{GaussianSR}\\
		\end{minipage}
		\begin{minipage}{0.19\textwidth}
			\includegraphics[width=1\linewidth]{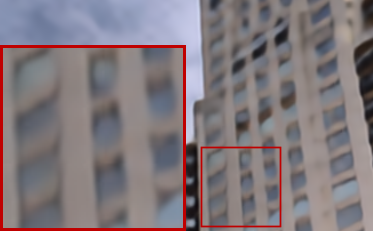}\\
			\centering{SRNO}\\
			\includegraphics[width=1\linewidth]{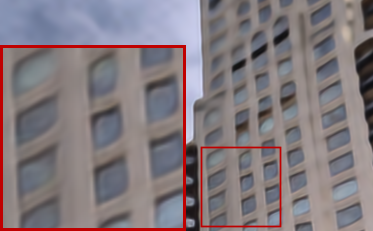 }
			\centering{GSASR (Ours)}\\
		\end{minipage}
		\begin{minipage}{0.19\textwidth}
			\includegraphics[width=1\linewidth]{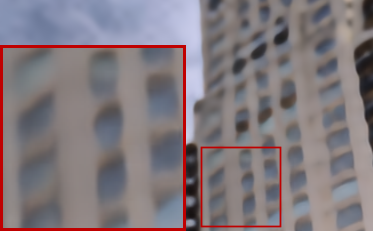}\\
			\centering{LINF}\\
			\includegraphics[width=1\linewidth]{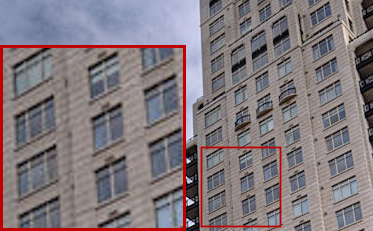}
			\centering{GT} \\
		\end{minipage}
		\captionsetup{font={small}}
		\caption{Visualization of GSASR and other methods under $\times 8$ scaling factor with EDSR \cite{lim2017enhanced} feature extraction backbone. }
		\label{fig:visualization7}
	\end{figure*}

\begin{figure*}[h]
		\small
		\centering
		\begin{minipage}{0.19\textwidth}
			\includegraphics[width=1\linewidth]{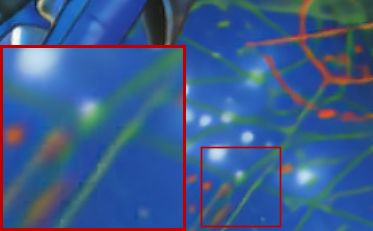}\\
			\centering{Meta-SR}\\
			\includegraphics[width=1\linewidth]{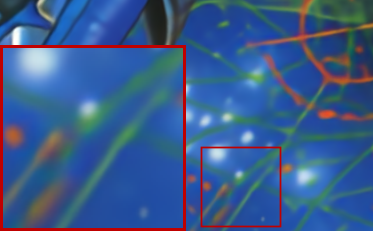 }
			\centering{LMF}\\
		\end{minipage}
		\begin{minipage}{0.19\textwidth}
			\includegraphics[width=1\linewidth]{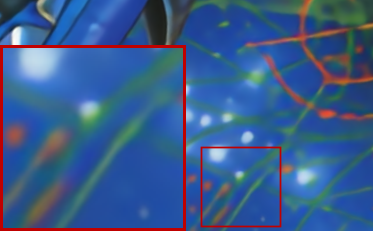}\\
			\centering{LIIF}\\
			\includegraphics[width=1\linewidth]{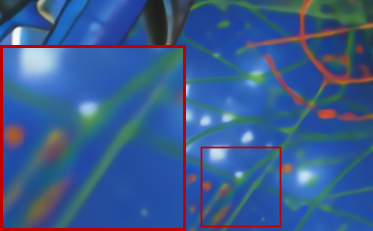 }
			\centering{CiaoSR}\\
		\end{minipage}
		\begin{minipage}{0.19\textwidth}
			\includegraphics[width=1\linewidth]{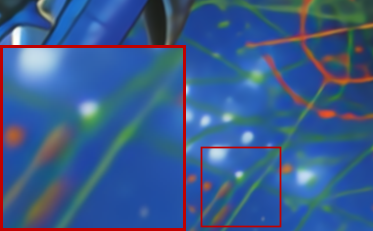}\\
			\centering{LTE}\\
			\includegraphics[width=1\linewidth]{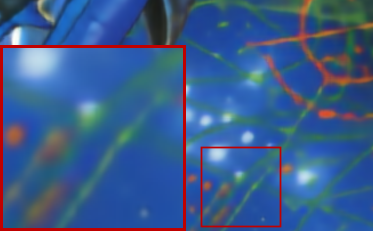 }
			\centering{GaussianSR}\\
		\end{minipage}
		\begin{minipage}{0.19\textwidth}
			\includegraphics[width=1\linewidth]{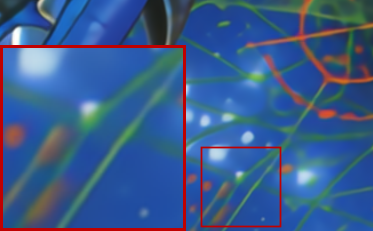}\\
			\centering{SRNO}\\
			\includegraphics[width=1\linewidth]{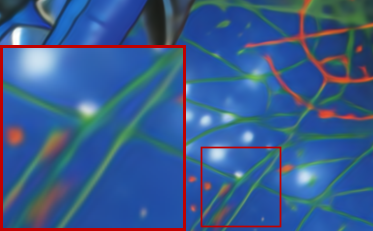 }
			\centering{GSASR (Ours)}\\
		\end{minipage}
		\begin{minipage}{0.19\textwidth}
			\includegraphics[width=1\linewidth]{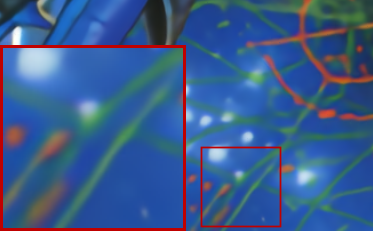}\\
			\centering{LINF}\\
			\includegraphics[width=1\linewidth]{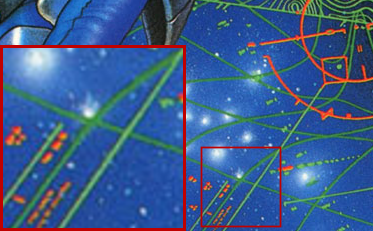}
			\centering{GT} \\
		\end{minipage}
		
		\captionsetup{font={small}}
		\caption{Visualization of GSASR and other methods under $\times 8$ scaling factor with RDN \cite{zhang2018residual} feature extraction backbone. }
		\label{fig:visualization8}
	\end{figure*}

\begin{figure*}[h]
		\small
		\centering
		\begin{minipage}{0.19\textwidth}
			\includegraphics[width=1\linewidth]{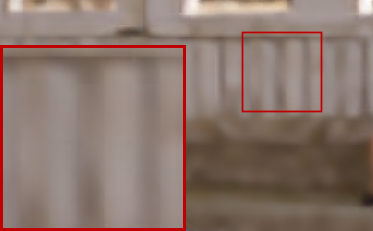}\\
			\centering{Meta-SR}\\
			\includegraphics[width=1\linewidth]{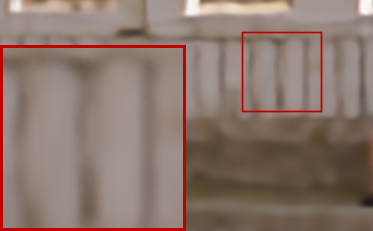 }
			\centering{LMF}\\
		\end{minipage}
		\begin{minipage}{0.19\textwidth}
			\includegraphics[width=1\linewidth]{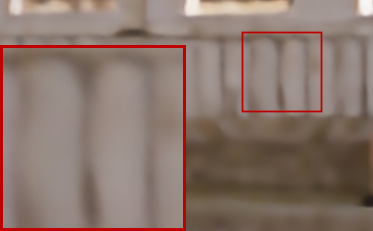}\\
			\centering{LIIF}\\
			\includegraphics[width=1\linewidth]{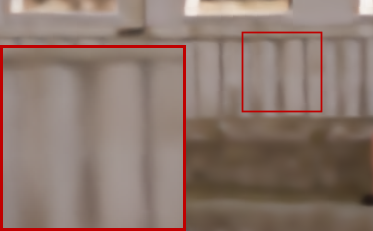 }
			\centering{CiaoSR}\\
		\end{minipage}
		\begin{minipage}{0.19\textwidth}
			\includegraphics[width=1\linewidth]{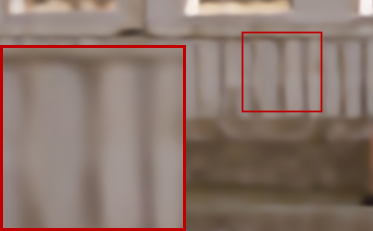}\\
			\centering{LTE}\\
			\includegraphics[width=1\linewidth]{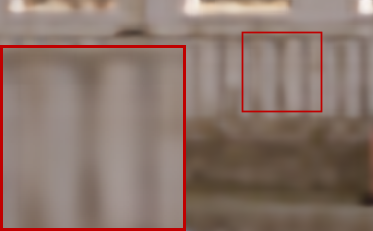 }
			\centering{GaussianSR}\\
		\end{minipage}
		\begin{minipage}{0.19\textwidth}
			\includegraphics[width=1\linewidth]{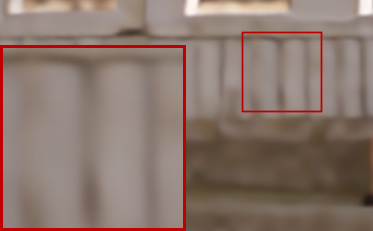}\\
			\centering{SRNO}\\
			\includegraphics[width=1\linewidth]{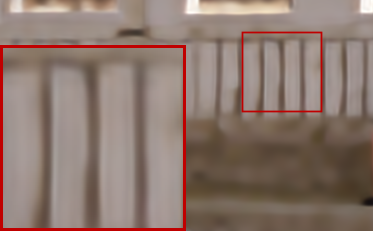 }
			\centering{GSASR (Ours)}\\
		\end{minipage}
		\begin{minipage}{0.19\textwidth}
			\includegraphics[width=1\linewidth]{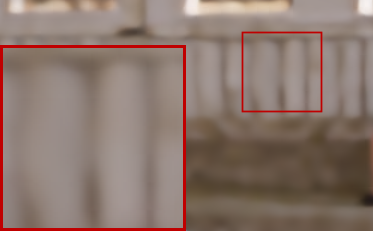}\\
			\centering{LINF}\\
			\includegraphics[width=1\linewidth]{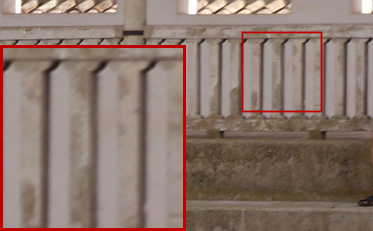}
			\centering{GT} \\
		\end{minipage}
		
		\captionsetup{font={small}}
		\caption{Visualization of GSASR and other methods under $\times 12$ scaling factor with EDSR \cite{lim2017enhanced} feature extraction backbone. }
		\label{fig:visualization9}
	\end{figure*}

\begin{figure*}[h]
		\small
		\centering
		\begin{minipage}{0.19\textwidth}
			\includegraphics[width=1\linewidth]{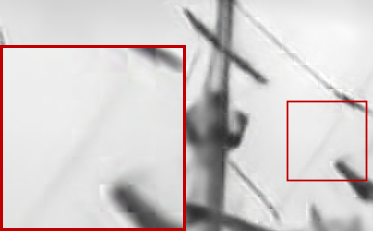}\\
			\centering{Meta-SR}\\
			\includegraphics[width=1\linewidth]{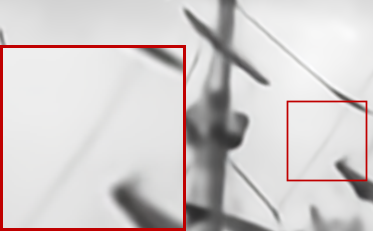 }
			\centering{LMF}\\
		\end{minipage}
		\begin{minipage}{0.19\textwidth}
			\includegraphics[width=1\linewidth]{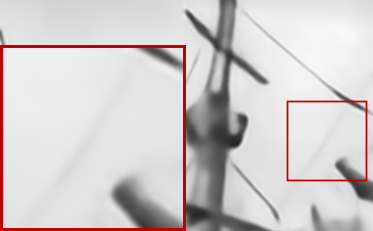}\\
			\centering{LIIF}\\
			\includegraphics[width=1\linewidth]{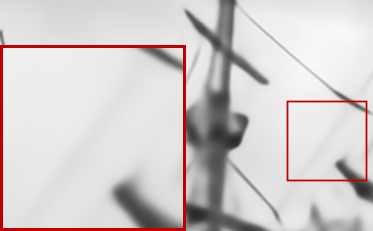 }
			\centering{CiaoSR}\\
		\end{minipage}
		\begin{minipage}{0.19\textwidth}
			\includegraphics[width=1\linewidth]{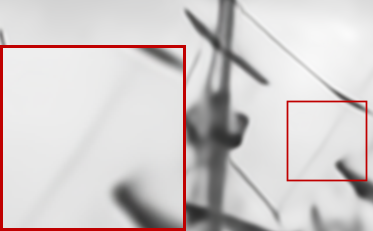}\\
			\centering{LTE}\\
			\includegraphics[width=1\linewidth]{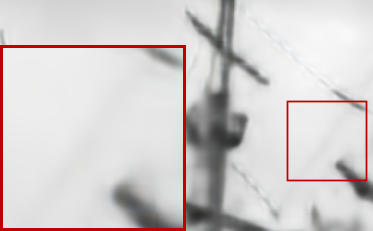 }
			\centering{GaussianSR}\\
		\end{minipage}
		\begin{minipage}{0.19\textwidth}
			\includegraphics[width=1\linewidth]{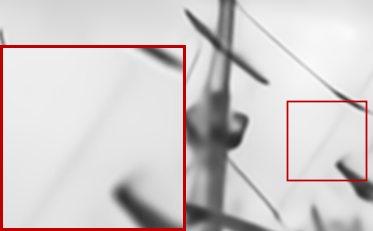}\\
			\centering{SRNO}\\
			\includegraphics[width=1\linewidth]{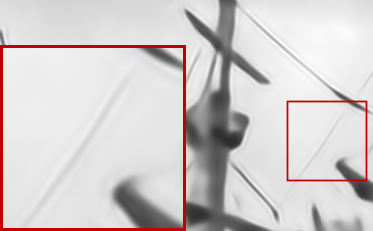 }
			\centering{GSASR (Ours)}\\
		\end{minipage}
		\begin{minipage}{0.19\textwidth}
			\includegraphics[width=1\linewidth]{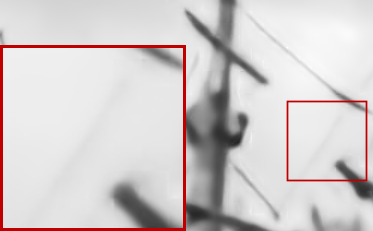}\\
			\centering{LINF}\\
			\includegraphics[width=1\linewidth]{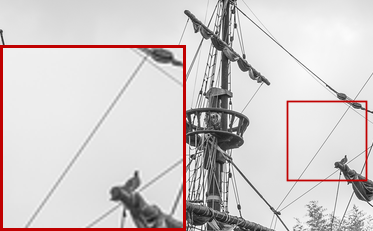}
			\centering{GT} \\
		\end{minipage}
		
		\captionsetup{font={small}}
		\caption{Visualization of GSASR and other methods under $\times 12$ scaling factor with RDN \cite{zhang2018residual} feature extraction backbone. }
		\label{fig:visualization10}
	\end{figure*}

\begin{figure*}[h]
		\small
		\centering
		\begin{minipage}{0.32\textwidth}
			\includegraphics[width=1\linewidth]{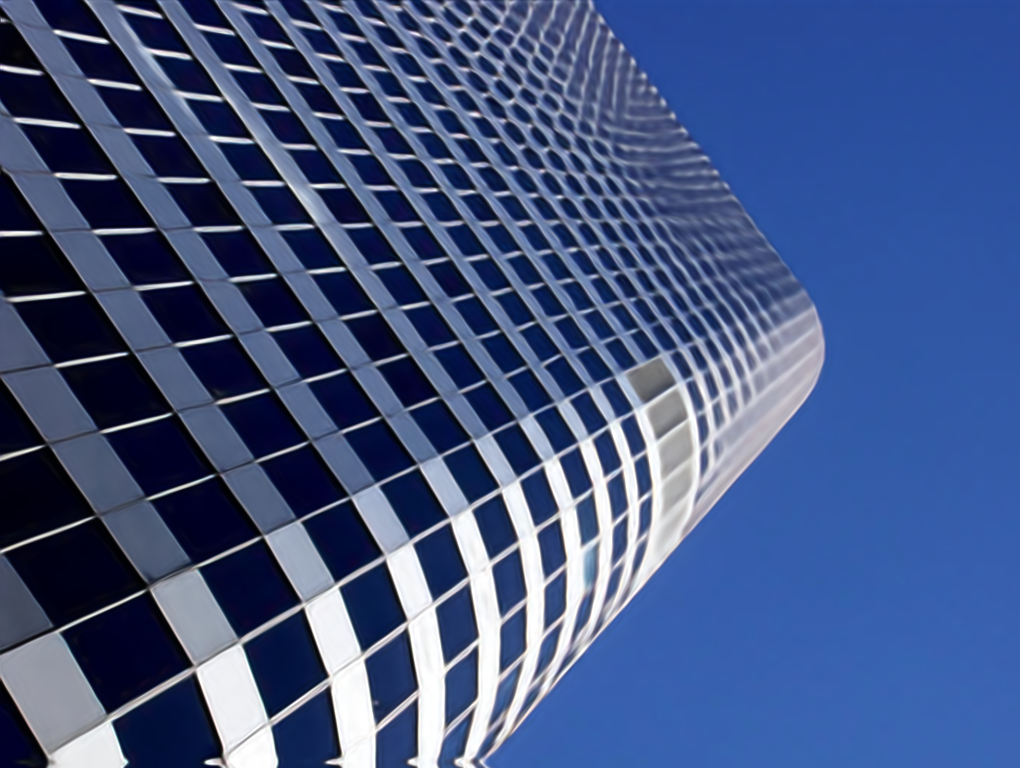}\\
			\includegraphics[width=1\linewidth]{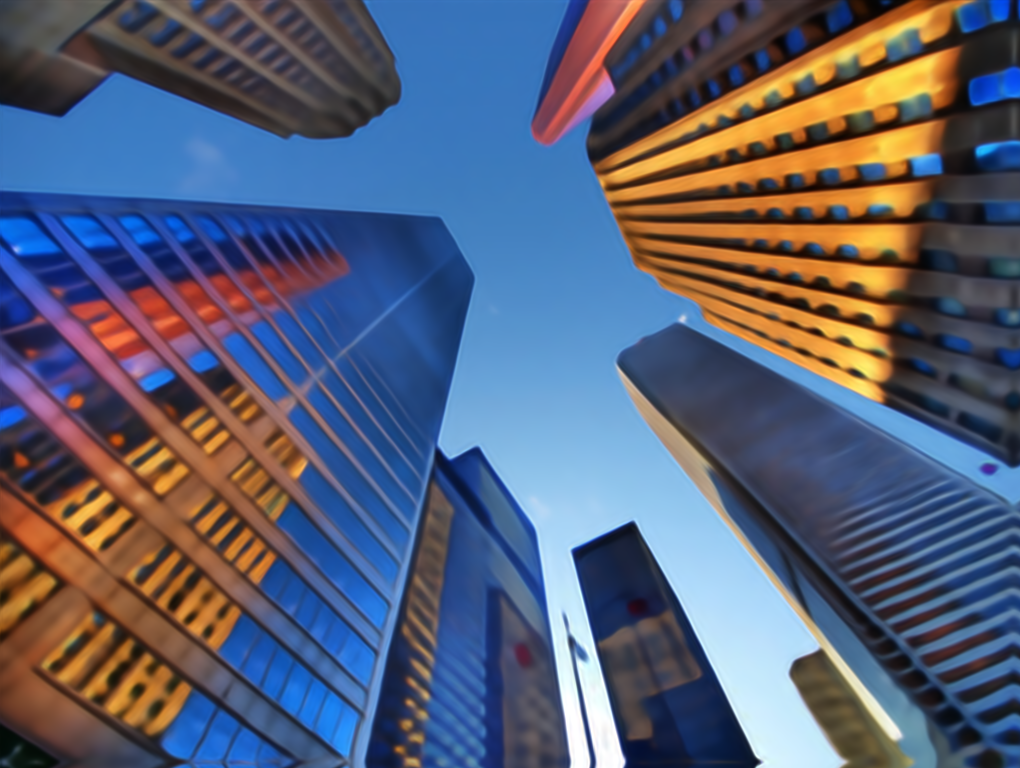}\\
			\includegraphics[width=1\linewidth]{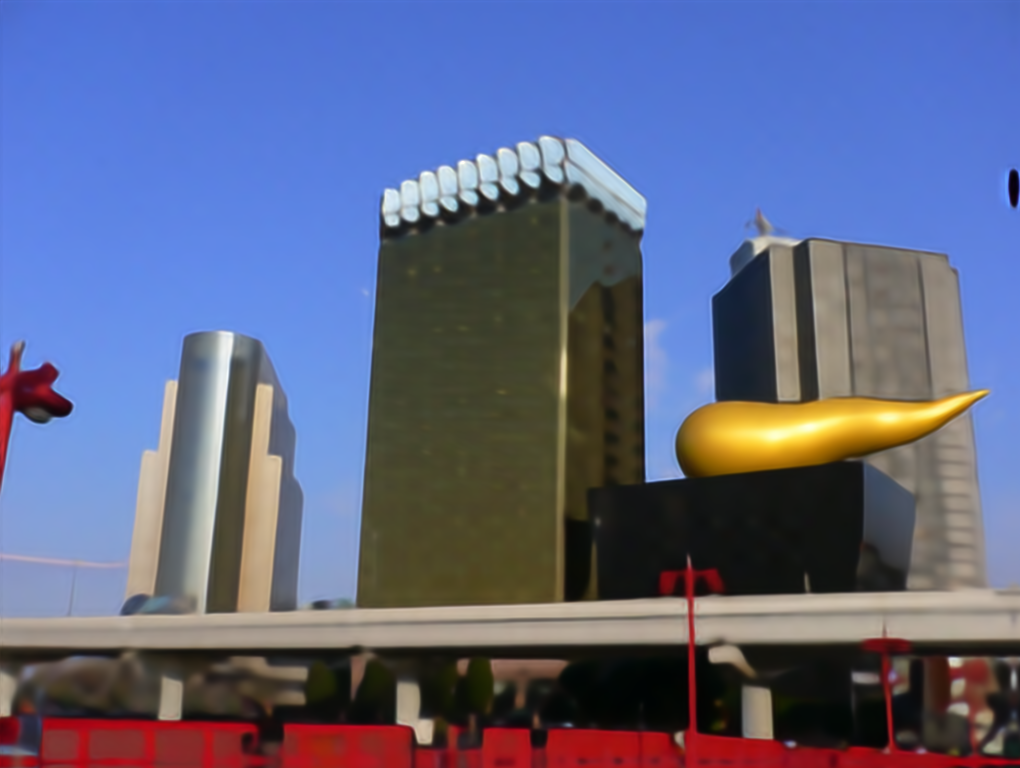}\\
			\includegraphics[width=1\linewidth]{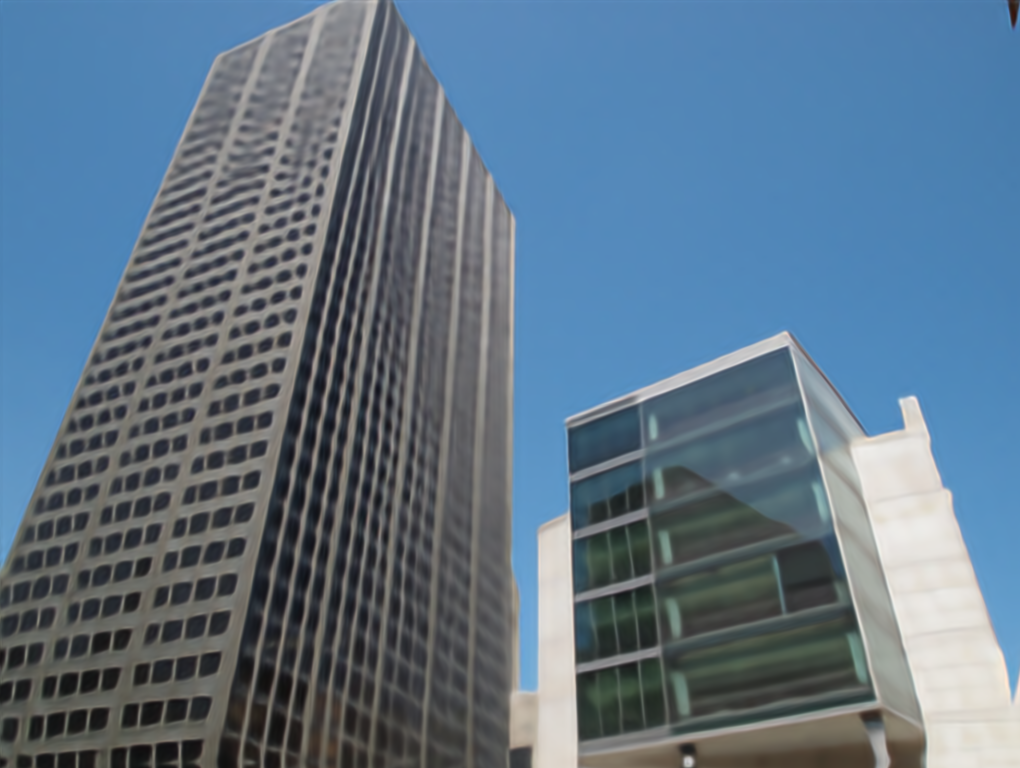}\\
			\centering{SR Image}\\
		\end{minipage}
		\begin{minipage}{0.32\textwidth}
			\includegraphics[width=1\linewidth]{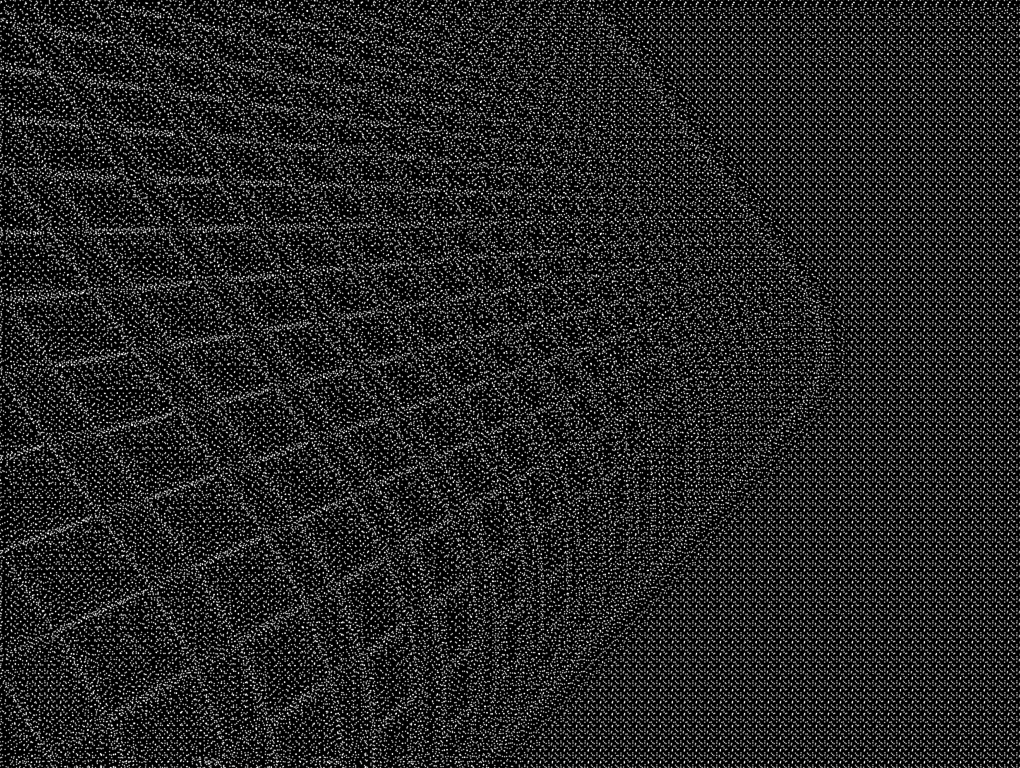}\\
			\includegraphics[width=1\linewidth]{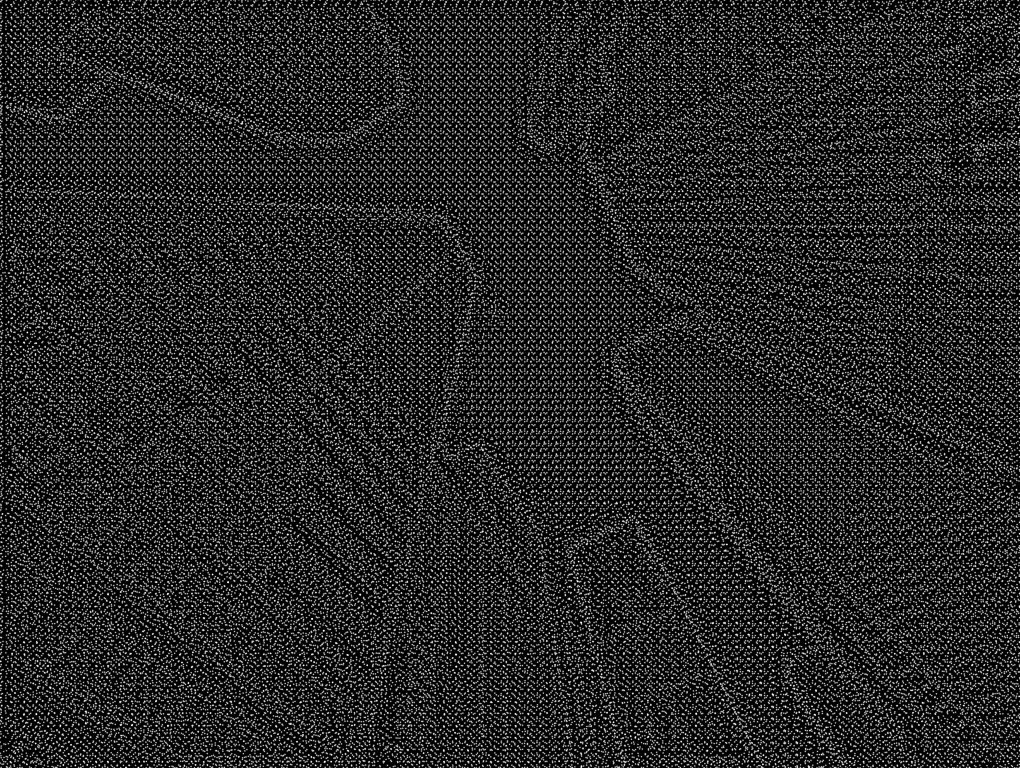}\\
			\includegraphics[width=1\linewidth]{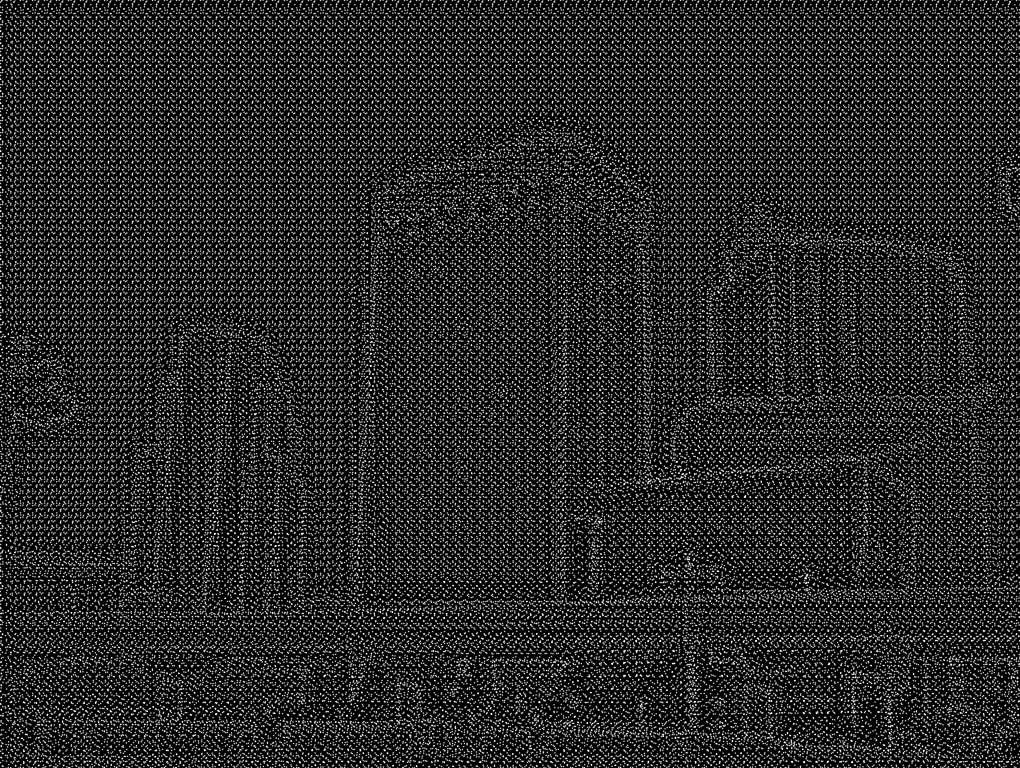}\\
			\includegraphics[width=1\linewidth]{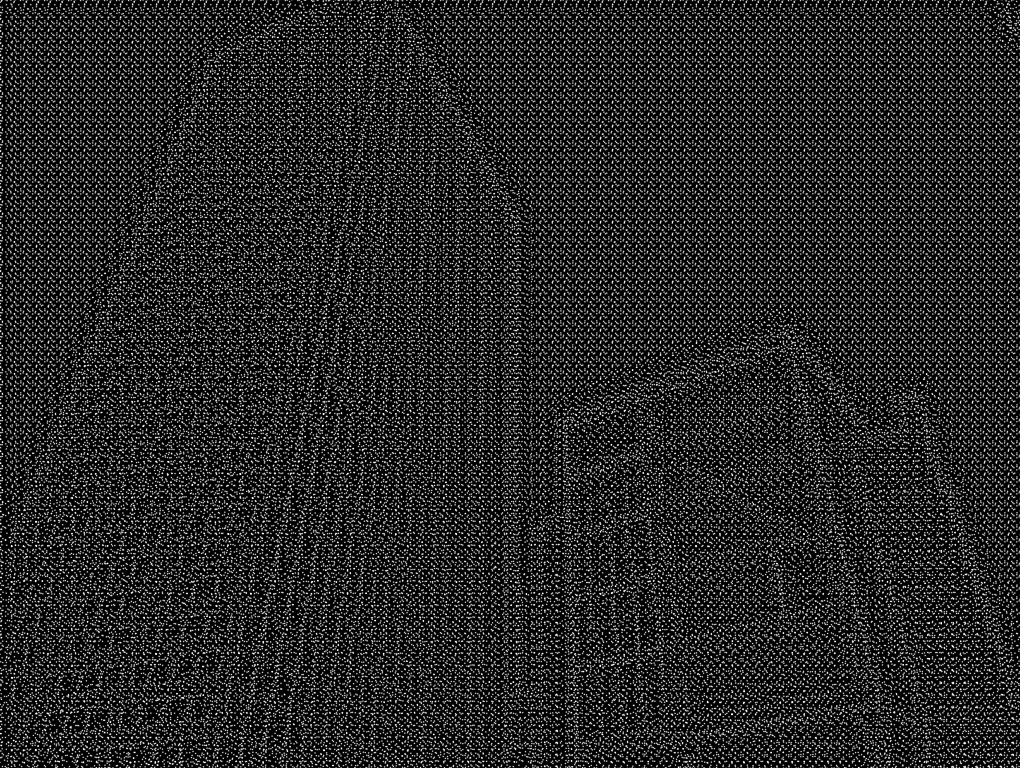}\\
			\centering{Position Distribution of Gaussians}\\
		\end{minipage}
		\begin{minipage}{0.32\textwidth}
			\includegraphics[width=1\linewidth]{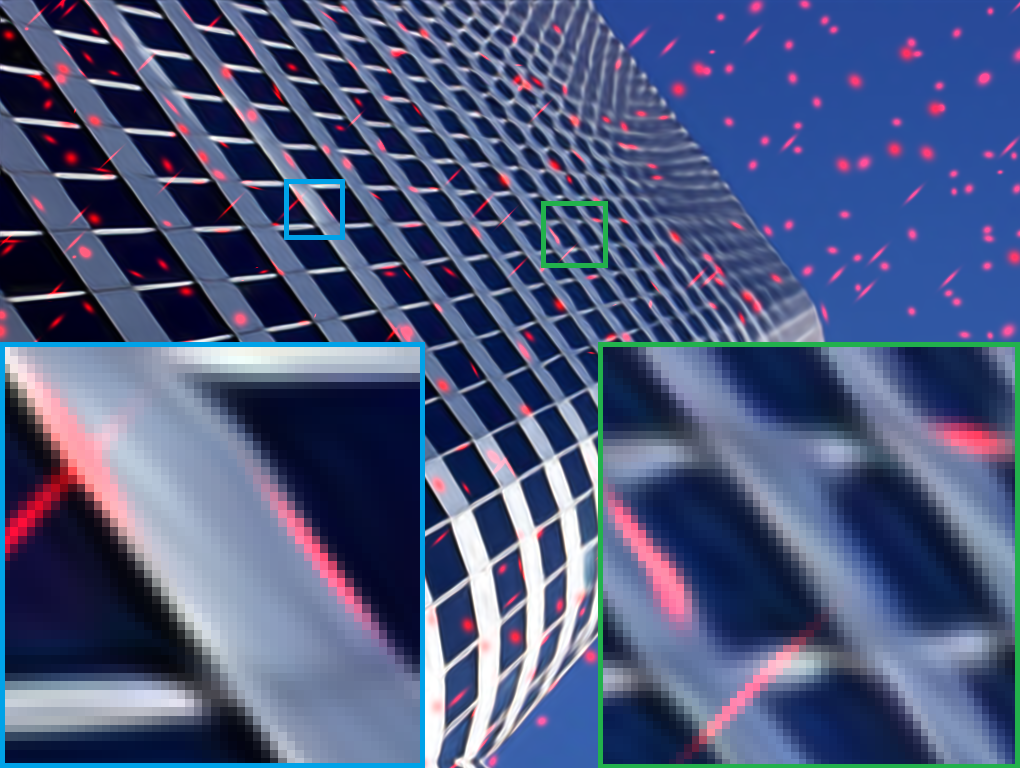}\\
			\includegraphics[width=1\linewidth]{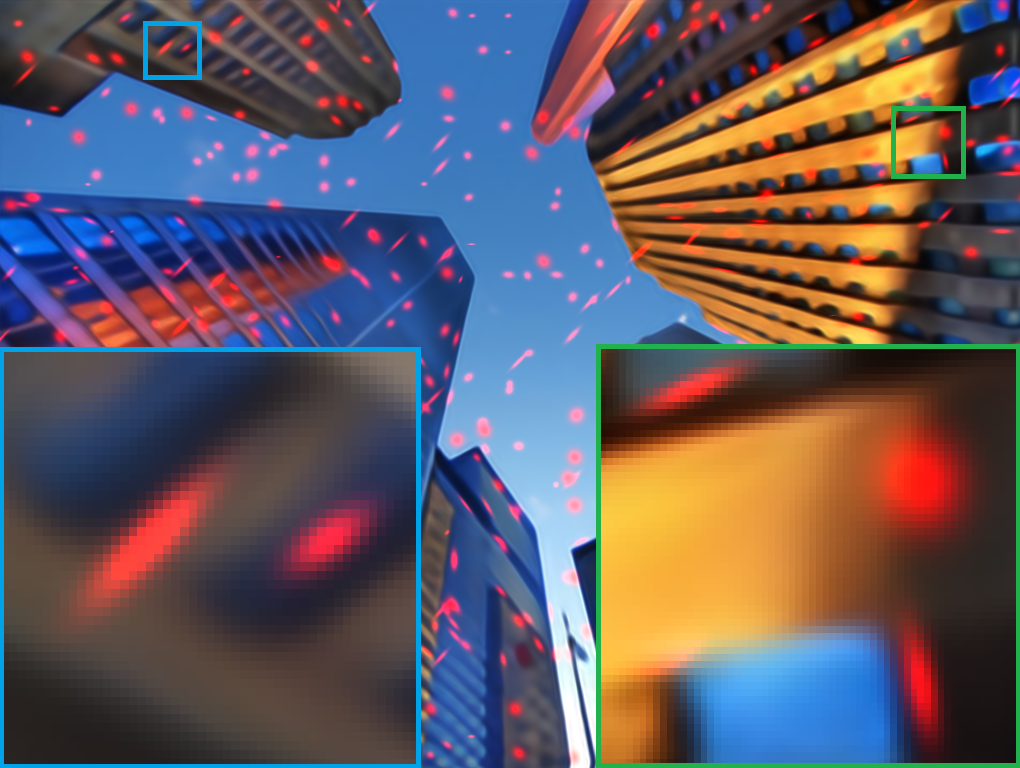}\\
			\includegraphics[width=1\linewidth]{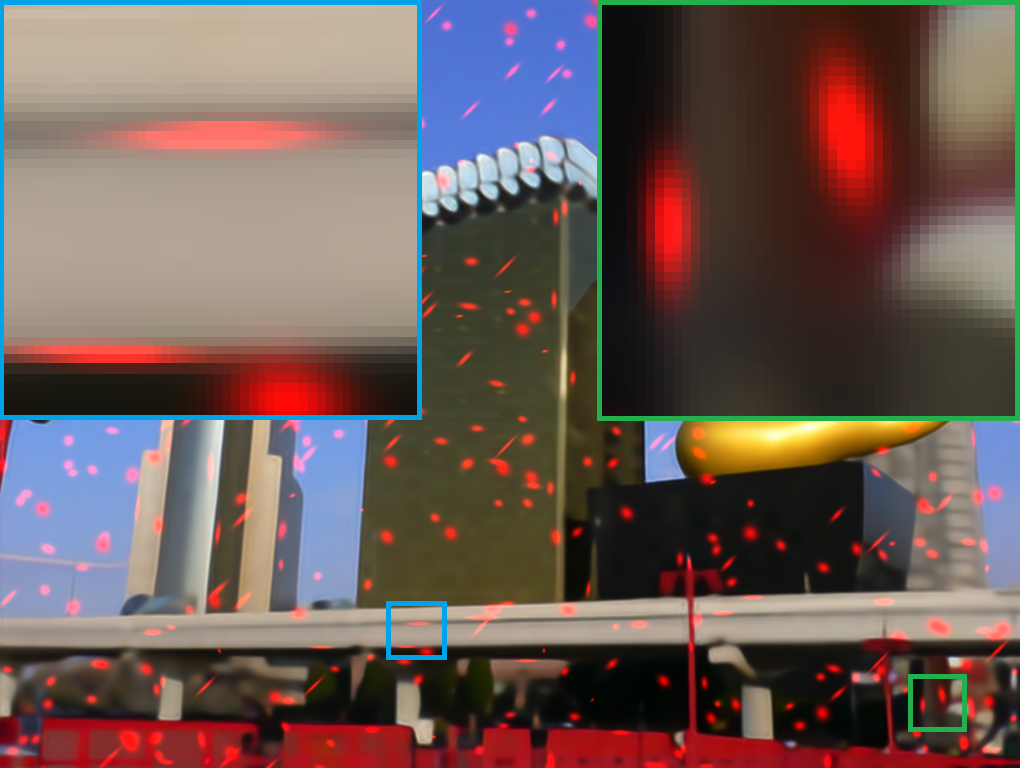}\\
			\includegraphics[width=1\linewidth]{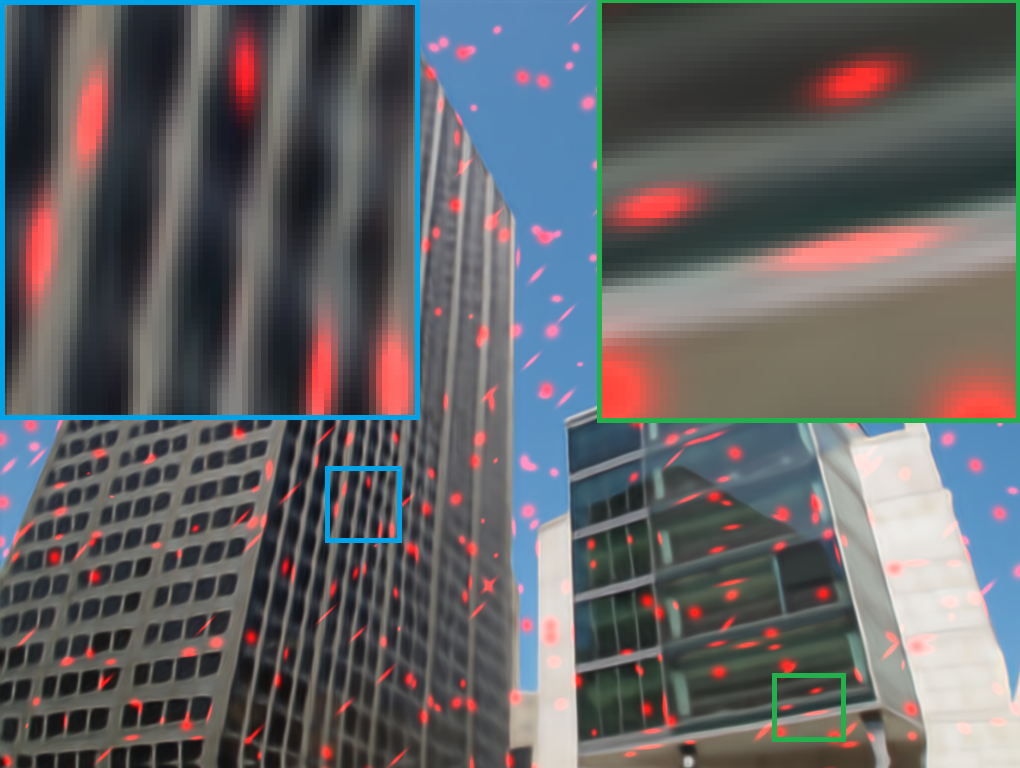}\\
			\centering{Colorized Gaussians}\\
		\end{minipage}
		\caption{Demonstration of the rich expressiveness of Gaussians for ASR. The position distribution is obtained by setting $\{\sigma, \rho, c\}$ to fixed values. One can observe that Gaussians are evenly distributed in regions with simple textures, while their positions are adjusted in regions with complex textures to fit details. (\textbf{Please zoom in for better observation}.) In the right image, we randomly select parts of Gaussians and highlight their colors to red. One can see that 2D Gaussians can learn to fit the different object shapes ($e.g.,$ the edge of window). }
		\label{fig:visualization of position and shape of Gaussians}
	\end{figure*}

	We provide more qualitative results of our proposed GSASR and the compared methods, including Meta-SR \cite{hu2019meta}, LIIF \cite{chen2021learning}, LTE \cite{lee2022local}, SRNO \cite{wei2023super}, LINF \cite{yao2023local}, CiasoSR \cite{cao2023ciaosr}, LMF \cite{he2024latent} and GaussianSR \cite{hu2025gaussiansr}. The visual comparisons are shown in Figs.~\ref{fig:visualization3}, ~\ref{fig:visualization4}, ~\ref{fig:visualization5}, ~\ref{fig:visualization6}, ~\ref{fig:visualization7}, ~\ref{fig:visualization8}, ~\ref{fig:visualization9}, ~\ref{fig:visualization10}.
	One could find that GSASR generates much clearer details than all the other methods, especially for complex textures. For example, in Fig.~\ref{fig:visualization3}, the textures of the windows are very clear in GSASR, while the other methods lack lateral textures. Similar results could be observed from other visualization examples. In GSASR, Gaussians could move freely across the whole image, and automatically concentrate on areas with complex textures, benefiting the super-resolution of details. In the first row of Fig.~\ref{fig:visualization of position and shape of Gaussians}, one could find that Gaussians are inclined to concentrate more on the edge pixels, such as the building textures, while distributing uniformly on areas with smooth textures, such as the sky.
	
	\section{Rendering Cost Comparison between GSASR and GaussianSR}
	
	We compare the computational cost between the Pytorch-based rendering pipeline in GaussianSR \cite{hu2025gaussiansr} and the CUDA-based one in our GSASR in Table~\ref{tab:computational burden between GaussianSR and GSASR}. One could find that our CUDA-based rendering consumes much less computational resources.

	\section{Computational Costs}
	
	We provide the computational costs of different ASR methods with RDN \cite{zhang2018residual} as backbone. We report the average inference time (ms) and GPU memory (MB) usage on $100$ images with $720 \times 720$ size, which are cropped from the DIV2K \cite{timofte2017ntire} dataset. The detailed results are shown in Table~\ref{tab:computational cost on RDN}. The inference time and GPU memory cost are computed for the whole SR pipeline, including the encoder, decoder and rendering parts. The best results are highlighted in \textcolor{red}{red}.

	From Table~\ref{tab:computational cost on RDN}, one could find that GSASR runs much faster than the SOTA method, CiaoSR \cite{cao2023ciaosr}, with comparable GPU memory usage on all scaling factors. When the scaling factor increases, GSASR could run with competitive speed with other methods, while keeping much better performance than them. For example, under the $\times 8$ scaling factor, the inference time of GSASR, LIIF \cite{chen2021learning}, CiaoSR \cite{cao2023ciaosr} is $206$ms, $187$ms, $633$ms, respectively, while GSASR could still maintain a balanced performance. Advantaged by the efficient 2D GPU/CUDA-based rasterization, ours method could render a super-resolved image with high efficiency and fidelity.

	\begin{table*}[!h]
		\centering
		\caption{Computational costs between the Pytorch-based rendering pipeline in GaussianSR \cite{hu2025gaussiansr} and CUDA-based one in GSASR.}
		\begin{tabular}{c|c|c|c} 
			\toprule[2pt]
			Scale & Method & Time (ms) & Memory (MB) \\ 
			\midrule[1pt] \midrule[1pt]
			\multirow{2}{*}{x4} & GaussianSR & 249 & 2685 \\
			& GSASR & \textbf{168} & \textbf{172} \\ 
			\midrule[1pt]
			\multirow{2}{*}{x8} & GaussianSR & 234 & 2650 \\
			& GSASR & \textbf{71} & \textbf{135} \\
			\bottomrule[2pt]
		\end{tabular}
		\label{tab:computational burden between GaussianSR and GSASR}
	\end{table*}
	
	\begin{table*}[h]
		\centering
		\caption{Comparison of computational costs. We report the results using RDN \cite{zhang2018image} as image encoder. PSNR and SSIM index are computed on Y channel of Ycbcr space. Apart from the PSNR/SSIM/LPIPS/DISTS metrics, we report the average inference time (ms) as well as the GPU memory (MB). The inference time/GPU memory cost is computed for the whole SR pipeline, including
			the encoder, decoder and rendering parts.. The best results are highlighted in \textcolor{red}{red}. }
		\scalebox{0.9}{
			\resizebox{1.0\linewidth}{!}{
				\begin{tabular}{c|c|ccccccccc} 
					\toprule[2pt]
					\multirow{3}{*}{Scale} & \multirow{3}{*}{\begin{tabular}[c]{@{}c@{}}Computational Cost\\and Performance\end{tabular}} & \multicolumn{9}{c}{Backbone: RDN} \\ 
					\cline{3-11}
					&  & \multicolumn{9}{c}{Testing GT Size: 720 * 720} \\
					&  & Meta-SR & LIIF & LTE & SRNO & LINF & LMF & CiaoSR & GaussianSR & GSASR \\ 
					\midrule[1pt] \midrule[1pt]
					\multirow{6}{*}{$\times$2} & PSNR & 37.76 & 37.41 & 37.44 & 37.64 & 37.74 & 37.61 & 37.85 & 37.54 & \textcolor{red}{37.92} \\
					& SSIM & 0.9510 & 0.9509 & 0.9511 & 0.9521 & 0.9511 & 0.9515 & 0.9518 & 0.9512 & \textcolor{red}{0.9526} \\
					& LPIPS & 0.0722 & 0.0734 & 0.0723 & 0.0695 & 0.0727 & 0.0715 & 0.0710 & 0.0735 & \textcolor{red}{0.0668} \\
					& DISTS & 0.0709 & 0.0712 & 0.0703 & 0.0693 & 0.0706 & 0.0705 & 0.0687 & 0.0711 & \textcolor{red}{0.0670} \\
					& Inference Time & \textcolor{red}{178} & 518 & 254 & 236 & 202 & 311 & 23711 & 1280 & 1679 \\
					& GPU Memory & 8775 & \textcolor{red}{1246} & 1970 & 6381 & 3732 & 7096 & 49231 & 5278 & 13447 \\ 
					\midrule[1pt]
					\multirow{6}{*}{$\times$3} & PSNR & 33.94 & 33.71 & 33.75 & 33.90 & 33.93 & 33.85 & 34.06 & 33.82 & \textcolor{red}{34.12} \\
					& SSIM & 0.8979 & 0.8978 & 0.8982 & 0.8998 & 0.8981 & 0.8992 & 0.8996 & 0.8983 & \textcolor{red}{0.9015} \\
					& LPIPS & 0.1692 & 0.1686 & 0.1678 & 0.1655 & 0.1687 & 0.1648 & 0.1667 & 0.169 & \textcolor{red}{0.1611} \\
					& DISTS & 0.1224 & 0.1223 & 0.1206 & 0.1195 & 0.1221 & 0.1203 & 0.1180 & 0.1219 & \textcolor{red}{0.1172} \\
					& Inference Time & \textcolor{red}{104} & 486 & 181 & 163 & 128 & 172 & 2064 & 955 & 857 \\
					& GPU Memory & 8454 & \textcolor{red}{610} & 930 & 6361 & 3571 & 6108 & 10081 & 5216 & 6080 \\ 
					\midrule[1pt]
					\multirow{6}{*}{$\times$4} & PSNR & 31.90 & 31.70 & 31.76 & 31.90 & 31.88 & 31.86 & 32.04 & 31.81 & \textcolor{red}{32.07} \\
					& SSIM & 0.8505 & 0.8498 & 0.8508 & 0.8530 & 0.8503 & 0.8519 & 0.8531 & 0.8505 & \textcolor{red}{0.8549} \\
					& LPIPS & 0.2399 & 0.2396 & 23.88 & 0.2350 & 0.2403 & 0.2380 & 0.2363 & 0.2403 & \textcolor{red}{0.2334} \\
					& DISTS & 0.1585 & 0.1602 & 0.1585 & 0.1571 & 0.1604 & 0.1585 & 0.1572 & 0.1591 & \textcolor{red}{0.1528} \\
					& Inference Time & \textcolor{red}{82} & 220 & 153 & 150 & 101 & 113 & 1202 & 824 & 572 \\
					& GPU Memory & 8344 & \textcolor{red}{389} & 566 & 6354 & 3516 & 5764 & 3390 & 5130 & 3500 \\ 
					\midrule[1pt]
					\multirow{6}{*}{$\times$6} & PSNR & 29.49 & 29.44 & 29.49 & 29.62 & 29.57 & 29.56 & 29.73 & 29.49 & \textcolor{red}{29.76} \\
					& SSIM & 0.7782 & 0.7806 & 0.7814 & 0.7841 & 0.7806 & 0.7825 & 0.7848 & 0.7792 & \textcolor{red}{0.7867} \\
					& LPIPS & 0.3274 & 0.3238 & 0.3317 & 0.3275 & 0.3296 & 0.3305 & 0.3217 & 0.3358 & \textcolor{red}{0.3210} \\
					& DISTS & 0.2088 & 0.2129 & 0.2120 & 0.2105 & 0.2147 & 0.2119 & 0.2115 & 0.2129 & \textcolor{red}{0.2072} \\
					& Inference Time & \textcolor{red}{69} & 201 & 136 & 118 & 84 & 94 & 736 & 751 & 284 \\
					& GPU Memory & 8265 & 345 & \textcolor{red}{315} & 6350 & 3475 & 5517 & 1625 & 5301 & 1658 \\ 
					\midrule[1pt]
					\multirow{6}{*}{$\times$8} & PSNR & 28.06 & 28.08 & 28.15 & 28.26 & 28.20 & 28.22 & \textcolor{red}{28.35} & 28.02 & 28.32 \\
					& SSIM & 0.7305 & 0.7353 & 0.7366 & 0.7390 & 0.7350 & 0.7375 & 0.7400 & 0.7309 & \textcolor{red}{0.7414} \\
					& LPIPS & 0.3877 & 0.3816 & 0.3958 & 0.3910 & 0.3911 & 0.3946 & 0.3814 & 0.4039 & \textcolor{red}{0.3799} \\
					& DISTS & \textcolor{red}{0.2425} & 0.2489 & 0.2500 & 0.2477 & 0.2514 & 0.2499 & 0.2478 & 0.2521 & 0.2433 \\
					& Inference Time & \textcolor{red}{55} & 187 & 128 & 107 & 77 & 80 & 633 & 702 & 206 \\
					& GPU Memory & 8236 & 328 & \textcolor{red}{302} & 6348 & 3462 & 5431 & 1583 & 5092 & 1132 \\ 
					\midrule[1pt]
					\multirow{6}{*}{$\times$12} & PSNR & 26.25 & 26.33 & 26.41 & 26.49 & 26.42 & 26.44 & 26.56 & 26.08 & \textcolor{red}{26.58} \\
					& SSIM & 0.6746 & 0.6828 & 0.6838 & 0.6857 & 0.6817 & 0.6842 & 0.6875 & 0.6737 & \textcolor{red}{0.6884} \\
					& LPIPS & 0.4716 & 0.4701 & 0.4896 & 0.4845 & 0.4837 & 0.4868 & 0.4691 & 0.5162 & \textcolor{red}{0.4665} \\
					& DISTS & \textcolor{red}{0.2935} & 0.3022 & 0.3062 & 0.3038 & 0.3071 & 0.3060 & 0.3019 & 0.3141 & 0.2983 \\
					& Inference Time & \textcolor{red}{58} & 178 & 129 & 114 & 78 & 110 & 546 & 701 & 98 \\
					& GPU Memory & 8216 & 317 & \textcolor{red}{293} & 6347 & 3451 & 5369 & 1549 & 5291 & 553 \\
					\bottomrule[2pt]
				\end{tabular}
			}
		}
		\label{tab:computational cost on RDN}
	\end{table*}

	\section{Ablation Study}
	
	We conduct ablation studies on five factors that will affect the final performance of our proposed GSASR: (1) the number of Gaussians $N$, (2) the functionality of the reference position, (3) the rasterization ratio $r$, (4) the dimension $d$ of Gaussian embedding, and (5) the window size $k$ in condition injection block and Gaussian interaction block.
	
	\subsection{Number of Gaussians}
	
	In our implementation, we set the number of Gaussians $N$ be proportional to LR image size ($H \times W$) with $N = m \times (H \times W)$, where $m$ indicates the density of 2D Gaussians. To study the effect of the number of Gaussians $N$ on super-resolution, we set the ratio $m$ to different values: $m=\{1, 4, 9, 16\}$. The experiments are conducted on DIV2K \cite{timofte2017ntire} and the results are shown in Table~\ref{tab:ablation study on Gaussian numbers}. One could find that, with the increase of ratio $m$, our method could obtain better super-resolution results. Those results indicate that introducing more Gaussians could enhance the representation ability of our model. However, considering the computational cost, we set $m=16$ by default.
	
	\begin{table*}[h]
		\centering
		\caption{Ablation study on the number of Gaussians $N$. We utilize the EDSR-backbone \cite{lim2017enhanced} to extract the deep features. DIV2K \cite{timofte2017ntire} dataset is utilized as the testing set. The best results are highlighted in \textcolor{red}{red}.} 
		\scalebox{1.0}{
			\resizebox{1.0\linewidth}{!}{
				\setlength{\extrarowheight}{1pt} 
				\begin{tabular}{c|cccc|cccc|cccc} 
					\toprule[2pt]
					\multirow{4}{*}{\begin{tabular}[c]{@{}c@{}}The Value\\ of ratio m\end{tabular}} & \multicolumn{12}{c}{Backbone:
						EDSR-baseline} \\ 
					\cline{2-13}
					& \multicolumn{12}{c}{Testing Datasets:
						DIV2K} \\ 
					\cline{2-13}
					& \multicolumn{4}{c|}{x2} & \multicolumn{4}{c|}{x4} & \multicolumn{4}{c}{x8} \\
					& PSNR & SSIM & LPIPS & DISTS & PSNR & SSIM & LPIPS & DISTS & PSNR & SSIM & LPIPS & DISTS \\ 
					\midrule[1pt] \midrule[1pt]
					1 & 36.36 & 0.9474 & 0.0818 & 0.0537 & 30.61 & 0.8418 & 0.2689 & 0.1335 & 27.00 & 0.7234 & 0.4378 & 0.2292 \\
					4 & 36.54 & 0.9488 & 0.0780 & 0.0520 & 30.80 & 0.8466 & 0.2561 & 0.1307 & 27.14 & 0.7292 & 0.4166 & 0.2227 \\
					9 & 36.55 & 0.9488 & 0.0781 & 0.0521 & 30.80 & 0.8465 & 0.2555 & 0.1310 & 27.16 & 0.7299 & 0.4138 & 0.2232 \\
					16 & \textcolor{red}{36.65} & \textcolor{red}{0.9495} & \textcolor{red}{0.0767} & \textcolor{red}{0.0514} & \textcolor{red}{30.89} & \textcolor{red}{0.8486} & \textcolor{red}{0.2518} & \textcolor{red}{0.1301} & \textcolor{red}{27.22} & \textcolor{red}{0.7321} & \textcolor{red}{0.4077} & \textcolor{red}{0.2214} \\
					\bottomrule[2pt]
				\end{tabular}
			}
		}
		\label{tab:ablation study on Gaussian numbers}
		\vspace{-0.3cm}
	\end{table*}

	\subsection{Reference Position}
\begin{figure*}[h]
		\centering
		\includegraphics[width=0.8\linewidth]{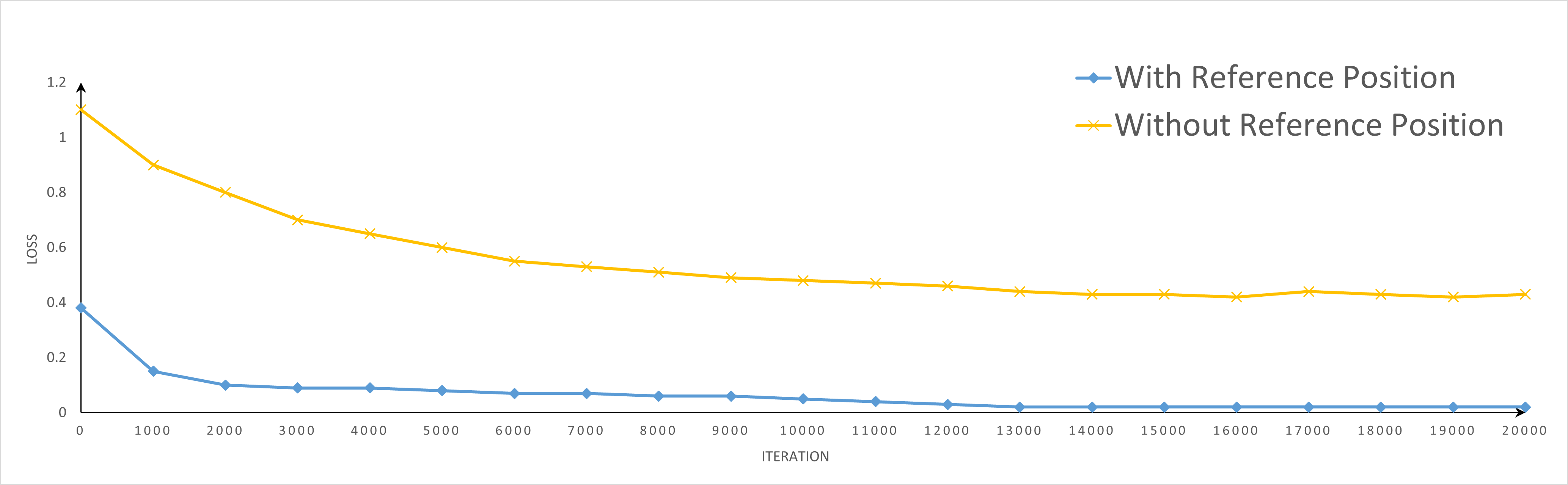}
		\caption{The training loss curves with and without reference position. }
		\label{fig:training loss}
	\end{figure*}

	As discussed in Section 3.2 from the main paper, we assign a reference position $p_i$ to the $i$-th Gaussian embedding $\mathbf{E}_i$, which predicts relative offset $o_i$ to obtain its final position with $\mu_i=p_i+o_i$. We plot and compare the training loss curves to demonstrate the importance of reference position.  
	As shown in Fig.~\ref{fig:training loss}, if we drop the reference position $p$ and make all Gaussian embeddings regress their final positions $\mu$ directly, the training loss will not decrease and stop at around $4e^{-1}$. As the Gaussian embeddings $\mathbf{E}$ are duplicated from the same $\mathbf{E}_{base}$, those embeddings lack the ability to interpret contents at different areas, causing optimization challenges. In contrast, the reference positions can guide different $\mathbf{E}_{base}$ to handle different regions so that the network can be properly optimized with loss decreasing to around $1e^{-2}$.
	
	\subsection{Rasterization Ratio}
	During rasterization, the most straight-forward strategy to render an SR image is querying each pixel from all 2D Gaussians, avoiding missing any Gaussian responses. However, this will lead to a huge computational cost of $\mathcal{O}(s^2HWN)$, where $H \times W$ denotes the size of the input LR image, $s$ indicates the scaling factor, and $N$ means the number of 2D Gaussians. Theoretically, a Gaussian is infinite on the 2D space; however, its response will decrease rapidly from the center to the neighborhood. Actually, each Gaussian contributes to a local area with limited sizes, while the responses could be neglected when the distance to the center is large enough. To accelerate the inference speed and save GPU memory cost, we introduce a rasterization ratio $r \leq 1$ to decrease the computational costs. More specifically, we set $r = 0.1$ to ignore the minor responses, and process all Gaussians in parallel to render pixels which are close to the center. In this way, we speed up the rasterization significantly. The ablation study results of rasterization ratio $r$ could be found in Table~\ref{tab:ablation study of r}. We can see that, by simply setting $r=0.1$, the final performance has little drop compared with $r=1.0$ but with much faster speed. However, if setting $r$ to a too small value ($r=0.01$), one will find a significant performance drop. To balance between the efficiency and performance, we set $r=1.0$ cautiously.

	\subsection{Dimension of Gaussian Embedding}
	The dimension $d$ of Gaussian embedding determines the capacity to capture complex relationships between different embeddings. We conduct ablation study on $d$, and the results could be found in Table~\ref{tab:ablation study of GS width}. As can be observed, the performance becomes saturated when $d \geq 180$. Although there is a slight promotion by setting the dimension $d$ to $256$, it will lead to higher computation cost in the MLP layers of Gaussian interaction block, since the computational complexity is $O(d^{2}N)$, where $N$ is the number of Gaussians. Therefore, We set $d=180$ to balance accuracy and efficiency.
	
	\subsection{Window Size}
	As mentioned in Section 3.2 of the main paper, to dynamically support arbitrary LR size ($H \times W$), we split the LR feature $\mathbf{F}\in \mathbb{R}^{C \times H \times W}$ into $\frac{H}{k}\times \frac{W}{k}$ windows with window size $k\times k$, then we duplicate base Gaussian embedding $\mathbf{E}_{base}\in \mathbb{R}^{mk^2\times d}$  for $\frac{H}{k}\times \frac{W}{k}$ times so as to cover all windows. The windows size in the condition injection block and Gaussian interaction block keeps the same as $k$. The window size determines the ability to capture long range dependency. We conduct ablation study on $k$, and the results could be found in Table~\ref{tab:ablation study of window size}. As can be seen, the larger kernel size, the better performance. However, setting a too high $k$ value will lead to huge computational cost in self attention layers, to balance between accuracy and efficiency, we set it to $12$ in our implementation.

	\begin{table*}[h]
		\centering
		\caption{Ablation study on rasterization ratio $r$. We test  with EDSR-baseline on DIV2K with $720 \times 720$ size. PSNR/SSIM is calculated on Y channel of Ycbcr space.}
		\scalebox{0.9}{
			\resizebox{1.0\linewidth}{!}{
				\begin{tabular}{c|ccccc|ccccc} 
					\bottomrule[2pt]
					Scale & \multicolumn{5}{c|}{x4} & \multicolumn{5}{c}{x8} \\ 
					\midrule[1pt] \midrule[1pt]
					\multirow{2}{*}{\begin{tabular}[c]{@{}c@{}}Performance\\and Cost\end{tabular}} & \multicolumn{5}{c|}{Rasterization Ratio r} & \multicolumn{5}{c}{Rasterization Ratio r} \\
					& 0.01 & 0.1 & 0.4 & 0.8 & 1 & 0.01 & 0.1 & 0.4 & 0.8 & 1 \\ 
					\midrule[1pt]
					PSNR/SSIM & 29.25/0.8202 & 32.01/0.8536 & 32.01/0.8536 & 32.01/0.8536 & 32.01/0.8536 & 15.97/0.5126 & 28.25/0.7397 & 28.25/0.7397 & 28.25/0.7397 & 28.25/0.7397 \\
					Inference Time (ms) & 383 & 543 & 2490 & 6986 & 9419 & 134 & 195 & 888 & 2451 & 3285 \\
					GPU Memory (MB) & 3419 & 3420 & 3421 & 3421 & 3421 & 1051 & 1051 & 1052 & 1052 & 1052 \\
					\bottomrule[2pt]
				\end{tabular}
			}
		}
		\label{tab:ablation study of r}
	\end{table*}

	\begin{table*}[h]
		\centering
		\caption{Ablation study on the dimension of Gaussian embeddings. We test with EDSR-baseline on DIV2K and LSDIR. PSNR/SSIM is calculated on Y channel of Ycbcr space.}
		\scalebox{0.9}{
			\resizebox{1.0\linewidth}{!}{
				\setlength{\extrarowheight}{-5pt} 
				\begin{tabular}{c|cccc|cccc} 
					\toprule[2pt]
					\multirow{3}{*}{Scale} & \multicolumn{4}{c|}{DIV2K} & \multicolumn{4}{c}{LSDIR} \\ 
					& \multicolumn{4}{c|}{Dimension of
						Gaussian Embedding d} & \multicolumn{4}{c}{Dimension of Gaussian Embedding d} \\
					& 64 & 128 & 180 & 256 & 64 & 128 & 180 & 256 \\ 
					\midrule[1pt] \midrule[1pt]
					x4 & 30.66/0.8435 & 30.74/0.8454 & 30.89/0.8486 & \textcolor{red}{30.92/0.8494} & 26.42/0.7689 & 26.49/0.7714 & 26.65/0.7774 & \textcolor{red}{26.69/0.7788} \\
					x8 & 27.06/0.7257 & 27.11/0.7280 & 27.22/0.7321 & \textcolor{red}{27.24/0.7331} & 23.45/0.6182 & 23.49/0.6211 & 23.58/0.6269 & \textcolor{red}{23.60/0.6280} \\
					\bottomrule[2pt]
				\end{tabular}
			}
			\label{tab:ablation study of GS width}
		}
	\end{table*}
	
	\subsection{The Functionality of Learnable Position Parameter $o$}
	
	The position parameter $o$ determines the position of Gaussians in the image space. In our implementation, we set it as a learnable parameter so that the Gaussians can automatically concentrate on complex textures. Here we validate the effectiveness of learnable $o$ by conducting ablation study on fixed parameters. From Table~\ref{tab:ablation study on position parameter}, one could find that if we fix the position parameters, the performance will drop. Therefore, setting learnable position parameters will improve the representation capability.
	
	\subsection{The Functionality of Learnable Standard Deviation Parameter $\sigma$}
	
	The standard deviation parameter $\sigma$ controls the shape of the Gaussian. In our implementation, we set it as a learnable parameter so that the Gaussians can fit the shape of complex textures. Here we validate the effectiveness of learnable $\sigma$ by conducting ablation study on fixed parameters. From Table~\ref{tab:ablation study on standard deviation parameter}, one could find that if we set the standard deviation to fixed values, the performance will drop a lot. Therefore, setting learnable standard deviation parameters will equip the Gaussians with a stronger fitting capability.
	
	\begin{table*}[h]
		\centering
		\caption{Ablation study on the window size in the Gaussian interaction block. We test with EDSR-baseline on DIV2K and LSDIR. PSNR/SSIM is calculated on Y channel of Ycbcr space.}
		\scalebox{0.8}{
			\resizebox{1.0\linewidth}{!}{
				\begin{tabular}{c|ccc|ccc} 
					\toprule[2pt]
					\multirow{3}{*}{Scale} & \multicolumn{3}{c|}{DIV2K} & \multicolumn{3}{c}{LSDIR} \\ 
					\cline{2-7}
					& \multicolumn{3}{c|}{Kernel Size k} & \multicolumn{3}{c}{Kernel Size k} \\
					& 4 & 8 & 12 & 4 & 8 & 12 \\ 
					\midrule[1pt] \midrule[1pt]
					x4 & 30.80/0.8467 & 30.84/0.8478 & \textcolor{red}{30.89/0.8486} & 26.55/0.7739 & 26.60/0.7758 & \textcolor{red}{26.65/0.7774} \\
					x8 & 27.13/0.7297 & 27.19/0.7310 & \textcolor{red}{27.22/0.7321} & 23.51/0.6241 & 23.56/0.6256 & \textcolor{red}{23.58/0.6269} \\
					\bottomrule[2pt]
				\end{tabular}
			}
			\label{tab:ablation study of window size}
		}
	\end{table*}
	
	\section{Performance of GSASR with Larger Backbone}
	In order to explore the performance upper-bound of our proposed GSASR, we substitute the encoder backbone with HAT-L \cite{chen2023activating}. Besides, we utilize the SA-1B \cite{kirillov2023segment} dataset to train our model. To reduce computational burden, we substitute the vanilla self-attention layers from both HAT-L \cite{chen2023activating} and our Gaussian decoder with Flash Attention \cite{dao2022flashattention}. Since the relative position embedding \cite{liu2021swin} in self-attention layers cannot be directly embedded in Flash Attention \cite{dao2022flashattention}, we substitute it with the rotary position embedding \cite{heo2024rotary}. To further lower the computational cost, we train and test our HAT-L based model with Automatic Mixed Precision (AMP) in bfloat16 precision. 
	We train our model with $16$ NVIDIA A100 GPUs for $500,000$ iterations. The mini batch size on a single GPU is set to 8. The input image size is set to $64 \times 64$. We randomly sample the scaling factor $s$ from $[1, 16]$. The initial learning rate is $2e^{-4}$, and it halves at iterations $250,000$, $400,000$, $450,000$, $475,000$. The Adam \cite{kingma2014adam} optimizer is utilized.
	
	The quantitative results are shown in Table~\ref{tab:upper-bound GSASR}. One could find that, towards the PSNR metric, our HAT-L based GSASR outperforms the second best model, \ie, RDN based GSASR, by 1.29dB on the Urban100 dataset under the $\times 4$ scaling factor. Such a great promotion validates that GSASR could fit complex textures with high accuracy.

	\begin{table*}[h]
		\centering
		\caption{Ablation study on the parameters of position $o$.}
		\begin{tabular}{c|cc|cc|cc} 
			\toprule[2pt]
			\multirow{2}{*}{Scale} & \multicolumn{2}{c|}{DIV2K} & \multicolumn{2}{c|}{Urban100} & \multicolumn{2}{c}{LSDIR} \\
			& Fixed & Learnable & Fixed & Learnable & Fixed & Learnable \\ 
			\midrule[1pt] \midrule[1pt]
			x4 & 30.86/0.8481 & \textcolor{red}{30.89/0.8486} & 26.91/0.8118 & \textcolor{red}{27.01/0.8142} & 26.62/0.7767 & \textcolor{red}{26.65/0.7774} \\
			x8 & 27.18/0.7303 & \textcolor{red}{27.22/0.7321} & 22.97/0.6414 & \textcolor{red}{23.09/0.6480} & 23.55/0.6243 & \textcolor{red}{23.58/0.6269} \\
			\bottomrule[2pt]
		\end{tabular}
		\label{tab:ablation study on position parameter}
	\end{table*}

	\begin{table*}[h]
		\centering
		\caption{Ablation study on learnable parameters of standard deviation $\sigma$.}
		\scalebox{1.0}{
			\resizebox{1.0\linewidth}{!}{
				\begin{tabular}{c|cccc|cccc} 
					\toprule[2pt]
					\multirow{3}{*}{Scale} & \multicolumn{4}{c|}{DIV2K} & \multicolumn{4}{c}{LSDIR} \\
					& \multicolumn{4}{c|}{Standard deviation} & \multicolumn{4}{c}{Standard deviation} \\
					& 0.1 & 0.4 & 0.8 & Learnable & 0.1 & 0.4 & 0.8 & Learnable \\ 
					\midrule[1pt] \midrule[1pt]
					x4 & 20.87/0.5932 & 30.76/0.8044 & 29.93/0.8307 & \textcolor{red}{30.89/0.8486} & 20.87/0.5932 & 30.76/0.8044 & 29.93/0.8307 & \textcolor{red}{30.89/0.8486} \\
					x8 & 9.50/0.0213 & 27.06/0.7271 & 26.45/0.7093 & \textcolor{red}{27.22/0.7321} & 9.50/0.0213 & 23.45/0.6187 & 22.96/0.5931 & \textcolor{red}{23.58/0.6269} \\
					\bottomrule[2pt]
				\end{tabular}
			}
			\label{tab:ablation study on standard deviation parameter}
		}
	\end{table*}
	
	\begin{table*}[h]
		\centering
		\caption{Performance of the upper-bound version of GSASR. PSNR/SSIM is calculated on Y channel of Ycbcr space.}
		\scalebox{1.0}{
			\resizebox{1.0\linewidth}{!}{
				\begin{tabular}{c|c|c|c|c|c} 
					\toprule[2pt]
					\multirow{2}{*}{\begin{tabular}[c]{@{}c@{}}Encoder \\Backbone\end{tabular}} & \multirow{2}{*}{Method} & \multirow{2}{*}{\begin{tabular}[c]{@{}c@{}}Training \\Dataset\end{tabular}} & \multicolumn{3}{c}{PSNR/SSIM/LPIPS/DIST (x4 scaling factor)} \\ 
					\cline{4-6}
					&  &  & DIV2K & LSDIR & Urban100 \\ 
					\midrule[1pt] \midrule[1pt]
					\multirow{4}{*}{\begin{tabular}[c]{@{}c@{}}EDSR-\\baseline\end{tabular}} & LIIF & DIV2K & 30.43/0.8388/0.2662/0.1403 & 26.21/0.7614/0.2978/0.1678 & 26.14/0.7885/0.2271/0.1738 \\
					& GaussianSR & DIV2K & 30.46/0.8389/0.2684/0.1406 & 26.23/0.7615/0.3007/0.1679 & 26.19/0.7893/0.2283/0.1730 \\
					& CiaoSR & DIV2K & 30.67/0.8431/0.2585/0.1370 & 26.42/0.7681/0.2865/0.1631 & 26.69/0.8091/0.2078/0.1659 \\
					& GSASR & DIV2K & 30.89/0.8486/0.2518/0.1301 & 26.65/0.7774/0.2777/0.1554 & 27.01/0.8142/0.1987/0.1552 \\ 
					\midrule[1pt]
					\multirow{4}{*}{RDN} & LIIF & DIV2K & 30.71/0.8449/0.2566/0.1354 & 26.48/0.7714/0.2838/0.1603 & 26.71/0.8055/0.2062/0.1562 \\
					& GaussianSR & DIV2K & 30.76/0.8457/0.2570/0.1347 & 26.53/0.7727/0.2837/0.1595 & 26.77/0.8064/0.2069/0.1610 \\
					& CiaoSR & DIV2K & 30.91/0.8481/0.2525/0.1327 & 26.66/0.7770/0.2768/0.1563 & 27.10/0.8142/0.1966/0.1559 \\
					& GSASR & DIV2K & 30.96/0.8500/0.2505/0.1288 & 26.73/0.7801/0.2752/0.1533 & 27.15/0.8177/0.1953/0.1515 \\ 
					\midrule[1pt]
					HAT-L & GSASR & SA-1B & \textcolor{red}{31.31/0.8570/0.2381/0.1268} & \textcolor{red}{27.17/0.7948/0.2548/0.1470} & \textcolor{red}{28.44/0.8493/0.1580/0.1394} \\
					\bottomrule[2pt]
				\end{tabular}
			}
			\label{tab:upper-bound GSASR}
		}
	\end{table*}

	\clearpage

	{
		\small
		\bibliographystyle{ieeenat_fullname}
		\bibliography{supp}
	}